\pdfoutput=1

\documentclass{siamltex}

\usepackage{thumbpdf}
\usepackage[%
  colorlinks%
  ,bookmarks={false}%
  ,pdftitle={}%
  ,pdfsubject={}%
  ,pdfkeywords={}%
  ,pdfauthor={Vladyslav Gapyak <vladyslav.gapyak@h-da.de>}%
  ]{hyperref}

\usepackage[numbers]{natbib}
\usepackage{a4wide}
\usepackage[a4paper, left=2.5cm, right=2.5cm, top=3cm, bottom=3cm]{geometry}
\usepackage{amsmath}
\usepackage{amssymb}
\usepackage{mathpazo}
\usepackage{graphicx}
\usepackage{enumerate}
\usepackage[utf8,latin1]{inputenc}
\usepackage[T1]{fontenc} 

\usepackage{mathtools}
\usepackage{xcolor}
\usepackage{caption}
\usepackage{subcaption}
\usepackage{tabularx}
\usepackage{multirow}
\usepackage{booktabs}
\usepackage{array}
\usepackage[percent]{overpic}
\usepackage{bm}
\usepackage{algpseudocode}
\usepackage{algorithm}
\usepackage{pgffor}
\usepackage{placeins}
\newcommand{\beginsupplement}{%
	\setcounter{subsection}{0}
	\renewcommand{\thesubsection}{S\arabic{subsection}}%
	\setcounter{table}{0}
	\renewcommand{\thetable}{S\arabic{table}}%
	\setcounter{figure}{0}
	\renewcommand{\thefigure}{S\arabic{figure}}%
}

\newcommand{\rom}[1]{%
	\textup{\uppercase\expandafter{\romannumeral#1}}%
}

\title{An $\ell^1$-Plug-and-Play Approach for MPI Using a Zero Shot Denoiser with Evaluation on the 3D Open MPI Dataset}

\author{Vladyslav Gapyak\footnotemark[1] \and Corinna E. Rentschler\footnotemark[1] \and Thomas M\"{a}rz\footnotemark[1] \and Andreas Weinmann\footnotemark[1]\ }

\begin{document}
\maketitle
\renewcommand{\thefootnote}{\fnsymbol{footnote}}
\footnotetext[1]{Department of Mathematics and Natural Sciences, Hochschule Darmstadt, Sch\"{o}fferstra{\ss}e 3, 64295, Darmstadt, Germany. Emails: vladyslav.gapyak@h-da.de, thomas.maerz@h-da.de, andreas.weinmann@h-da.de}
\renewcommand{\thefootnote}{\arabic{footnote}}

\begin{abstract}
    Objective: Magnetic Particle Imaging (MPI) is an emerging medical imaging modality which has gained increasing interest in recent years.
    Among the benefits of MPI are its high temporal resolution, and that the technique does not expose the specimen to any kind of ionizing radiation. 
    It is based on the non-linear response of magnetic nanoparticles to an applied magnetic field. 
    From the electric signal measured in receive coils, the particle concentration has to be reconstructed.
    Due to the ill-posedness of the reconstruction problem, various regularization methods have been proposed for reconstruction ranging from early stopping methods, via classical Tikhonov regularization and iterative methods to modern machine learning approaches. In this work, we contribute to the latter class: we propose a Plug-and-Play approach based on a generic zero-shot denoiser with an $\ell^1$-prior.
    
    Approach: We validate the reconstruction parameters of the method on a hybrid dataset and compare it with the baseline Tikhonov, ART, DIP and the previous PP-MPI, which is a Plug-and-Play method with denoiser trained on MPI-friendly data.
    
    Main results: We derive a Plug-and-Play reconstruction method based on a generic zero-shot denoiser. Addressing (hyper)parameter selection, we perform an extended parameter search on a hybrid validation dataset we produced and apply the derived parameters for reconstruction on the 3D Open MPI Dataset. We offer a quantitative and qualitative evaluation of the zero-shot Plug-and-Play approach on the 3D Open MPI dataset with the validated parameters. Moreover, we show the quality of the approach with different levels of preprocessing of the data.
    
    Significance: The proposed method employs a zero-shot denoiser which has not been trained for the MPI reconstruction task and theqrefore saves the cost for training. Moreover, it offers a method that can be potentially applied in future MPI contexts.
\end{abstract}


\section{Introduction}
\label{sec:introduction}
Magnetic Particle Imaging (MPI) is an emerging medical imaging modality that images the distribution of superparamagnetic nanoparticles by exploiting their non-linear magnetization response to dynamic magnetic fields. 
The principles behind MPI have been introduced in the Nature paper by Gleich and Weizenecker~\cite{GleichWeizenecker2005}. 
Later, Weizenecker \emph{et al.}~\cite{Weizenecker_etal2009} produced a video (3D+time) of a distribution of a clinically approved MRI contrast agent flowing through the beating heart of a mouse, thereby taking an important step towards medical applications of MPI. 
Further, Gleich, Weizenecker and their team won the European Inventor Award in 2016 for their contributions to MPI. 
Since then, interest in the technology has been on the rise both with view towards applications and with view towards improving the technology on the hardware side as well as on the algorithmic side.

There are various medical applications such as cancer detection \cite{Song2018} and cancer imaging~\cite{Du2019,Tay2021,Yu2017}, stem cell tracing~\cite{connell2015advancedcellTherapies,JUNG2018139,Lemaster2018}, inflammation tracing and lung perfusion imaging~\cite{Zhou_2017}, drug delivery and monitoring~\cite{Zhu2019}, cardiovascular~\cite{Bakenecker2018MPIvascular,Tong2021,Vaalma2017} and blood flow~\cite{Franke2020BloodFlow} imaging, tracking of medical instruments~\cite{haegele2012instrumentvisualization}
as well as  brain injury detection~\cite{Orendorff2017} and MPI-assisted stroke detection and brain monitoring based on recent human-sized MPI scanners~\cite{Graeser2019humanbrain}. MPI can be also used within multimodal imaging~\cite{aramiMultimodalImagingMPI}. Benefits of MPI are quantifyability and high sensitivity, and compared with other imaging modalities such as CT~\cite{Buzug2008CTbook}, MRI~\cite{schamelsafety2015}, PET~\cite{PET} and SPECT~\cite{SPECT}, its ability to provide higher spatial resolution in a shorter acquisition time and the absence of ionizing radiation or radioactive tracers~\cite{BuzugKnopp2012}. Further details concerning comparison with existing imaging modalities and further examples of state-of-the-art applications can be found in~\cite{billingsMPIapplications,Yang2022}.

In brief, the principle and the procedure behind MPI are as follows: 
(i) the specimen to be scanned is injected with a tracer containing the magnetic nanoparticles (MNs) and placed in the scanner; 
(ii) a dynamic magnetic field with a field-free region (FFR), e.g., a field-free point, is applied and the FFR is spatially moved; 
(iii) the FFR acts as a sensitive region whose motion causes the particles to induce a voltage in the receive coils. The imaging task of MPI is the reconstruction of the spatial distribution of the particles from the voltage signal acquired during a scan.
There are currently two main types of approaches to reconstructing particle distributions: 
measurement-based approaches and model-based approaches. 
Measurement-based approaches typically acquire a system matrix: 
for each column of the matrix, one performs scanning of a delta probe located in a corresponding voxel 
of a chosen 3D voxel grid~\cite{knopp2011prediction,WeizeneckerBorgertGleich2007,Rahmer_etal2012,Lampe_etal2012}. 
More precisely, a 3D grid of the volume to be scanned inside the scanner's bore is considered; a probe with a reference concentration of tracer is iteratively positioned and scanned at each voxel of the grid. 
This way the response of the scanning system to discrete delta impulses at each voxel position is collected and stored in the system matrix $A$. 
Assuming linearity, the acquired signal $f$ of the scan of a specimen is then obtained as a superpositon of the scans of the delta impulses.
For reconstruction, one has to solve a corresponding system of equations $A u = f, $ where the symbol $u$ denotes the desired concentration distribution; loosely speaking, one has to invert the system matrix for the measured signal $f$. Due to the presence of noise and the ill-conditioning of the system matrix (e.g., \cite{Knopp_etal2010ec,storath2016edge}), regularization techniques are needed for the inversion as discussed later.
The development of model-based reconstruction techniques is a direction of research aiming at avoiding or significantly reducing the calibration procedures needed to measure the system matrix; for details, cf. for instance~\cite{Rahemeretal2009,Schomberg2010,GoodwillConolly2010,GoodwillConolly2011,Gruettner_etal2013, marz2016model, bringout2020new,maass2024equilibriumanysotropy}. 
In this work, we consider the measurement approach and focus on regularized machine learning-based reconstruction in that scenario.

\paragraph{Related Work.} 

Due to the ill-conditioning of the system matrix
and the related ill-posedness of the reconstruction problem in MPI \cite{knopp2008singular,marz2016model,kluth2018degree},
small perturbations in the data can lead to significant errors in the reconstruction;
thus, regularization is needed  \cite{bertero2021introduction}.
Classical regularization approaches in MPI include Tikhonov regularization 
\cite{Weizenecker_etal2009,Knopp_etal2010ec}
using conjugate gradients and the Algebraic Reconstruction Technique (ART), also known as the Kaczmarz method. 
To improve noise suppression capabilities and enhance the reconstruction of features such as shape, edges, corners etc.,
structural priors based on the $\ell^1$-norm have been used within iterative schemes~\cite{storath2016edge,nawwas2021reduction}, as well as for sparse recovery of system matrices~\cite{grosser2021sparse}.
For a broader overview we refer to~\cite{Daubechies2003AnIT}.  

In recent years, there is increasing interest in reconstruction methods involving machine learning approaches. 
Among these are methods learning the reconstruction by training a neural network, e.g. \cite{hatsuda2015basic,vonGladiss2022reconstruction}.
Further, reconstruction based on the Deep Image Prior (DIP) has been proposed~\cite{dittmer2020deep,ulyanov2018deep}.
Recently, Deep Equilibrium Models have been employed in the context of MPI~\cite{gungor2023deqmpi, bai2019deep}.
Deep learning approaches have been further used to improve the spatial resolution of the reconstructions~\cite{Shang_2022},
and to reduce the calibration time, for example by partially acquiring a system matrix on a down-sampled grid and use deep-learning for super-resolution of the system matrix~\cite{gungor2021superresolution,Schrank2022superresolution,gungor2022tranSMS,Yin2023dipSM}. 

A particular class of machine learning-based approaches are Plug-and-Play (PnP) approaches first introduced by Venkatakrishnan et al.~\cite{venkatakrishnan2013pnp}.
They are inspired by iterative schemes realizing energy minimization: from a Bayesian perspective, the reconstruction of an image $x$ from (noisy) measurements $y$ can be achieved through Maximum a Posterior Estimate (MAP), i.e., by maximizing the conditional probability $\mathbb{P}( x | y)$; this maximization problem is equivalent to the minimization of the an energy functional of the type $\ell (y,x)+\lambda\mathcal{R} (x)$ where $\ell (y,x)$ is the log-likelihood and $\mathcal{R}(x)$ is a prior. By considering the log-likelihood in terms of a variable $x_1$ and the prior in terms of a variable $x_2$ and adding the constraint $x_1 = x_2$, it is possible to formulate an equivalent but constrained minimization problem \cite{venkatakrishnan2013pnp}, which can be solved with a chosen splitting scheme in an alternated fashion, alternating between a ridge regression problem and a Gaussian denoising problem;
instead of classical denoising, typically machine learning-based denoisers are employed~\cite{Zhang2022pnp} in the Gaussian denoising subproblems of the splitting scheme.
PnP approaches have been used for a variety of imaging modalitites, e.g.,~\cite{Ahmad2020pnpMRI, Sreehari2016pnpelectrontomography}
as well as for computer vision tasks, e.g., \cite{Zhang2022pnp, sun2019online}.
These approaches are conservative in the sense that machine learning only enters the scheme via iterative denoising and that well-known and well-understood components from iterative schemes and classical Tikhonov regularization are still integrated into the reconstruction scheme.  

Most closely related to the present work is the paper~\cite{askin2022pnp} which also considers a Plug-and-Play prior for MPI reconstruction. Our approach differs in the underlying iterative scheme and the denoising method used, which is at the heart of a PnP approach, and in the employment of an additional $\ell^1$-prior. We further provide both quantitative and qualitative evaluation on the full 3D Open MPI data set \cite{knopp2020openmpidata}. A detailed discussion concerning the novelties compared to \cite{askin2022pnp} can be found in the discussion in Section~\ref{sec:discussion}.

\paragraph{Contribution.} 
In this work we consider a PnP approach for MPI reconstruction
based on a generic (zero-shot) denoiser, a half quadratic splitting scheme plus an additional $\ell^1$-prior.
Moreover, we validate the parameters on a hybrid dataset and evaluate the proposed methods  
on the publicly available Open MPI dataset.
In detail, we make the following contributions:
\begin{itemize}
	\item[(i)] In the proposed PnP method we employ a generic (zero-shot) image denoiser. 
	This means the denoiser is not transfer-learned on MPI or similar MRI data. 
	(To apply the 2D image denoiser on a 3D volume, we use slicing.) 	We further employ a shrinkage/$\ell^1$-prior which is motivated by the works \cite{storath2016edge,kluth2020l1data,nawwas2021reduction,grosser2021sparse} which use a corresponding prior in an energy minimization framework in MPI reconstruction. 
	\item[(ii)] We create a hybrid 3D MPI dataset by applying the system matrix in the Open MPI Dataset~\cite{knopp2020openmpidata} to simulated phantoms as in~\cite{knopp2023adeeplearningapproach,zhao2024mpigan,gungor2023deqmpi}; we use this dataset to choose the reconstruction parameters that we later apply to real MPI data and hence provide a way to fix reconstruction parameters for real MPI data. Avoiding fine tuning of the hyperparameters on the data is beneficial in real world applications, as the ground truth is not available. In particular, we show here that the ZeroShot-$\ell^1$-PnP and ZeroShot-PnP algorithm obtain competitive results without tuning of the hyperparmeters on the data and with one common set of hyperparameters.
	\item[(iii)] We quantitatively and qualitatively evaluate our approach on the full 3D Open MPI dataset \cite{knopp2020openmpidata} 
	which presently represents the only publicly available real 3D MPI dataset. In particular, we quantitatively and qualitatively compare our method with the DIP approach~\cite{dittmer2020deep} using the preprocessing pipeline in~\cite{kluth2019enhancedrec}, with the classical Tikhonov regularization, the ART method~\cite{kluth2019enhancedrec,Knopp_etal2010ec}, and the previous PnP approach~\cite{askin2022pnp}. Moreover, we investigate the influence of the preprocessing on the reconstruction quality. 
\end{itemize}
Both quantitative and qualitative evaluation of a PnP approach (including comparison with another machine learning approach) on the full 3D Open MPI dataset has not been done before.
(We note that quantitative evaluation is possible on the Open MPI dataset since a reference CAD model is available in the dataset and serves the purpose of a ground truth~\cite{knopp2020openmpidata}; we remark however, that such CAD ground truths are used exclusively for quantitative evaluation of the results and not for the choice of reconstruction parameters, which we perform on the hybrid dataset.) Furthermore, employing a zero-shot denoiser~\cite{Zhang2022pnp} (trained on a large set of natural images) in the context of MPI is new. 

\paragraph{Outline.} 
In Section~\ref{sec:Methods} we develop the method: we describe the formal MPI reconstruction problem (Section~\ref{sec:RecSys}), the preprocessing pipeline of the MPI data (Section~\ref{sec:Preproc}), we present the proposed ZeroShot-PnP approach with and without the additional $\ell^1$-prior (Section~\ref{sec:PnPprop}), we consider parameter selection in Section~\ref{sec:param:selection} and describe the hybrid dataset created for the validation of the parameters in Section~\ref{sec:simul:dataset}. In Section~\ref{sec:Results} we quantitatively and qualitatively evaluate the proposed ZeroShot-$\ell^1$-PnP algorithm on the OpenMPI Dataset: we compare  with the DIP, Tikhonov regularization, the ART method and the previous Plug-and-Play approach PP-MPI (Section~\ref{eq:EvalOnOpenMPI} and~\ref{sec:exp1bis}),
investigate the influence of different levels of preprocessing (Section~\ref{sec:exp2})
and consider a setup with most basic preprocessing consisting only of background removal and suppressing of excitation crosstalk in Section~\ref{sec:exp4}.
We conclude with Section~\ref{sec:discussionAndconclusion} which contains a discussion and a conclusion part.

\section{Method}\label{sec:Methods}

We first recall the formal problem of MPI reconstruction based on a system matrix approach in Section~\ref{sec:RecSys}.
Then we discuss the employed preprocessing in Section~\ref{sec:Preproc}.
Finally, we derive the proposed ZeroShot-$\ell^1$-PnP approach using a zero-shot denoiser in Section~\ref{sec:PnPprop}
and consider parameter selection in Section~\ref{sec:param:selection}.

\subsection{The Reconstruction Problem in MPI Using a System Matrix} \label{sec:RecSys}

The task of MPI reconstruction is to determine the concentration $u$ of iron
oxide nanoparticles contained in an object given the measured signal $f$ during a scan. The relation between the concentration $u$ and the measured data $f$ is modeled as a linear mapping
called the system function or system matrix $A$. In order to determine the system matrix/function both model and measurement-based approaches have been applied; cf. \cite{Rahmer_etal2012,KnoppBiederer_etal2010} for instance.
At present, the measurement-based approach is the most popularly used in real data scenarios.
We consider MPI reconstruction in three dimesional space using real measurement data from the Open MPI dataset \cite{knopp2020openmpidata}. 
Hence, the reconstructed 3D concentration $u_\mathrm{rec}$
lives in the space  $X = \mathbb R^{n_1}\times\mathbb{R}^{n_2}\times\mathbb{R}^{n_3},$
where $n_1=n_2=n_3=19$ for the Open MPI data set.
We denote the total number of voxels by $N = n_1n_2n_3 = 19^3,$ and the number of frequency components by $M$ (possibly upon transformation of the data if provided in time domain into the Fourier domain); 
here $M$ depends on the preprocessing applied to the MPI data (cf. Section~\ref{sec:Preproc}). In particular, a scan in the Open MPI dataset contains the data coming from three different channels, i.e., from two couples of perpendicular receiving coils coaxial to the y and z dimensions, a solenoid in the x direction, and the calibration data is Fourier transformed if provided in time domain. This means that for each of the three channels the real and imaginary parts of the Fourier coefficients are considered and stacked, yielding a real-valued submatrix $A_l$, where $l=1,2,3$ is the channel index. Finally, the full matrix $A$ is formed by stacking the submatrices $A_l$ for each channel $l$ (for a detailed explanation we refer the interested reader to~\cite{kluth2019enhancedrec}).
In order to avoid notational issues, we use the following conventions: $u_{\mathrm{GT}} \in \mathbb R^N$ denotes the (columnwise) vectorization 
of the ground truth spatial distribution, we will use a tilde to denote the 3D or 2D reshaping of $u_\mathrm{GT}$, i.e., $\tilde u \in \mathbb R^{n_1 \times n_2 \times n_3}$. We identify the adjoint $A^\ast$ of $A$ with the matrix $A^T$. Using this notation, we have the relation 
\begin{equation}\label{eq:ForwardMPISysMat}
	f = A u_{\mathrm{GT}} + \eta ,
\end{equation}
where $f$ is the measured data, and the symbol $\eta$ denotes the noise on the measurements which affects both $f$ and each column of the system matrix $A$. The reconstruction task consists of determining a distribution $u_{\mathrm{rec}}$ approximating $u_{\mathrm{GT}}$ from \eqref{eq:ForwardMPISysMat} given measurements $f$ (and calibration system matrix $A$). This is challenging due to noise and the ill-conditioning of $A$, and requires regularization.

\subsection{The Preprocessing of the Data} \label{sec:Preproc}

We briefly discuss the preprocessing of the Open MPI data employed.
It is inspired but not identical to the preprocessing proposed in~\cite{kluth2019enhancedrec}.
For this reason, and since we discuss our scheme w.r.t. different levels of preprocessing in Section~\ref{sec:Results}, we briefly explain the used method. 

Preprocessing is applied to both calibration and phantom scan data.
Since the calibration data are stored in the Fourier domain we Fourier transform the scan data (given as a time series). 
We further split the real and imaginary parts of the Fourier coefficients for both calibration and phantom data. Although each of the preprocessing steps may decrease the number of rows of $A$, we always use the same symbol $M$ to denote these numbers. 

\paragraph{Bandpass Filtering and Frequency Selection.} During the collection of scan data, an analog filter is in place to suppress excitation crosstalk; consequently, the Signal-to-Noise ratio of frequencies below about 80 kHz is reduced~\cite{Rahmer_etal2012}. For this reason, we always exclude frequencies lower than the 80 kHz threshold.
Further frequency selection avoiding frequencies with systematic strong noise or with low signal-to-noise ratio (SNR) has been considered in \cite{kluth2019enhancedrec} and are part of the possible preprocessing steps. To reduce the number of variables at play and to avoid estimations of the SNR we consider no SNR-based thresholding in the preprocessing here.

\paragraph{Background Correction.} The noisy measured phantom data $f$ is of the form $f = f_{\mathrm{cl.}}+b_f + \eta_f$, with clean signal $f_{\mathrm{cl.}}$ containing the information on $u,$
the background signal $b_f$ produced by the drive field and perturbations in the measurement chain, and noise $\eta_f$ (assuming it is additive in nature). 
Accordingly, we decompose the system matrix as $A = A_{\mathrm{cl}} + b_A +\eta_A,$ 
with clean signals $A_{\mathrm{cl}},$ background $b_A$ and noise $\eta_A.$
To exclude the background signals $b_f,$ $b_A$, background measurements with no phantom in the scanner are performed after the scan and during the calibration procedure. To account for the background signal to be dynamic, in the calibration procedure a background scan is repeated after every $19$th delta probe scan (after running through the $x$-axis) and a corresponding convex combination is subtracted.  Each phantom and delta scan is performed $1000$ times and the average is taken as input for the reconstruction. By linearity, the averaging and subtraction is in fact carried out in the Fourier domain. We refer to~\cite{kluth2019enhancedrec} for more details.

\paragraph{Whitening Estimation and Low Rank Approximation of the System Matrix.} Following~\cite{dittmer2020deep}, we consider a whitening matrix $W$ obtained from the diagonal covariance matrix of multiple background scans, i.e., we consider the diagonal matrix $W\in\mathbb{R}^{M\times M}$ whose diagonal entries are $1/\sigma_b$ where $\sigma_b^2$ is the variance of the background scans provided. When whitening is applied, we consider the whitened system matrix $WA$ and the whitened data $Wf$ instead of $A$ and $f$.

The low rank approximation is obtained by selecting $K\leq\min\lbrace M,N\rbrace$ and by performing a randomized Singular Value Decomposition (rSVD)~\cite{kluth2019enhancedrec,halko2011rSVD}, yielding the triple of matrices $(\tilde U_K ,\tilde\Sigma_K ,\tilde V_K )$ and considering for the reconstruction the transformed matrix $\tilde U_K^\ast A \in\mathbb{R}^{K\times N}$ and the transformed data $\tilde f = \tilde U_K^\ast f\in\mathbb{R}^{K}$. Because we aim at using the rSVD obtained we store $\tilde V_K$ and $\mathrm{diag}(\tilde \Sigma_K)\in\mathbb{R}^K$ as well as $\tilde U_K^\ast A$ and $\tilde U_K^\ast f$. 

\subsection{ZeroShot-PnP: The Proposed Plug-and-Play Approach with a Zero-Shot Denoiser.}  \label{sec:PnPprop}

We now describe the proposed ZeroShot-$\ell^1$-PnP approach and its variant, the ZeroShot-PnP approach. The regularized inversion of the ill-posed MPI reconstruction problem in Equation~\eqref{eq:ForwardMPISysMat}
can be formulated as a minimization problem of the form
\begin{equation}\label{eq:Min:Functional:std}
	u_\mathrm{rec} = \arg\min_{u}\lVert f - Au\rVert_2^2 + \lambda\mathcal{R}(u), 
\end{equation}
where $\mathcal{R}$ is a regularizer and $\lambda >0$ is a parameter controlling the strength of the regularization. In the context of MPI, adding positivity constraints is motivated by the fact that the solutions $u_\mathrm{rec}$ in Eq.~\eqref{eq:Min:Functional:std} represent particle distribution, and additional $\ell^1$-priors have turned out to be effective in MPI thanks to their sparsifying effect on the solution~\cite{storath2016edge,kluth2020l1data,nawwas2021reduction,jin2012sparsity}. Adding positivity and sparsity constraints  results in a minimization problem of the form
\begin{equation}\label{eq:Min:Functional:priors}
	u_\mathrm{rec} = \arg\min_{u}\lVert f - Au\rVert_2^2 + \lambda\mathcal{R}(u) + \alpha \lVert u\rVert_1 + \iota_+ (u), 
\end{equation}
where $\iota_+$ is the indicator function defined as $\iota_+ (u)=0$ if the components $u_i$ of $u$ are all $u_i >0$ and $\iota_+ (u)=+\infty$ otherwise, and $\alpha > 0$ is an additional regularization parameter controlling strength of the $\ell^1$-prior.

\paragraph{The Iterative Scheme.}

Choosing auxiliary variables $u_1$, $u_2$ and $u_3$, we decouple the quantities in Equation~\eqref{eq:Min:Functional:priors}, in particular, we choose the following equivalent constrained minimization problem in consensus form:
\begin{align}\label{eq:Min:Functional:consensus}
	\min_{u_1, u_2 ,u_3} \quad & \underbrace{\lVert f - Au_1\rVert_2^2 + \lambda\mathcal{R}(u_2) + \alpha \lVert u_3\rVert_1 + \iota_+ (u_2 )}_{=: E(u_1 ,u_2 ,u_3 )}\\
	\text{s.t.} \quad & u_1 - u_2 = 0 \quad\text{and}\quad u_1 - u_3 = 0 .
\end{align}
We now perform a Half Quadratic Splitting (HQS) by considering the Lagrangian of the following form $\mathcal{L}_\mu (u_1 ,u_2 ,u_3 ) = E(u_1 ,u_2 ,u_3 ) + \frac{\mu}{2}\lVert u_1 -u_2\rVert_2^2 + \frac{\mu}{2}\lVert u_1 -u_3\rVert_2^2$ with parameter $\mu$. The HQS here chosen in its scaled dual form yields the following iterative scheme:
\begin{align}
	u_1^{k+1} & = \arg\min_{u_1}\lVert f-Au_1\rVert_2^2 + \mu_k\left\lVert u_1-\frac{u_2^k + u_3^k}{2}\right\rVert_2^2 \label{eq:PnP:split:data}\\
	u_2^{k+1} & = \arg\min_{u_2} \frac{1}{2(\lambda /\mu_k )}\rVert u_2-u_1^{k+1}\rVert^2 + \mathcal{R}(u_2 ) + \iota_+ (u_2 ) \label{eq:PnP:split:denoise} \\
	u_3^{k+1} & = \arg\min_{u_3} \frac{\alpha}{\mu_k}\lVert u_3\rVert_1 + \frac{1}{2}\lVert u_1^{k+1}-u_3\rVert_2^2 \quad .
	\label{eq:PnP:split:shrink}
\end{align}
We follow~\cite{Zhang2022pnp} and employ a continuation strategy, i.e., a varying $\mu_k$ in Equation~\eqref{eq:PnP:split:data}; using parameters varying in each iteration have already been employed in non MPI-related contexts, for example in~\cite{chan2017pnp}. The continuation strategy is one of the differences with the previous PP-MPI approach~\cite{askin2022pnp}. Concerning the HQS splitting, it is possible to assign a helpful interpretation to the subproblems: at each iteration $k$, the term in Equation~\eqref{eq:PnP:split:data} is in charge of the regularized inversion of the system matrix whereas the term in Equation~\eqref{eq:PnP:split:denoise} performs regularized Gaussian denoising with positivity constraint; if we define the combined parameter $\sigma_{k+1}^2\coloneq \lambda /\mu_k$, because Equation~\eqref{eq:PnP:split:denoise} is equivalent in form to a Gaussian denoising problem, the parameter $\sigma_{k+1}^2$ can be interpreted as the variance of the noise of $u_1^{k+1}$ and $\sigma_{k+1}=\sqrt{\lambda /\mu_k}$ as its noise level.
Finally, the term in Equation~\eqref{eq:PnP:split:shrink} is the proximal mapping of the $\ell^1$-norm, i.e., it results in a soft-thresholding of parameter $\alpha / \mu_k$.

\paragraph{The Denoising Step.}

The idea behind the PnP approach is to substitute the denoising problem in Equation~\eqref{eq:PnP:split:denoise} by any other denoiser with positive output and w.r.t.\ noise level $\sqrt{\lambda /\mu_k}$, yielding the following algorithm:
\begin{align}
	u_1^{k+1} & = \arg\min_{u_1}\lVert f-Au_1\rVert_2^2 +  \mu_k\left\lVert u_1-\frac{u_2^k + u_3^k}{2}\right\rVert_2^2 \label{eq:PnP:split:data2}\\
	u_2^{k+1} & = \mathrm{Denoiser}\left (u_1^{k+1}\, , \sqrt{\lambda /\mu_k}\right )  \label{eq:PnP:split:denoise2} \\
	u_3^{k+1} & = \mathrm{prox}_{\frac{\alpha}{\mu_k}\lVert\bullet\rVert_1}(u_1^{k+1}). \label{eq:PnP:split:st}
\end{align}
The choice of a suitable denoiser in Equation~\eqref{eq:PnP:split:denoise2} is key because the performance of the algorithm depends on it. 
In Computer Vision, machine learning based 2D denoisers have gained a lot of interest in recent years. 
They are typically trained on large datasets and possess a higher generalization ability than smaller models~\cite{Zhang2022pnp}.

The denoiser we employ in this work is the benchmark {\em deep denoiser prior}~\cite{Zhang2022pnp}. 
It is based on a deep CNN architecture (DRUNet) which combines a U-Net~\cite{Ronneberger2015unet} and ResNet~\cite{He2016resnet}. The denoiser is publicly available and trained on a large dataset that combines various datasets including BSD~\cite{Chen2017bsd}, Waterloo Exploration Database~\cite{Ma2017Waterloo}, DIV2K~\cite{Agustsson2017ntire} and Flick2K~\cite{Lim2017Flick2k}. During training of the deep denoiser prior, the authors in \cite{Zhang2022pnp} considered 16 patches of size $128\times 128$ randomly cropped out of the training dataset (image in the range $[0,255]$) and added to them additive white Gaussian noise with noise level $\sigma$ randomly chosen from $[0,50]$, to account for big variations in the noise level. In particular, the deep denoiser prior takes as input the noisy image and a noise level map, which is a uniform map filled with the value $\sigma$ and with the same size as the image. In the PP-MPI prior in \cite{askin2022pnp}, the denoiser does not take a noise level map as input; here lies the main difference between the PP-MPI approach and our ZeroShot-PnP approach: the possibility of inputting a noise level map in the deep denoiser prior allows to couple the parameter $\mu_k$ in Eq.~\eqref{eq:PnP:split:data2} with the noise level $\sqrt{\lambda /\mu_k}$ in Eq.~\eqref{eq:PnP:split:denoise2} and design an automatic parameter selection strategy that we will describe in Section~\ref{sec:param:selection}. Moreover, we do not retrain the deep denoiser prior on MPI or MPI-related data, but employ it as a zero-shot denoiser (in the sense that it has not seen the domain before). 

We perform the denoising of the 3D target in a slicewise fashion strategy~\cite{askin2022pnp}, i.e., at each iteration $k$ we consider the 2D slices perpendicular to an axis and perform the denoising of each 2D slice with the 2D denoiser. This slicewise denoising is performed at step $k$ along each axis and the three resulting volumes are averaged. The overall algorithm is given in Algorithm~\ref{alg:pnp}.

If we consider no $\ell^1$-prior the same HQS yields the following iterative scheme, which can be regarded as a special case of Algorithm~\ref{alg:pnp}, which we call ZeroShot-PnP:
\begin{equation}\label{eq:HQS:pnp}
	\begin{split}
		u_1^{k+1} & = \arg\min_{u_1}\lVert f-Au_1\rVert_2^2 +  \mu_k\left\lVert u_1-u_2^k \right\rVert_2^2\\
		u_2^{k+1} & = \mathrm{Denoiser}\left (\tilde{u}_1^{k+1}\, , \sqrt{\lambda /\mu_k}\right ) .
	\end{split}
\end{equation}

In Section~\ref{sec:Results} we will see that both the ZeroShot-$\ell^1$-PnP and the ZeroShot-PnP have benefits, especially on the Open MPI dataset.

For comparison we provide more details also on the PP-MPI denoiser~\cite{askin2022pnp}: the authors have first generated a dataset of MPI-like images obtained from time-of-flight MRA images from the public ``ITKTubeTK - Bullitt - Healthy MR Dataset" by CASILab, by splitting the 95 healthy subjects into 77 training, 9 validation and 9 test sets; then, multiple $10\times 64\times 64$ 3D patches are cropped, thin-slab maximum-intensity projected along the first direction and finally downsampled to images of size $32\times 32$ and normalized to 1. Noise versions of these images have obtained by adding white Gaussian noise with standard deviation of 0.05. The PP-MPI denoiser uses a residual Dense network (RDN)~\cite{zhang2021rdn} as its backbone architecture, adapted to have a compact set of parameters to match the small size of the MPI images. Concerning the number of network parameters the deep denoiser prior has $32, 638, 656$ parameters and the PP-MPI-Denoiser has $58, 393$.

We conclude this section by observing that even though we have chosen to study and analyze the effect of the $\ell^1$-prior, the steps that led to the ZeroShot-$\ell^1$-PnP algorithm can be performed for any other prior $\mathcal{P}$ in place of the $\ell^1$ prior. In particular, the algorithm here proposed can be formulated to find minimizers of functionals of the form $\arg\min_{u}\lVert f - Au\rVert_2^2 + \lambda\mathcal{R}(u) + \alpha \mathcal{P}(u) + \iota_+ (u)$ for an additional convex prior $\mathcal{P}$; the convexity of $\mathcal{P}$ guarantees that its proximal mapping is well defined and can be directly substituted into Equation\eqref{eq:PnP:split:st}.

\begin{algorithm}
	\caption{The proposed ZeroShot-$\ell^1$-PnP algorithm.}\label{alg:pnp}
	\textbf{Input}: scan data $f$ and system matrix $A$, parameters $\alpha$, $n_{\mathrm{it}}$, $\mu_0$.\\
	\textbf{Output}: reconstructed volume $\tilde{u}_\mathrm{rec}$.\\
	\begin{algorithmic}[1]
		\State $u_2^0 ,u_3^0\gets 0 $;
		\State $k \gets 0$;
		\While{$k \leq n_{\mathrm{it}}$}
		\State $u_1^{k+1}\gets$ConjGrad$\left (A^{T} A + \mu_k\mathrm{Id}\, , A^{T} f + \mu_k \frac{u_2^k + u_3^k}{2}\right )$;
		\State $\sigma_{k+1} \gets \sqrt{\mathrm{Var}(u_1^{k+1})}$;
		\If {$k=0$}
		\State $\lambda \gets\mu_0 \cdot\sigma_0^2$
		\EndIf
		\State $u_2^{k+1}\gets$Slicewise-Denoise$\left (\tilde{u}_1^{k+1}\, ,\sigma_{k+1}\right ) $; \hfill\Comment{ZeroShot-Denoiser}
		\State $u_3^{k+1}\gets$Soft-Threshold$\left (u_1^{k+1}\, ,\frac{\alpha}{\mu_k}\right )$;
		\State $\mu_{k+1}\gets \lambda /\sigma_{k+1}^2$;
		\State $k\gets k+1$;
		\EndWhile
		\Return $\tilde{u}_2^{k+1}$
	\end{algorithmic}
\end{algorithm}

\subsection{Parameter Selection Strategies}\label{sec:param:selection}

The ZeroShot-$\ell^1$-PnP approach proposed depends on the (hyper-)parameters $\mu_k$, $\lambda$ and $\alpha$ as well as on the number of iterations $k$ of the scheme (stopping criterion). We remark that the $\sigma_{k+1}$ are not free parameters, but related to $\lambda$ and $\mu_k$ via $\sigma_{k+1} =\sqrt{\lambda /\mu_{k}}$. Typically, in PnP approaches (e.g.,~\cite{Zhang2022pnp}) the corresponding parameters are set fixed, in particular the noise related parameter $\sigma_k$ fed into the denoiser is set fixed for each iteration $k$ for a fixed number of iterations $n_{\mathrm{it}}$. 
We now propose more adaptive ways to choose the above parameters, as motivated by the continuation strategies in the previous section.

\paragraph{Choice of $\lambda$ and of the $\mu_k$.}
We start out from the relation  $\sigma_{k+1}^2 =\lambda /\mu_k,$ noting that the $\sigma_{k+1}$ have an interpretation  as (Gaussian) noise levels of the $u_1^{k+1}$. Using this interpretation, we let
\begin{equation}\label{eq:noiselevel}
	\tilde \sigma_{k+1}^2 = \mathrm{Var}(u_1^{k+1})=\mathbb{E}\left [ (u_1^{k+1} - \bar{u}_1^{k+1})^2\right ] ,
\end{equation}
where $\bar{u}_1^{k+1}$ is the mean of the vector $u_1^{k+1}$.
The value $\tilde \sigma_{k+1}^2$ is an upper bound on the noise level (and potentially can be seen as a very coarse estimate which empirically turns out to be sufficient for our purposes; another interpretation can be as a normalization). Then, we set  the $\mu_{k+1}$ by 
\begin{equation}\label{eq:noiselevel2}
	\mu_{k+1} = \tilde \mu_{k+1} = \tilde \lambda / \tilde \sigma_{k+1}^2.
\end{equation}
To determine $\lambda = \tilde \lambda,$ we use the relation $\tilde \lambda = \tilde \mu_0 \tilde \sigma_{0+1}^2=\tilde \mu_0 \tilde \sigma_{1}^2$, where we define
$\tilde \sigma_{1}^2$ via Equation~\eqref{eq:noiselevel}, and  $\tilde \mu_0$ by the following observation:
initializing $u_2^0 =u_3^0 = 0$ in Algorithm~\ref{alg:pnp}, $u_1^{1}$ is the solution of the Tikhonov regularization problem (cf. Equation~\eqref{eq:PnP:split:data2} and Equation~\eqref{eq:Min:Functional:std} with $\mathcal{R}(u) =\lVert u\rVert_2^2$). Consequently, one can set the parameter $\mu_0$ to 
\begin{equation}\label{eq:Mu0asTik}
	\tilde \lambda = \tilde \mu_0 \tilde \sigma_1^2,   \qquad \tilde \mu_0 = \tilde \lambda_\mathrm{Tik}
\end{equation}
where $\tilde \lambda_\mathrm{Tik}$ denotes an estimate on the parameter $ \lambda_\mathrm{Tik}$ of the corresponding Tikhonov problem. 

Finally, the algorithm is an iterative scheme that stops after $n_{\mathrm{it}}$ iterations and whose parameters are automatically updated at each iteration by the estimated parameter $\tilde{\sigma}_{k+1}$ in Equation~\eqref{eq:noiselevel}. The only parameters that have to be provided are $\alpha$, $n_{\mathrm{it}}$ and the starting Tikhonov parameter $\mu_0 = \tilde{\lambda}_{\mathrm{Tik}}$. Even if $\mu_0$ can be formally interpreted as a Tikhonov parameter, in Section~\ref{sec:Results} we treat $\mu_0$ as a hyperparameter and we validate $n_{\mathrm{it}}$ and $\mu_0$ on a hybrid data set which we describe in Section~\ref{sec:simul:dataset}.

\paragraph{Choice of  $\alpha$.} In each iteration a soft-thresholding (proximal mapping of the $\ell^1$ norm) of parameter $\alpha /\mu_k$ is performed (cf. Equation~\eqref{eq:PnP:split:st}), and thus all values of magnitude below $\alpha /\mu_k$ are set to zero. Let $u$ be the ground truth particle distribution and $u_\mathrm{rec}$ its reconstruction using the system matrix $A$. The matrix $A$ is obtained by scanning a delta distribution with a known concentration of $c_\delta$ mmol. It follows that if $c_u$ is the maximum concentration level of the ground truth $u$, the $\ell^\infty$-norm of the reconstruction is bounded from above $\lVert u_\mathrm{rec}\rVert_\infty\leq c_u /c_\delta =: c_\mathrm{rec}$. Theqrefore, to avoid the all zeros reconstruction, it is necessary that $\alpha /\mu_k <c_\mathrm{rec}$, i.e., that $\alpha <c_\mathrm{rec}\cdot \mu_0$. In particular, if some knowledge about the concentration injected and the concentration of the delta distribution is available, it is possible to set a rough bound on the parameter $\alpha$. In the experiments in Section~\ref{sec:Results}, because in the Open MPI Dataset $c_\delta = 100$mmol and we know that for each phantom $c_u\leq 100$mmol, we can deduce that $c_\mathrm{rec}\leq1$ and that $\alpha\leq\mu_0$. Theqrefore, it is reasonable to choose $\alpha$ as some percentage of $\mu_0$. In this work we have set $\alpha = 0.005 \mu_0$ for all experiments.

\subsection{Hybrid Dataset for Parameter Validation}\label{sec:simul:dataset}

For the validation of the parameters of both the proposed methods and the baseline methods we have produced a hybrid MPI dataset. In more detail, we have produced concentrations $u_{\mathrm{GT}}$ on a $19\times 19\times 19$ grid that play the role of MPI phantoms. We have considered the system matrix $A$ provided in the Open MPI dataset and without preprocessing we have applied it to the ground truths $u_{\mathrm{GT}}$ to obtain a signal $f$ with realistic noise distribution (since $A$ is composed of real noisy delta scans) and added an additional Gaussian noise $\eta$ obtaining a dataset of scans of the form $f = Au_{\mathrm{GT}} + \eta$. The total dataset consists of 30 ground truths, generated to provide the same challenges for which the Open MPI dataset has been created, i.e., to test the reconstruction capabilities w.r.t. shape, different concentration levels and resolution. In particular, we have produced 10 cone-shaped phantoms, 10 graph-like phantoms and 10 additional phantoms composed of small dot-like regions with different concentration levels. The cones where produced with random sizes, shifts and orientations inside the grid. The graph like-phantoms were generated with the following procedure: a number of vertices $V$ between 4 and 6 is randomly chosen and $V-1$ distinct edges (pairs of vertices) are randomly selected; the voxels thorough which an edge passes or that contain a vertex get value $1$ whereas every other voxel stays zero. The resulting phantom is then blurred with a 3D normal Gaussian filter of variance 1 and thresholded to thicken the edges. Finally, the last set of phantoms is generated with the same procedure as for the graph-like phantoms but considering a randomly selected number of vertices between $6$ and $9$ but without edges. Each vertex gets a concentration level randomly selected between $0.05$ and $1$ and the Gaussian blurring and subsequent tresholding is applied to obtain regions bigger than a single voxel. Because the delta probe of known and fixed maximum particle concentration $c_{\delta}=100$ mmol is used to acquire the system matrix, the outputs of the reconstruction methods will be rescaled reconstructions that shall be converted back to the absolute size via multiplication by $c_{\delta}$. All phantoms upon generation have maximum value 1 and have been rescaled to a maximum value $\beta\sim U (0.5\, , 1.5)$. The range between 0.5 and 1.5 corresponds to phantoms with maximum concentration levels between 50 and 150 mmol, which is the correct order of magnitude for the phantoms in the Open MPI Dataset even if the Resolution Phantom does not fall precisely in this range. However, the quality of the reconstructions in Section \ref{sec:Results} shows that such validation dataset works and that if it is generated with phantoms in the correct order of magnitude of the maximum concentration of the target phantom, the precise number does not affect too much the quality of the reconstruction.

We point out that this validation set is all we need to validate the parameters before testing on the Open MPI dataset and that no training set is necessary because no training is performed: we use a general purpose denoiser~\cite{Zhang2022pnp} pre-trained on natural images and we do not train an \emph{ad hoc} denoiser.

\section{Results} \label{sec:Results}

In this section we quantitatively and qualitatively evaluate both the ZeroShot-$\ell^1$-PnP in Algorithm~\ref{alg:pnp} and the ZeroShot-PnP variant in Equation~\eqref{eq:HQS:pnp} in Section~\ref{sec:PnPprop} on the OpenMPI dataset~\cite{knopp2020openmpidata}.
Qualitative comparision using the well established quality measures PSNR and SSIM is possible by inferring the ground truth from a CAD model; details are given in Section~\ref{eq:ImQualMeasures}. 
We compare the ZeroShot-$\ell^1$-PnP and the ZeroShot-PnP schemes with the DIP, with the standard Tikhonov $\ell^2$-regularization and the ART method as a baseline; moreover, we perform additional comparison with the PP-MPI algorithm~\cite{askin2022pnp} which uses a denoiser trained on MPI-friendly data. We further investigate the performance of the PP-MPI denoiser when it is used in place of the ZeroShot-Denoiser in ZeroShot-$\ell^1$-PnP (Algorihtm~\ref{alg:pnp}, line 9).
In each experiment, we employ the parameters obtained by evaluation on the hybrid dataset to reconstruct the phantoms in the Open MPI Dataset. All these methods are described in Section~\ref{sec:MethComparisions}.

The evaluation is conducted in Section~\ref{eq:EvalOnOpenMPI}-\ref{sec:exp4}. In particular, in Experiment~1 (Section~\ref{sec:Results}) we compare the results obtained with the DIP, Tikhonov, the ART and PP-MPI methods on the 3D Open MPI Dataset using typical preprocessing and the validated parameters. In Experiment~2 (Section~\ref{sec:exp1bis}) we focus more on the differences between the proposed ZeroShot-PnP method and the previous PP-MPI algorithm and the different performances of the denoisers employed. In Experiment~3 (Section~\ref{sec:exp2}) we test the ZeroShot-PnP with and without the $\ell^1$-prior with decreasing levels of preprocessing. Finally, in Experiment~4 (Section~\ref{sec:exp4}) we show that MPI reconstructions with the proposed methods are possible with most basic preprocessing (background removal and suppression of excitation crosstalk) and without the employment of the SVD, which becomes more and more cumbersome to compute as the dimensionality of the problem increases.

The preprocessing  was implemented both in Matlab and in Python while the ZeroShot-$\ell^1$-PnP and ZeroShot-PnP algorithms as well as the baseline algorithms for comparison were implemented in Python 3.9, using Numpy and PyTorch. For the ART method as well as the PP-MPI method we have used the code made available by the authors of~\cite{askin2022pnp,gungor2023deqmpi} at \href{https://github.com/icon-lab/PP-MPI}{https://github.com/icon-lab/PP-MPI}. The experiments were performed on a workstation with 13th Gen Intel(R) Core(TM) i9-13900KS, 128 GB of RAM, an NVIDIA RTX A6000 GPU and Windows 11 Pro.

\subsection{Image Quality Measures} \label{eq:ImQualMeasures}

In order to quantitatively assess the reconstruction quality, the Open MPI dataset provides CAD models of the phantoms. This allows to compare the reconstructions with the corresponding ground truth implied by the CAD model. We use both Peak Signal-to-Noise Ratio (PSNR) and the Structural Similarity Index Measure (SSIM)~\cite{ssim} as measures.
We recall that that the PSNR of the reconstruction $f$ with respect to groundtruth  $g$ is defined by
\begin{equation}\label{eq:psnr}
	\mathrm{PSNR}(f,g) = 10\cdot \log_{10} \left ( R^2 /\mathrm{MSE}(f,g)\right)
\end{equation}
where $R$ is the maximum value of $f$ and $\mathrm{MSE}(f,g)$ is the Mean Square Error defined either as
\begin{equation}
	\mathrm{MSE}(f,g)=\frac{1}{MN}\sum_{j=1}^{M}\sum_{i=1}^{N}(f_{ij}-g_{ij})^2
\end{equation}
or as
\begin{equation}
	\mathrm{MSE}(f,g)=\frac{1}{LMN}\sum_{k=1}^{L}\sum_{j=1}^{M}\sum_{i=1}^{N}(f_{ijk}-g_{ijk})^2 ,
\end{equation}
depending on if $f,g\in\mathbb{R}^{M\times N}$ are images or $f,g\in\mathbb{R}^{L\times M\times N}$ volumes.

For the $\mathrm{SSIM}$ measure we employ the following definition and choice of parameters: given two images $f,g$, the SSIM is defined as 
\begin{equation}
	\mathrm{SSIM}(f,g) = \left [l(f,g)\right ]^\alpha \cdot \left [c(f,g)\right ]^\beta\cdot \left [s(f,g)\right ]^\gamma
\end{equation}
where $l(f,g)=\frac{2\mu_f\mu_g +C_1}{\mu_f^2 + \mu_g^2 +C_1}$, $ c(f,g)=\frac{2\sigma_f\sigma_g +C_2}{\sigma_f^2 + \sigma_g^2 +C_2}$, $s(f,g)=\frac{\sigma_{fg}+C_3}{\sigma_f\sigma_g + C_3}$ are the luminance (l), contrast (c) and structure (s) ~\cite{ssim} and $\mu_f$ is the mean value, $\sigma_f$ the standard deviation of $f$; $\mu_g$ and $\sigma_g$ are analogously defined for $g$ and $\sigma_{fg}$ the covariance of $f$ and $g$. Following the implementation of~\cite{dittmer2020deep}, we set $\alpha =\beta =\gamma =1$ and $C_1 = (0.01\cdot R)^2$, $C_2 = (0.03\cdot R)^2$ and $C_3 = 0.5\cdot C_2$ where $R=100$, i.e., the concentration of the Delta Sample used during the calibration process. All reconstructions are multiplied by the concentration level before the computation of the quality measures to maintain consistency with the choice of $R$ in the definition of the SSIM.

Since there might be deviations between the position of the scanned phantom and the reference CAD model $u_{\mathrm{ref}}$, we follow~\cite{kluth2020l1data} and consider the quantity:
\begin{equation}\label{eq:psnr:max}
	\mathrm{PSNR}(u) =\mathrm{PSNR}_{\mathrm{max}}(u) = \max_{\Delta r\in\mathcal{R}}\mathrm{PSNR}(u,u_{\mathrm{ref},\Delta r}),
\end{equation}
where $\mathcal{R}$ is the set of all reference volumes $u_{\mathrm{ref},\Delta r}$ obtained by considering shifts $\Delta r$ of step size $0.5$mm in the neighborhood $[-3 \, \mathrm{mm}\, ,3 \, \mathrm{mm}]^3$. Analogously, we consider the positional-uncertainty-aware measure $\mathrm{SSIM}_{\mathrm{max}}$ as:
\begin{equation}
	\mathrm{SSIM}(u) = \mathrm{SSIM}_{\mathrm{max}}(u) = \max_{\Delta r\in\mathcal{R}}\mathrm{SSIM}(u,u_{\mathrm{ref},\Delta r}).
\end{equation}
In what follows we always display $\mathrm{PSNR}_\mathrm{max}$ and $\mathrm{SSIM}_\mathrm{max}$ when showing the reconstruction of the Open MPI dataset.

\subsection{Methods for Comparison and Parameter Validation}\label{sec:MethComparisions}

\paragraph{(Classical) Tikhonov Regularization.}
The inversion of the system matrix is performed considering the minimization problem in Equation ~\eqref{eq:Min:Functional:std} with $\mathcal{R}(u)=\lVert u\rVert_2^2$; the minimizer is then obtained solving the associated Euler-Lagrange equations using the Conjugated Gradient (CG) method when the SVD is not available. When the SVD is indeed available, we use it for faster direct solutions of the inversion problems, it is however not necessary as we point out in Experiment 4. For the validation, we have chosen its parameter $\lambda_\mathrm{Tik}$ optimally in the sense that we have validated it performing a grid search and selecting that value for which the average $\mathrm{PSNR}$ on the hybrid dataset is maximal. The grid search proceeds as follows: we first perform reconstruction with parameters $10^j$ for $j = -6, -5, \dots, 0,1, \cdots , 18$ and consider $j^\ast$ the exponent that maximizes the PSNR; then we consider the values $k\cdot 10^{j^\ast -1}$ and $k\cdot 10^{j^\ast}$ with $k=1, \dots ,9$ and again select the one maximizing the average PSNR.

\paragraph{Algebraic Reconstruction Technique (ART).} This is a method employed in MPI system-matrix-based reconstruction~\cite{kluth2019enhancedrec,Knopp_etal2010ec}: the reconstruction is obtained solving the following constrained minimization problem $\hat{u} = \arg\min_{u\geq 0}\lVert f - Au\rVert_2^2 + \lambda\lVert u\rVert_2^2$, using the ART method and incorporating a positivity constraint that is motivated by the fact that solution of the reconstruction a particle distributions. We have utilized the ART implementation provided in the PP-MPI author's git repository \cite{askin2022pnp,gungor2023deqmpi} available at \href{https://github.com/icon-lab/PP-MPI}{https://github.com/icon-lab/PP-MPI}. For the validation, we have performed the grid search for the regularization parameter $\lambda_{\mathrm{ART}}$ in the same fashion as for the Tikhonov method, performing 200 total iterations and considering selecting the regularization parameter and number of iterations $n_{\mathrm{it}}$ for which the PSNR is maximized after averaging on the hybrid validation set.

\paragraph{Deep Image Prior (DIP).}
The DIP has been used in MPI~\cite{dittmer2020deep}. 
It is, to our knowledge, the only machine learning approach
which has been both qualitatively and quantitatively  evaluated on the Open MPI dataset. 
It is based on the idea of dropping the explicit regularizer $\mathcal{R}$ in Equation ~\eqref{eq:Min:Functional:std} 
and of using the implicit prior captured by a neural network parametrization, 
i.e., if $\varphi_\theta$ is a neural network with parameters $\theta$ 
taking as input a noisy array $z$, we can train the network $\varphi_\theta$ 
by solving the following minimization problem
\begin{equation}\label{eq:dip}
	\theta^* = \arg\min_\theta \lVert A\varphi_\theta (z)-f\rVert_p^p ,
\end{equation}
to obtain a reconstruction $u=\varphi_{\theta^\ast}(z)$. The minimization of the functional in Equation~\eqref{eq:dip} is performed with $p=1$ as suggested, with the random input $z$ having entries uniformly distributed in $[0, 0.7]$ and the same shape of the output $(1,19,19,19)$; the regularizing architecture of $\varphi_\theta$ is an autoencoder obtained by not using skip connections in a U-Net with encoder steps that down-sample by a factor of 2 with 64, 128 and 256 channels respectively (the decoder is symmetric). (We refer to~\cite{dittmer2020deep} for details on the implementation.) The training of $\varphi_\theta$ has been performed minimizing the functional in Equation~\eqref{eq:dip} for 20000 iterations with Adam~\cite{kingma2014Adam} with different learning rates $\alpha_i = 10^{-i}$, for $i=3,4,5$ and momenta setting $\beta = (0.9, 0.999)$. The selection strategy of the DIP reconstruction is the one used in~\cite{dittmer2020deep}: we extract reconstructions after a number of iterations $s\in\lbrace$1,2,$\dots ,$ 10,12,$\dots ,$30,35,$\dots ,$50,60,$\dots ,$150,175,$\dots ,$500, 600,$\dots ,$2000,2500,$\dots ,$5000,6000,$\dots ,$20000$\rbrace$. For each iteration $s$ we have computed the $\mathrm{PSNR}$ and validated the optimal stopping iteration $s_\mathrm{opt.}$ and learning rate $\alpha_i$ on our hybrid dataset.

\paragraph{The PP-MPI Algorithm with Trained Denoiser.} We now briefly recall the PP-MPI algorithm in~\cite{askin2022pnp}. The authors consider the following equivalent formulation of the regularized inversion of the system matrix $A$:
\begin{equation}\label{eq:min:stat}
	\arg\min_u \sum_i \alpha_i g_i ( u ) \quad\text{s.t.}\quad \lVert Au - f\rVert_2 < \epsilon
\end{equation}
where $u$ is the unknown distribution of particles, the $\alpha_i$ are the parameters weighting each i-th regularization function $g_i$, the data fidelity term for given scan data $f$ and system matrix $A$ is imposed as a constraint; $\epsilon$ is the upper bound on the noise level.
The constrained minimization problem in Equation~\eqref{eq:min:stat} can be solved using an ADMM splitting scheme that yields the following algorithm:
\begin{equation}\label{eq:alg:ppmpi}
	\begin{split}
		u^{n+1} & = \left ( \mathbb{I}+A^{T} A\right )^{-1} \left ( A^{T} (z_{0}^n + d_{0}^n) + z_{1}^n + d_{1}^n\right ) \\
		z_{1}^{n+1} & = \mathrm{Denoiser}(u^{n+1}-d_{1}^{n}) \\
		d_{1}^{n+1} & = d_{1}^{n} + z_{1}^{n+1} - u^{n+1} \\
		z_{0}^{n+1} & = \Psi_{f,\epsilon} (Au^{n+1}-d_{0}^{n} ) \\
		d_{0}^{n+1} & = d_{0}^{n} + z_{0}^{n+1} - Au^{n+1} \\
	\end{split}
\end{equation}
where $\Psi_{f,\epsilon}$ is the projection function onto the ball of radius $\epsilon$ centered at $f$, taking care of the data fidelity constraint, and $z_{0}^{n}$, $z_{1}^{n}$, $d_{0}^{n}$, $d_{1}^{n}$ are auxiliary variables arising from the ADMM splitting that can be initialized as zero vectors at iteration $n=0$. The denoiser trained by the authors of~\cite{askin2022pnp} on MPI-friendly data will be called PP-MPI-Denoiser in what follows. For the PP-MPI algorithm we have used the algorithm and denoiser provided by the authors~\cite{askin2022pnp,gungor2023deqmpi} available at \href{https://github.com/icon-lab/PP-MPI}{https://github.com/icon-lab/PP-MPI}. We have validated the number of iterations and the variable $\epsilon$ in Equation~\eqref{eq:min:stat} on the hybrid dataset. Consistently with~\cite{askin2022pnp}, we consider $f = Au_{\mathrm{GT}} + \varepsilon \eta$ for $\eta\sim\mathcal{N}(0,\mathbb{I})$ with $\varepsilon = \sqrt{\lVert Au_{\mathrm{GT}}\rVert^2 / (\mathrm{SNR}\cdot \lVert \eta \rVert^2 )}$, and $\mathrm{SNR}_{\mathrm{db}}=10\log_{10}(\mathrm{SNR})$. We performed 3000 total iterations with $\epsilon$ with $\mathrm{SNR}_{\mathrm{db}}$ ranging from 20 to 30 with step size 1.

\paragraph{Validation of the Proposed ZeroShot-$\ell^1$-PnP and ZeroShot-PnP Methods.} We have validated the parameter $\mu_0$ and the optimal number of iterations $n_\mathrm{it}$  considering the optimal parameters that maximize the average $\mathrm{PSNR}$ on the hybrid dataset. For $\mu_0$ we have applied the same grid search strategy as for the Tikhonov parameter above.

\subsection{Experiment 1: Comparison on the Open MPI Dataset.}\label{eq:EvalOnOpenMPI}

The Open MPI Dataset~\cite{knopp2020openmpidata} contains three phantoms designed to offer different challenges for the reconstruction: the shape phantom, the resolution phantom and the concentration phantom. We present both quantitative and qualitative results organized along the phantoms. For the validation of the parameters we have used the dataset described in Section~\ref{sec:simul:dataset}.

\paragraph{Preprocessing.}
The scan data in this first experiment has been preprocessed as described in Section~\ref{sec:Preproc}, considering frequencies between $80$ and $625$ kHz, with whitening and the low-rank approximation with the rSVD and rank $K=2000$ (as in \cite{dittmer2020deep}).

\paragraph{Parameter Validation.}
\begin{table}[t]
	\footnotesize{
		\begin{center}
			\begin{tabular}{ |c|c|c|c|c| }
				\hline
				& \multicolumn{4}{c|}{Comparison Methods}\\
				\hline
				&Tikhonov & ART & DIP & PP-MPI~\cite{askin2022pnp} \\
				\hline
				Param.&  $\lambda_\mathrm{Tik.}=5\cdot 10^{5}$ & $\lambda_{\mathrm{ART}}=2\cdot 10^{4}$ & $\alpha = 10^{-3}$ & $\mathrm{SNR}_{\mathrm{db}} = $22 \\
				&  & it=200 & $ s = 14000$ & $n_\mathrm{it}=$1606 \\
				\hline
				PSNR &   $25.28\pm 3.28$ & $27.00 \pm 4.51$& $29.45 \pm 3.07$  & $28.13 \pm 3.75$ \\
				\hline
				SSIM &  $0.640 \pm 0.147$ & $0.482 \pm 0.198$ & $0.637\pm 0.248$ & $0.408 \pm 0.217$\\
				\hline
			\end{tabular}
		\end{center}
		
		\begin{center}
			\begin{tabular}{ |c|c|c| }
				\hline
				& \multicolumn{2}{c|}{Proposed ZeroShot-PnP}\\
				\hline
				& Without $\ell^1$-Prior & With $\ell^1$-Prior\\
				\hline
				Param.& $\mu_0 = 2\cdot 10^5$ & $\mu_0 = 3\cdot 10^5$\\
				& $n_\mathrm{it}=7$ & $n_\mathrm{it}=5$\\
				\hline
				PSNR & $30.45 \pm 5.04$ & $29.77 \pm 5.82$\\
				\hline
				SSIM & $0.788 \pm 0.141$ & $0.766 \pm 0.189$\\
				\hline
			\end{tabular}
		\end{center}
	}
	\caption{Validation results on the hybrid dataset of Section~\ref{sec:simul:dataset}. We observe that the proposed ZeroShot-PnP methods achieve the highest values w.r.t.~both PSNR and SSIM on the hybrid datset.
		\label{tab:exp1:validation}
	}
\end{table}
\def\imratio{0.24}
\begin{figure}[t]
	\centering
	\begin{subfigure}[t]{\imratio\linewidth}
		\includegraphics[width=\linewidth]{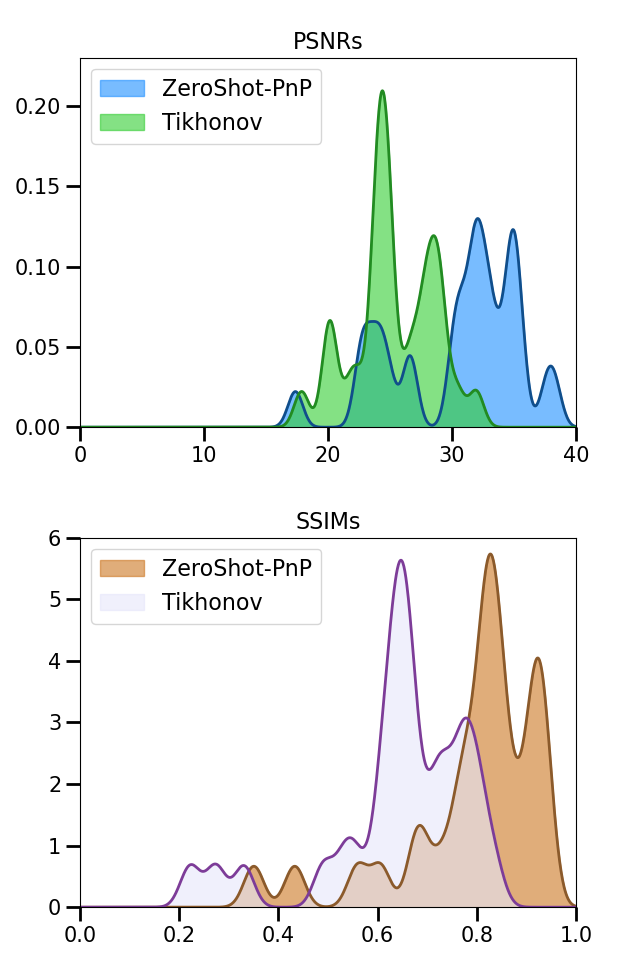}
		\caption{\centering\scriptsize ZeroShot-PnP vs. Tikhonov.}
		\label{subfig:hist:0:tik}
	\end{subfigure}
	\hfil
	\begin{subfigure}[t]{\imratio\linewidth}
		\includegraphics[width=\linewidth]{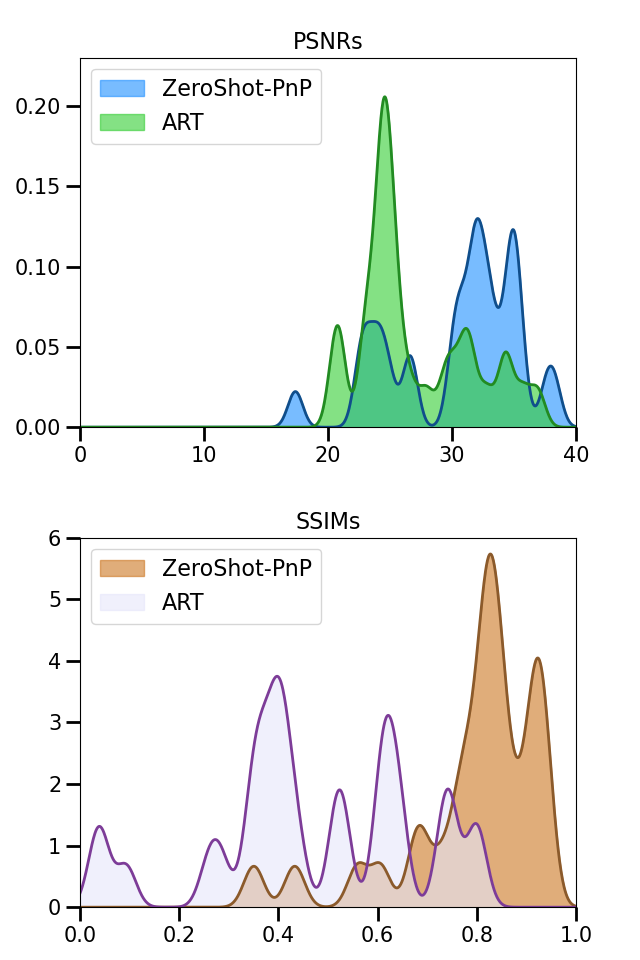}
		\caption{\centering\scriptsize ZeroShot-PnP vs. ART. }
		\label{subfig:hist:0:art}
	\end{subfigure}
	\hfil
	\begin{subfigure}[t]{\imratio\linewidth}
		\includegraphics[width=\linewidth]{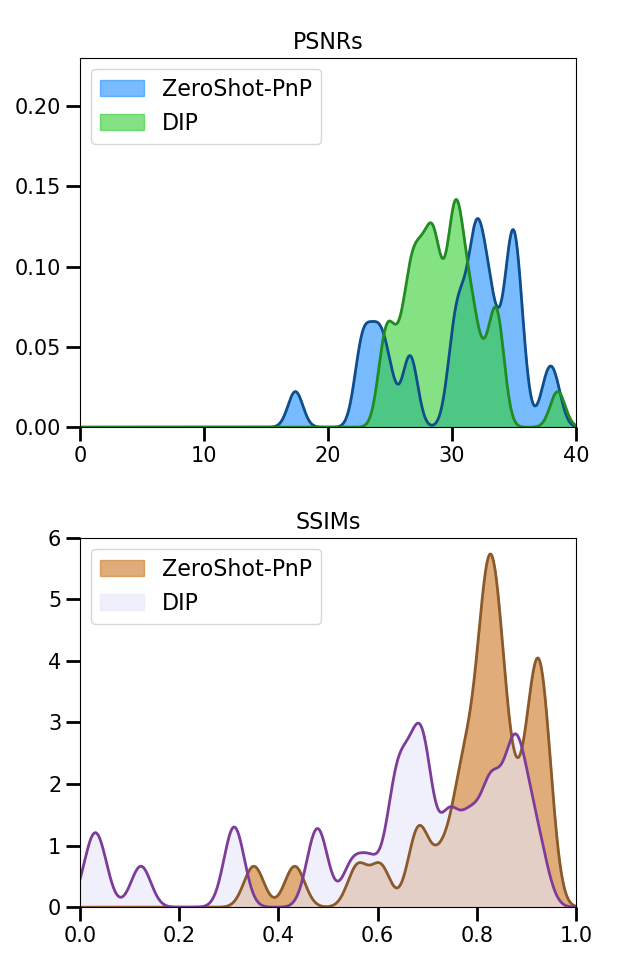}
		\caption{\centering\scriptsize ZeroShot-PnP vs. DIP.}
		\label{subfig:hist:0:dip}
	\end{subfigure}
	\hfil
	\begin{subfigure}[t]{\imratio\linewidth}
		\includegraphics[width=\linewidth]{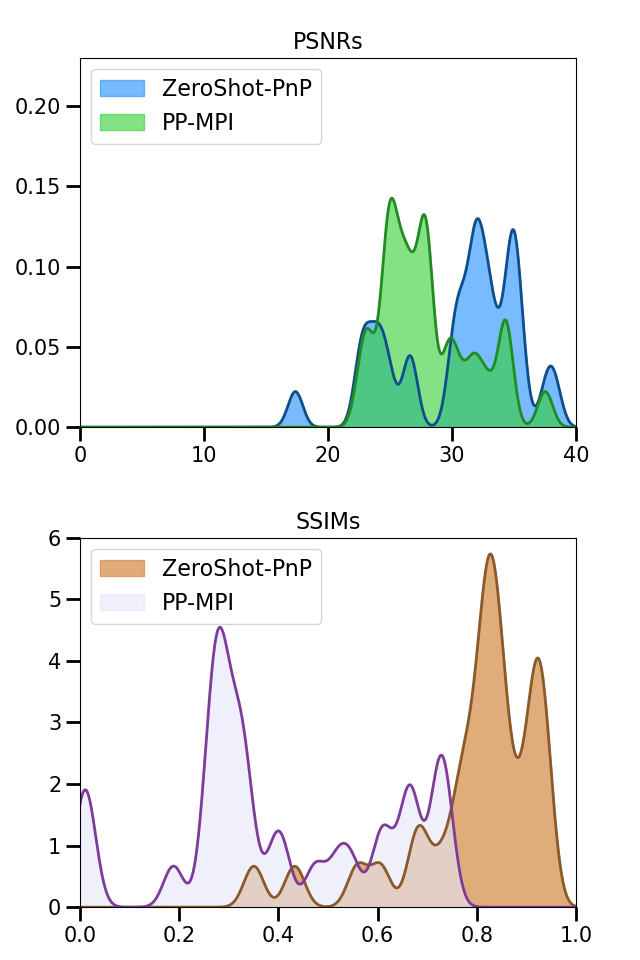}
		\caption{\centering\scriptsize ZeroShot-PnP vs. PPMPI.}
		\label{subfig:hist:0:ppmpi}
	\end{subfigure}
	\hfil
	\par\medskip
	\begin{subfigure}[t]{\imratio\linewidth}
		\includegraphics[width=\linewidth]{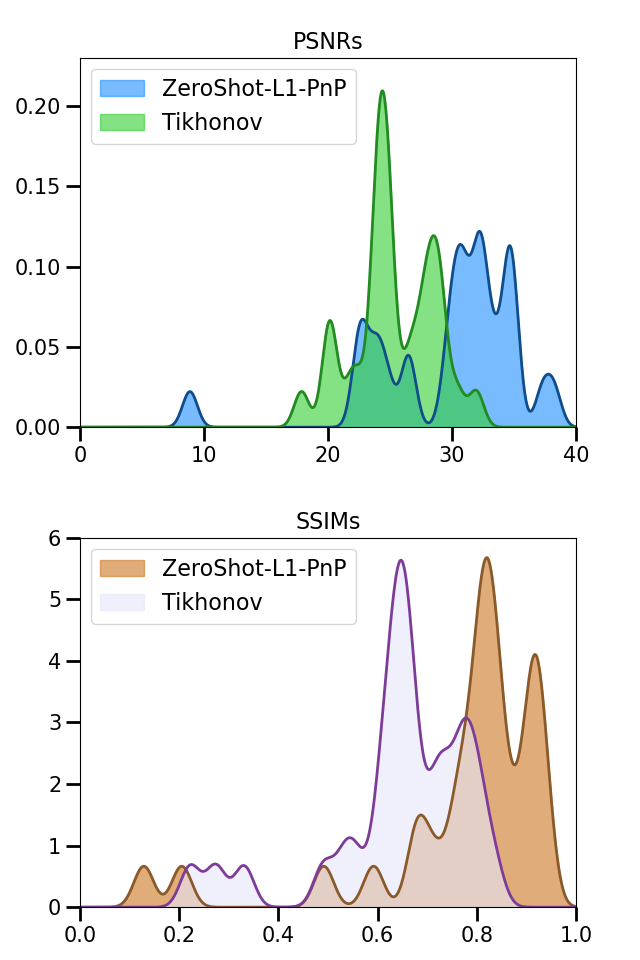}
		\caption{\centering\scriptsize ZeroShot-$\ell^1$-PnP vs. Tikhonov.}
		\label{subfig:hist:005:tik}
	\end{subfigure}
	\hfil
	\begin{subfigure}[t]{\imratio\linewidth}
		\includegraphics[width=\linewidth]{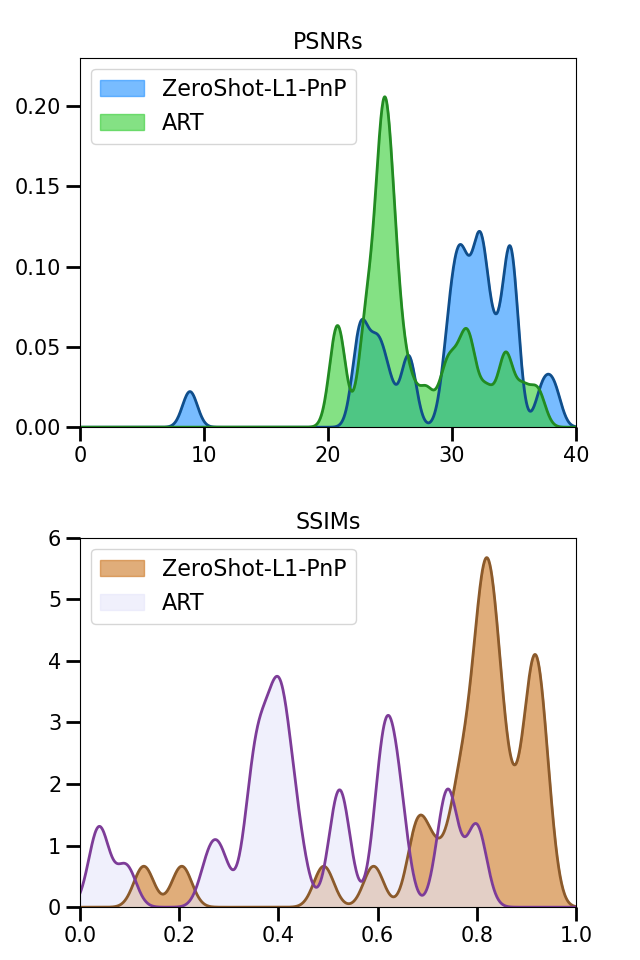}
		\caption{\centering\scriptsize ZeroShot-$\ell^1$-PnP vs. ART.}
		\label{subfig:hist:005:art}
	\end{subfigure}
	\hfil
	\begin{subfigure}[t]{\imratio\linewidth}
		\includegraphics[width=\linewidth]{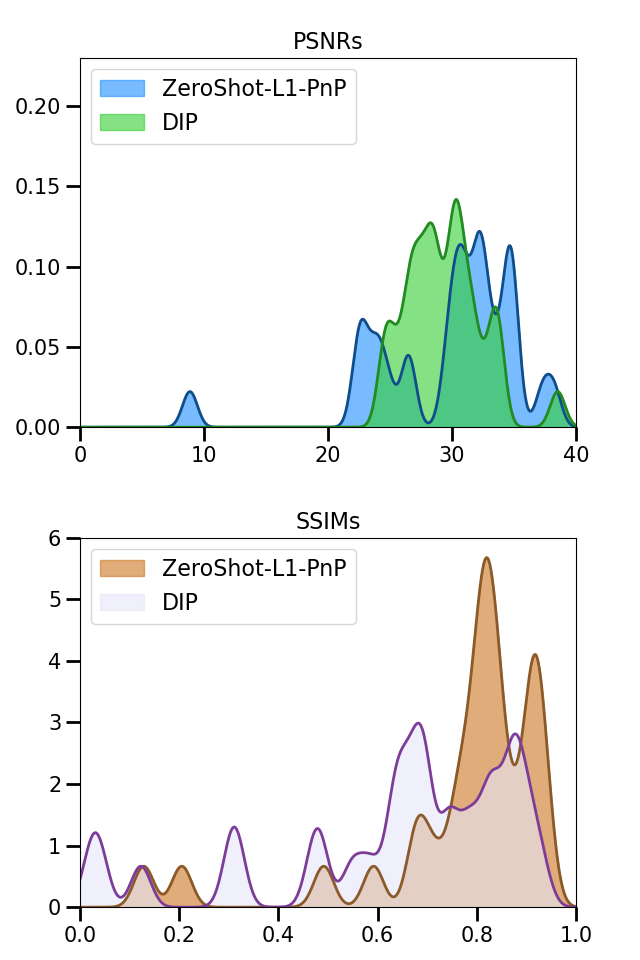}
		\caption{\centering\scriptsize ZeroShot-$\ell^1$-PnP vs. DIP.}
		\label{subfig:hist:005:dip}
	\end{subfigure}
	\hfil
	\begin{subfigure}[t]{\imratio\linewidth}
		\includegraphics[width=\linewidth]{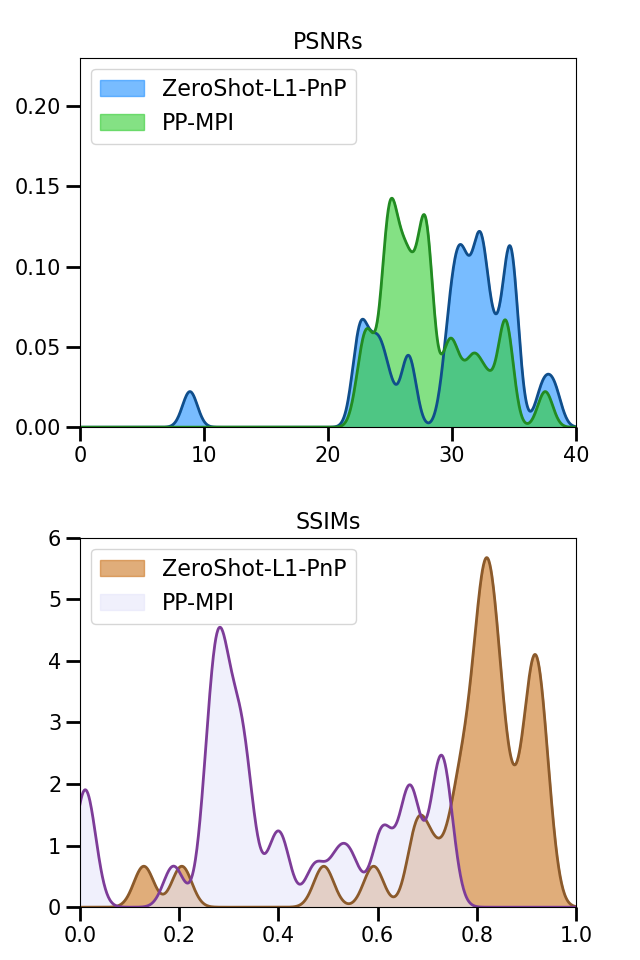}
		\caption{\centering\scriptsize ZeroShot-$\ell^1$-PnP vs. PPMPI.}
		\label{subfig:hist:005:ppmpi}
	\end{subfigure}
	\caption{Plot of the PSNR and SSIM distributions on the hybrid validation dataset representing the distribution of PSNR (in green and blue) and SSIM (in ocher and violet) in the ZeroShot-$\ell^1$-PnP (resp. ZeroShot-PnP) and the comparison methods. On the x-axis we display the ranges of the PSNR values (in $[0,40]$) and SSIM values (in $[0,1]$). the distributions have been estimated using Kernel Densities Estimations using Gaussian kernels.}
	\label{fig:exp1:densities}
\end{figure}
As dataset for parameter validation we employ the hybrid dataset of Section~\ref{sec:simul:dataset}. We display the found parameters for the proposed ZeroShot-$\ell^1$-PnP and the ZeroShot-PnP approach, the DIP as well as the Tikhonov regularization method, the ART method and the PP-MPI approach. An overview of the validated parameters is displayed in Table~\ref{tab:exp1:validation}. We observe that on the validation dataset both the ZeroShot-$\ell^1$-PnP and the ZeroShot-PnP proposed in this paper achieve higher $\mathrm{PSNR}$ and $\mathrm{SSIM}$ values. In order to better understand the variation of the PSNR and SSIM values of the proposed ZeroShot-PnP and ZeroShot-$\ell^1$-PnP algorithms and the comparison methods, we have plotted in Figure~\ref{fig:exp1:densities} the distributions of the PSNR and SSIM obtained with the optimal parameters on the validation dataset. In Figure~\ref{fig:exp1:densities} all possible comparisons between the proposed methods and the comparison methods are displayed; in particular, in blue and green we plot the PSNR distributions whereas in ocher and violet the SSIM distributions. We observe from the PSNR and SSIM distributions that the validation dataset contains phantoms on which all methods perform poorly (cf. for example Fig.~\ref{subfig:hist:0:dip}); because the validation dataset contains 30 phantoms in total, outliers may explain a higher variation of the PSNR values for our methods in Table~\ref{fig:exp1:densities}. If we consider trimmed mean and variances, i.e., excluding the lowest and highest 5\% results, we obtain that the trimmed PSNR values on the validation dataset are $31.3 \pm 4.24$ for the ZeroShot-PnP algorithm and $30.92 \pm 4.15$ for the ZeroShot-$\ell^1$-PnP algorithm. The trimmed values have a standard deviation in line with the comparison methods, which are $25.22\pm 3.80$ for the Tikhonov method, $26.78\pm 3.69$ for the ART method, $30.11\pm 2.40$ for the DIP and $28.21\pm 3.14$ for the PP-MPI.
Finally, we observe that the independent validation on the hybrid dataset of $\lambda_{\mathrm{Tik}}$ and of the parameters $\mu_0$ for the ZeroShot-$\ell^1$-PnP algorithms yields parameters $\lambda_{\mathrm{Tik}}=5\cdot 10^{5}$ and $\mu_0 = 2\cdot 10^{5}$ (resp. $\mu_0 = 3\cdot 10^{5}$) of the same order of magnitude and that are not too far away; this is interesting in view of the observation in Equation~\eqref{eq:Mu0asTik} where we noted that $\mu_0$ is interpretable as $\lambda_{\mathrm{Tik}}$, upon zero initialization of $u_2^0 = u_3^0 = 0$.

\def\imratio{0.23}
\begin{figure}[t]
	\centering
	\begin{subfigure}[t]{\imratio\linewidth}
		\includegraphics[width=\linewidth]{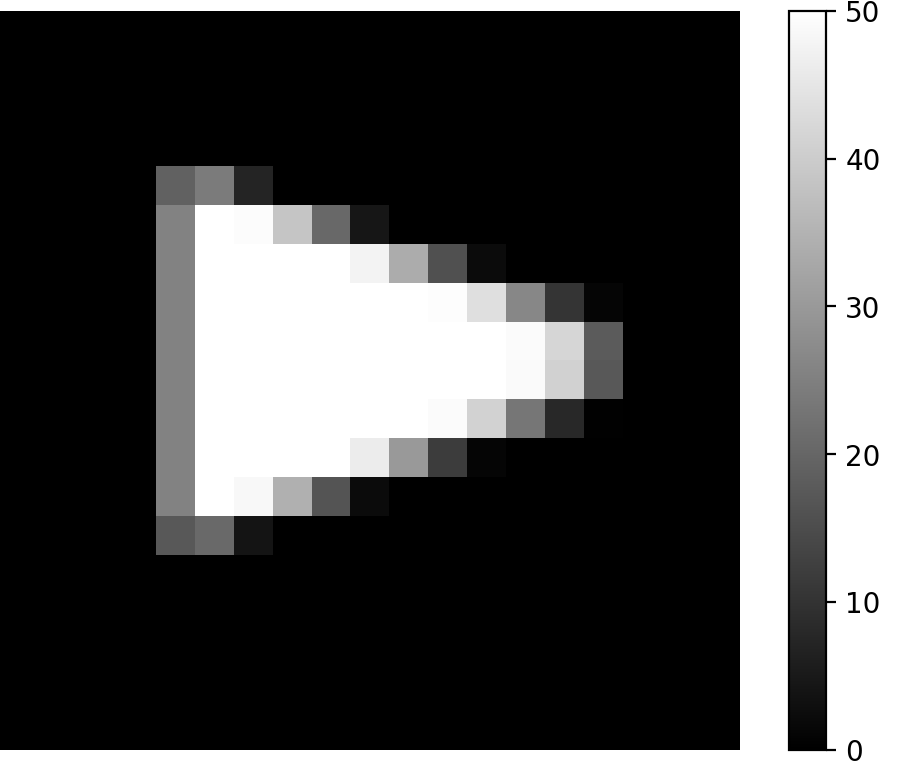}
		\caption{\centering\scriptsize CAD reference.}
		\label{subfig:shape:gt}
	\end{subfigure}
	\hfil
	\begin{subfigure}[t]{\imratio\linewidth}
		\includegraphics[width=\linewidth]{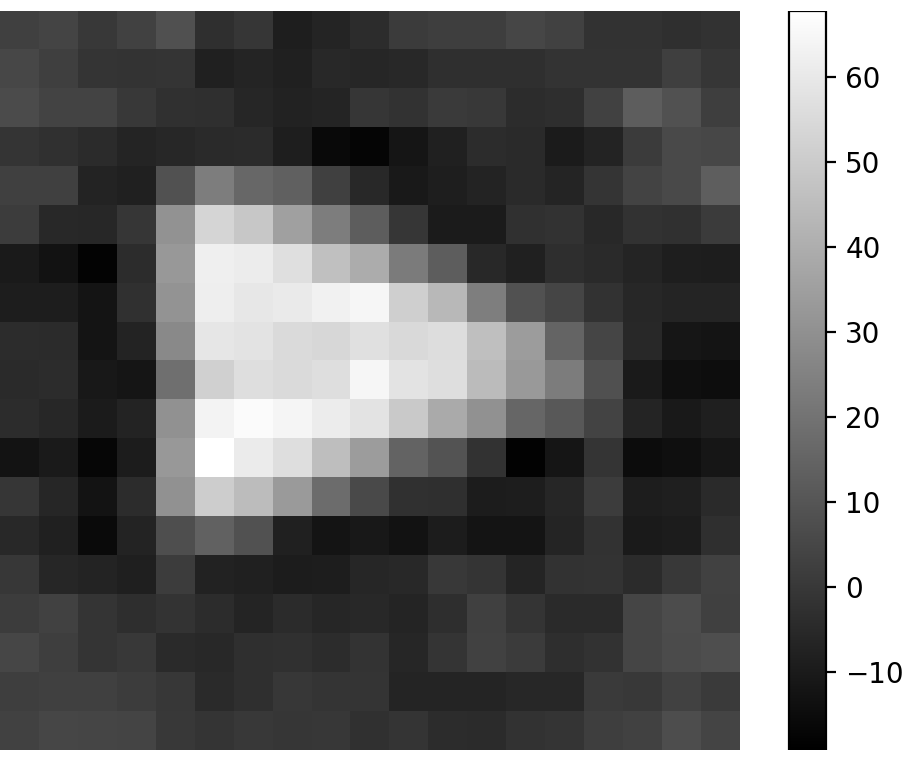}
		\caption{\centering\scriptsize Tikhonov.}
		\label{subfig:shape:tik}
	\end{subfigure}
	\hfil
	\begin{subfigure}[t]{\imratio\linewidth}
		\includegraphics[width=\linewidth]{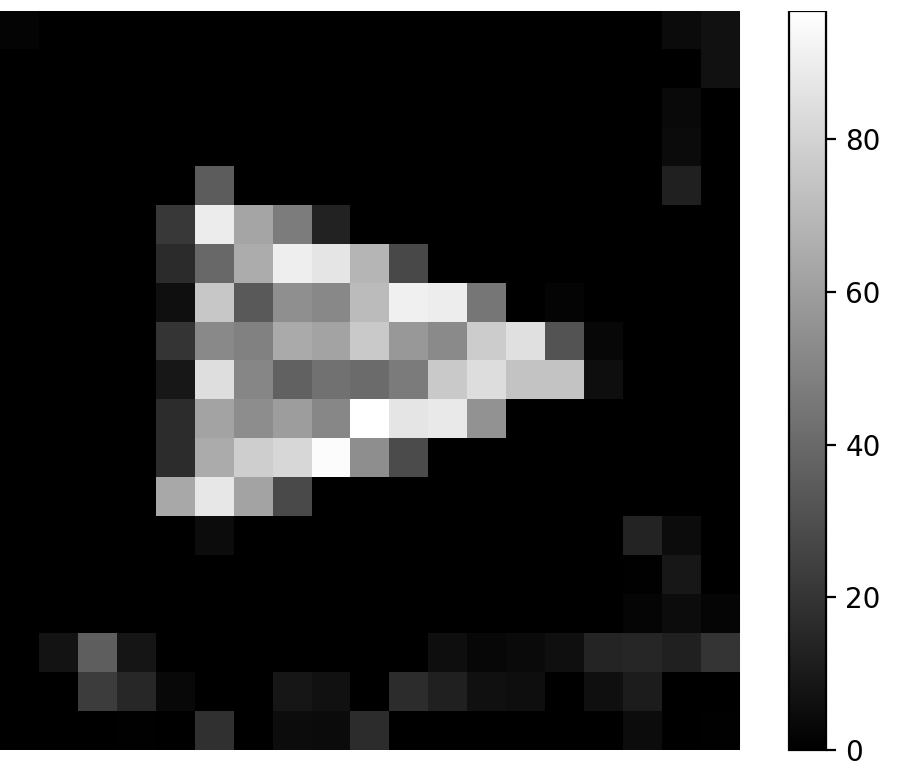}
		\caption{\centering\scriptsize ART.}
		\label{subfig:shape:ART}
	\end{subfigure}
	\hfil
	\begin{subfigure}[t]{\imratio\linewidth}
		\includegraphics[width=\linewidth]{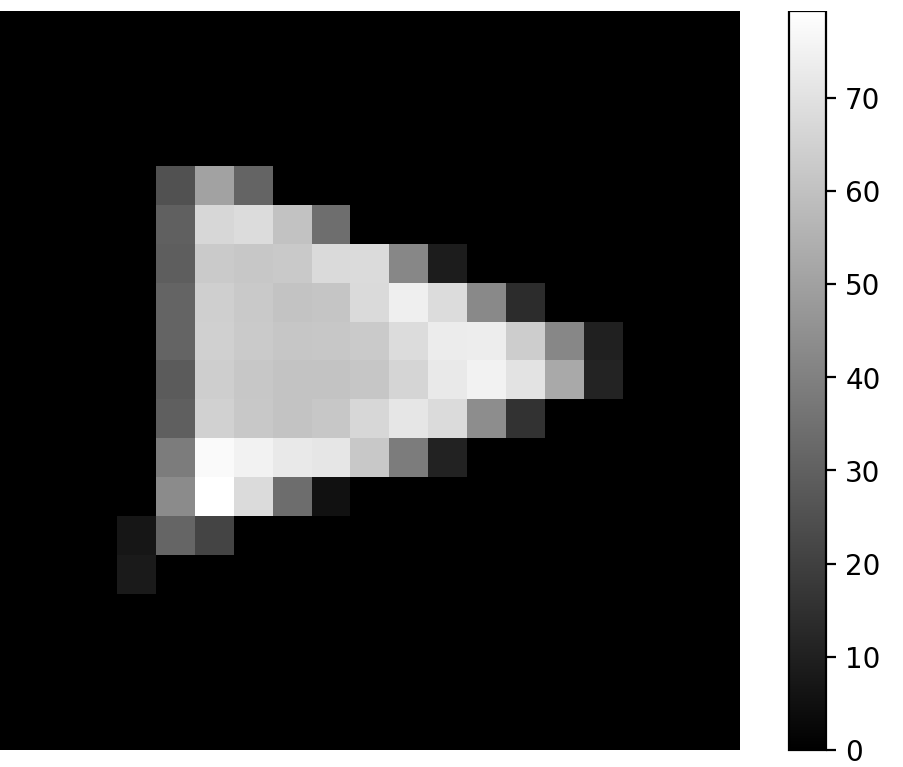}
		\caption{\centering\scriptsize DIP.}
		\label{subfig:shape:dip}
	\end{subfigure}
	\par\medskip
	\begin{subfigure}[t]{\imratio\linewidth}
		\includegraphics[width=\linewidth]{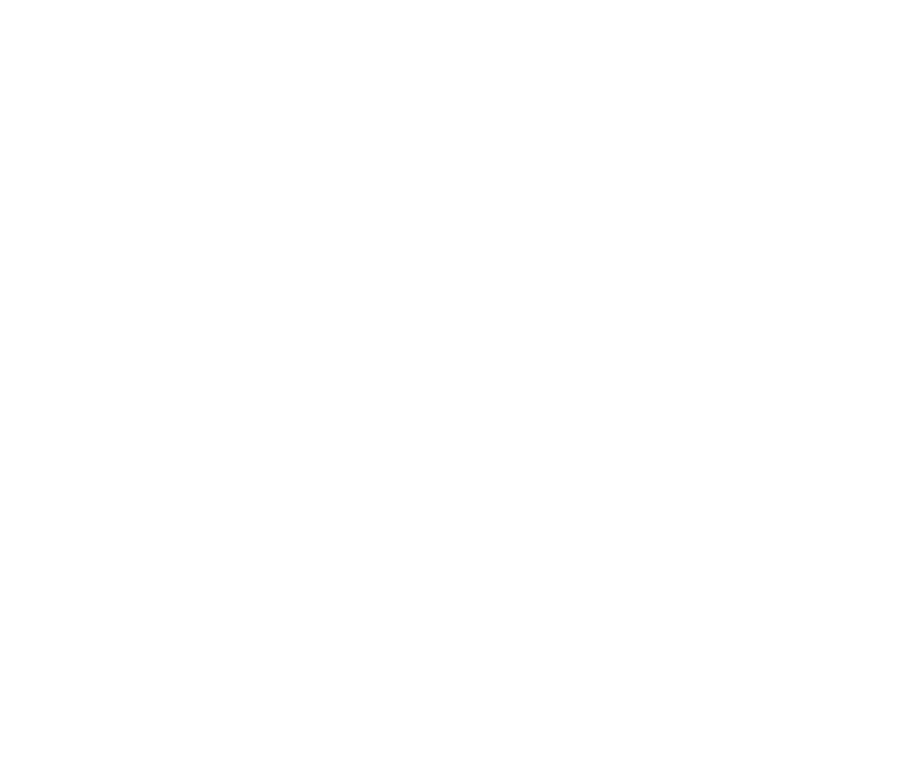}
	\end{subfigure}
	\hfil
	\begin{subfigure}[t]{\imratio\linewidth}
		\includegraphics[width=\linewidth]{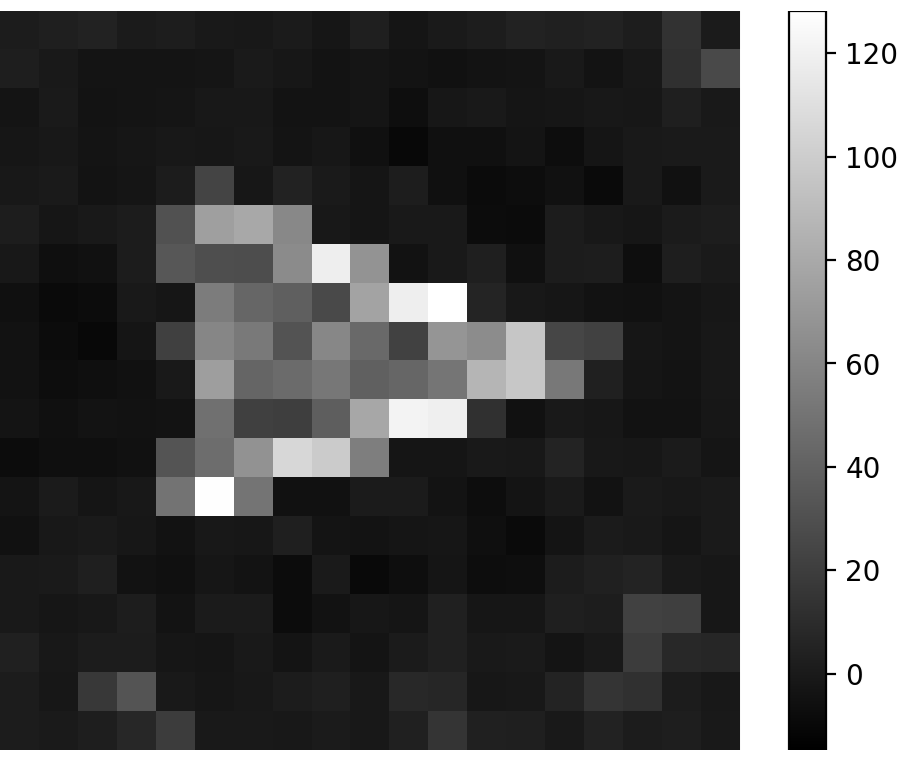}
		\caption{\centering\scriptsize PP-MPI.}
		\label{subfig:shape:ppmpi}
	\end{subfigure}
	\hfil
	\begin{subfigure}[t]{\imratio\linewidth}
		\includegraphics[width=\linewidth]{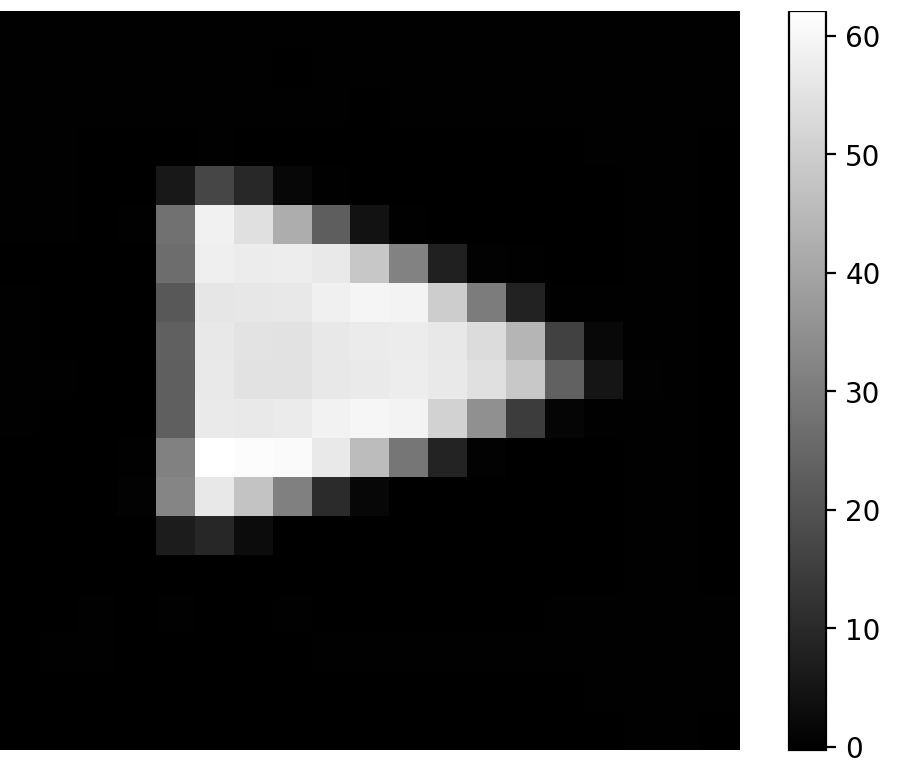}
		\caption{\centering\scriptsize ZeroShot-PnP.}
		\label{subfig:shape:hqs}
	\end{subfigure}
	\hfil
	\begin{subfigure}[t]{\imratio\linewidth}
		\includegraphics[width=\linewidth]{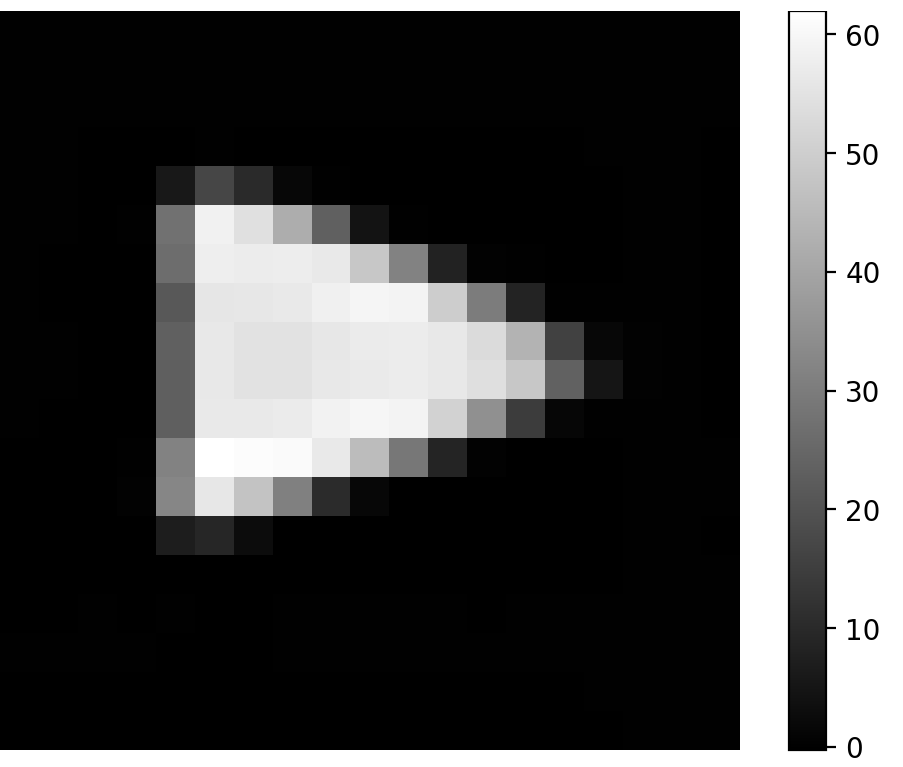}
		\caption{\centering\scriptsize ZeroShot-$\ell^1$-PnP.}
		\label{subfig:shape:l1pnp}
	\end{subfigure}	
	\caption{Reconstructions of the shape phantom. We display here the 10-th xz-slice. }
	\label{fig:exp1:shape}
\end{figure}

\def\imratio{0.23}
\begin{figure}[t]
	\centering
	\begin{subfigure}[t]{\imratio\linewidth}
		\includegraphics[width=\linewidth]{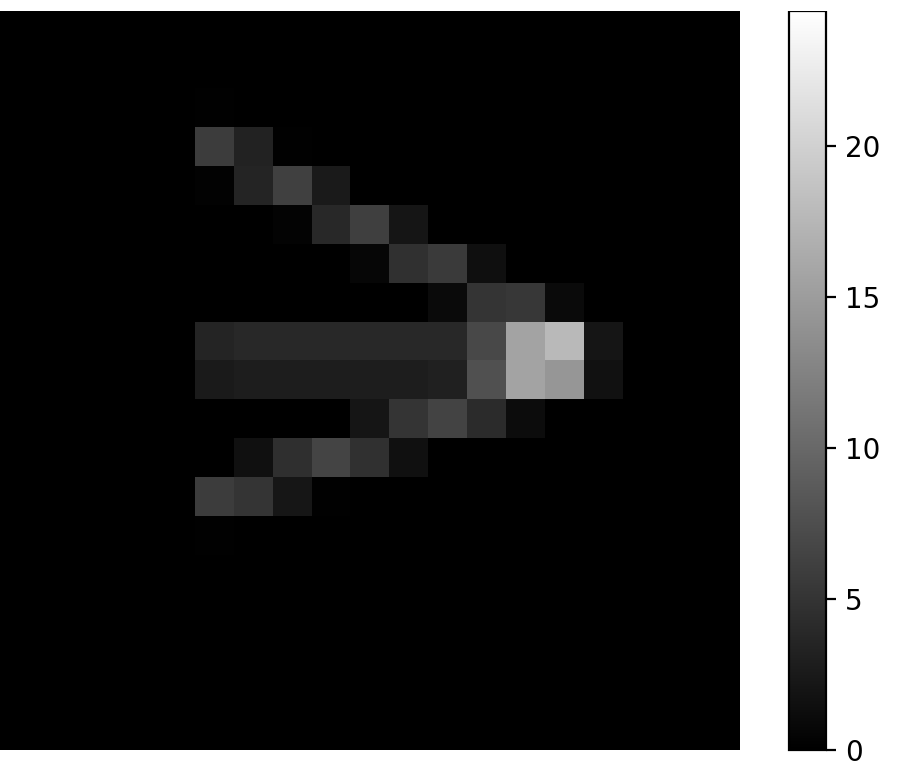}
		\caption{\centering\scriptsize CAD reference.}
		\label{subfig:res:gt}
	\end{subfigure}
	\hfil
	\begin{subfigure}[t]{\imratio\linewidth}
		\includegraphics[width=\linewidth]{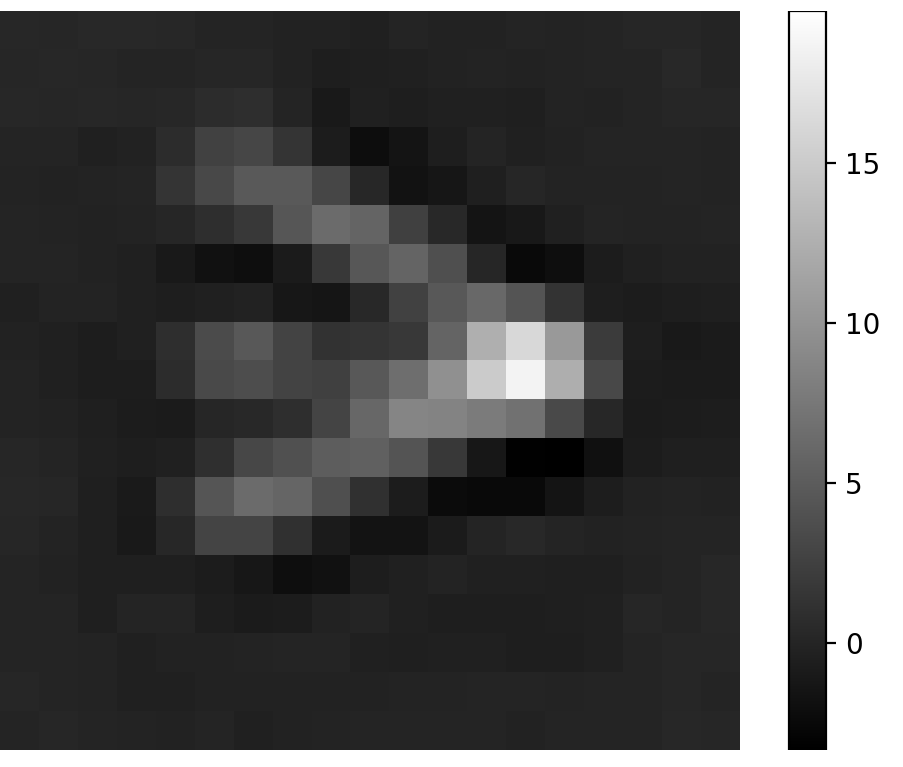}
		\caption{\centering\scriptsize Tikhonov.}
		\label{subfig:res:tik}
	\end{subfigure}
	\hfil
	\begin{subfigure}[t]{\imratio\linewidth}
		\includegraphics[width=\linewidth]{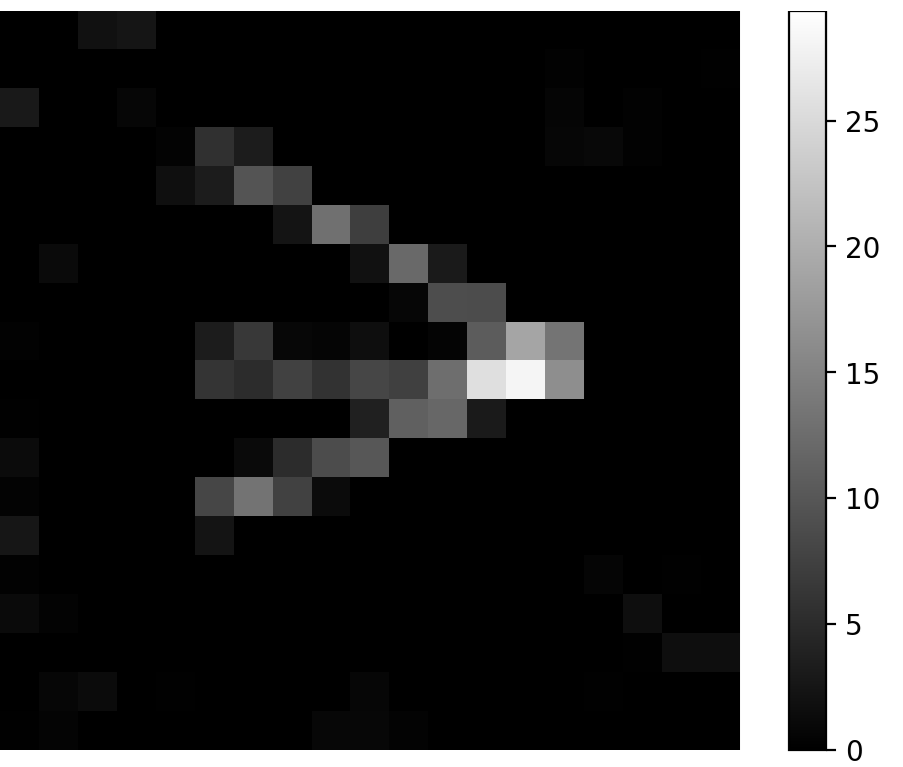}
		\caption{\centering\scriptsize ART.}
		\label{subfig:res:ART}
	\end{subfigure}
	\hfil
	\begin{subfigure}[t]{\imratio\linewidth}
		\includegraphics[width=\linewidth]{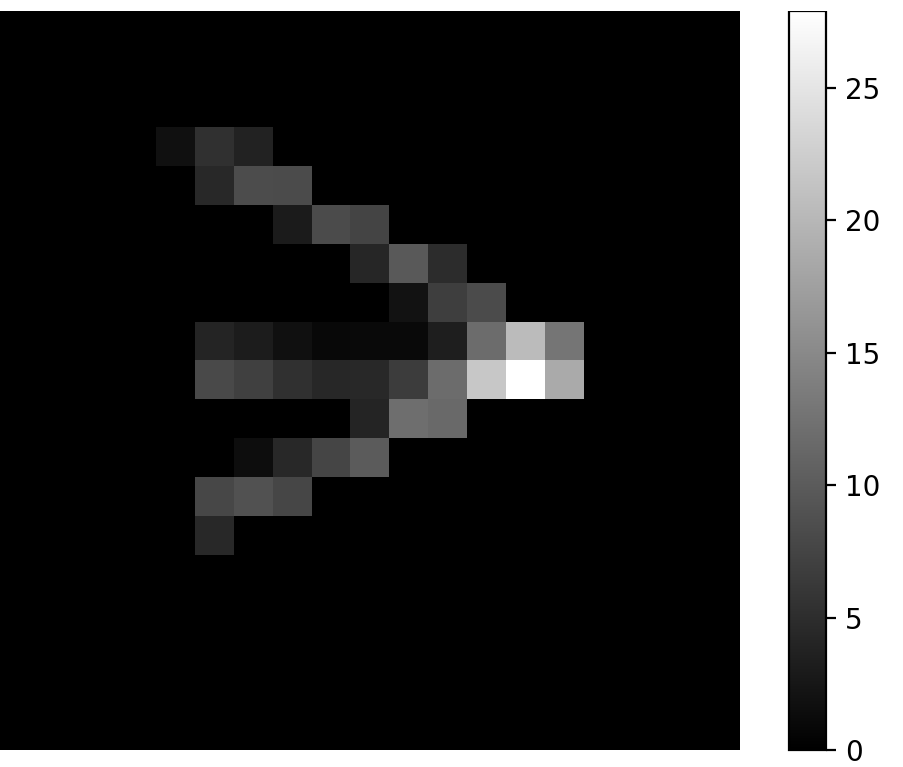}
		\caption{\centering\scriptsize DIP.}
		\label{subfig:res:dip}
	\end{subfigure}
	\par\medskip
	\begin{subfigure}[t]{\imratio\linewidth}
		\includegraphics[width=\linewidth]{results/empty.png}
	\end{subfigure}
	\hfil
	\begin{subfigure}[t]{\imratio\linewidth}
		\includegraphics[width=\linewidth]{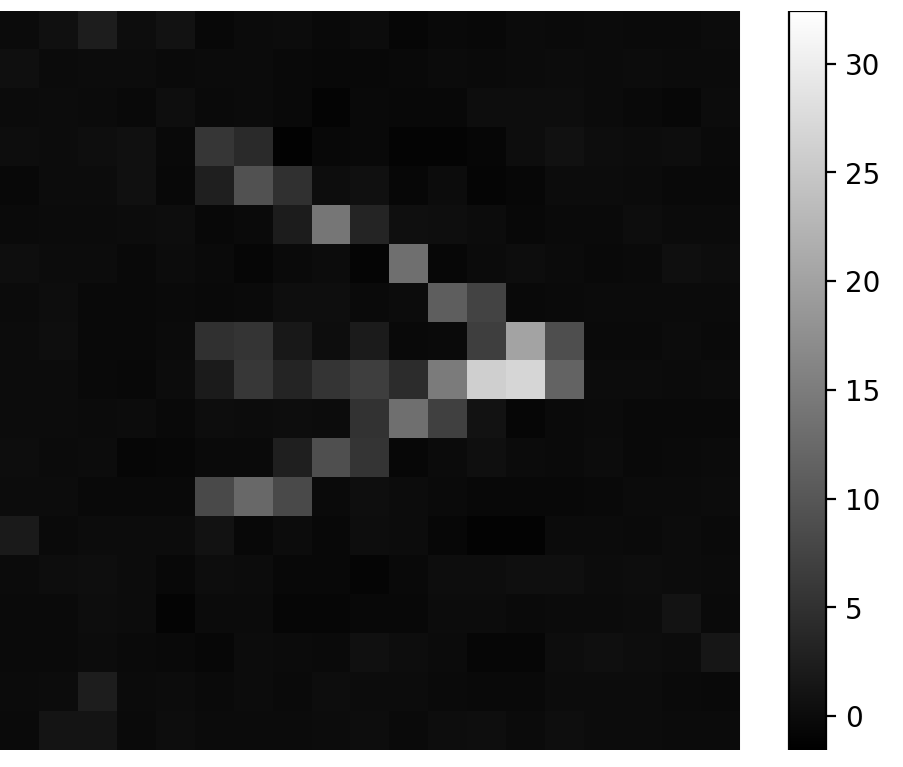}
		\caption{\centering\scriptsize PP-MPI.}
		\label{subfig:res:ppmpi}
	\end{subfigure}
	\hfil
	\begin{subfigure}[t]{\imratio\linewidth}
		\includegraphics[width=\linewidth]{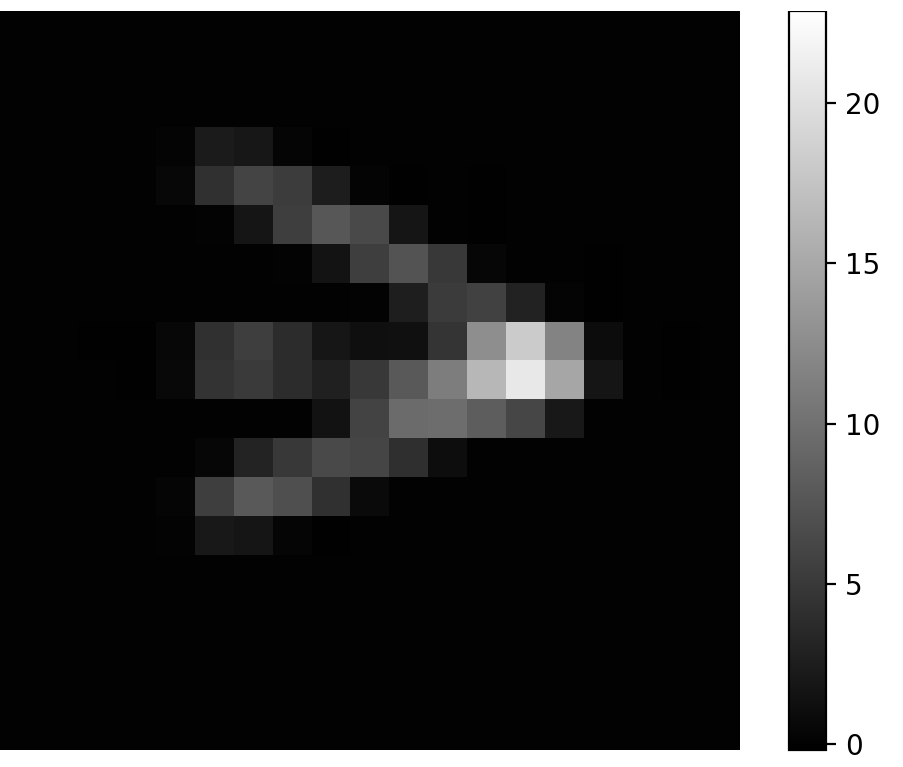}
		\caption{\centering\scriptsize ZeroShot-PnP .}
		\label{subfig:res:hqs}
	\end{subfigure}
	\hfil
	\begin{subfigure}[t]{\imratio\linewidth}
		\includegraphics[width=\linewidth]{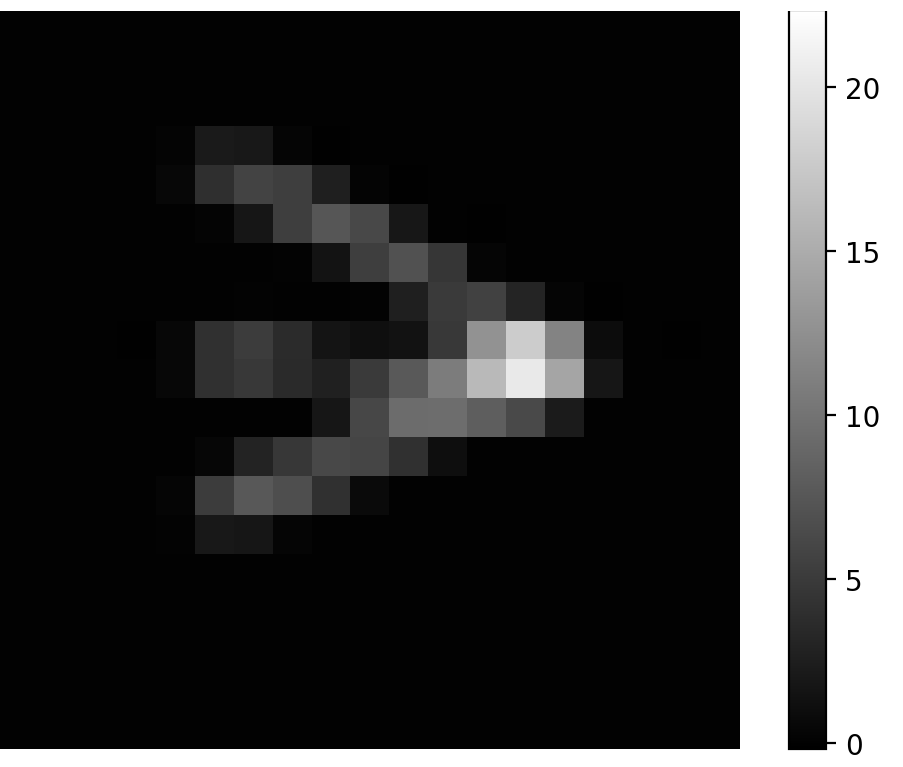}
		\caption{\centering\scriptsize ZeroShot-$\ell^1$-PnP.}
		\label{subfig:res:l1pnp}
	\end{subfigure}
	
	\caption{Reconstructions of the resolution phantom. We display the 10-th xy-slice. }
	\label{fig:exp1:resolution}
\end{figure}

\def\imratio{0.23}
\begin{figure}[t]
	\centering
	\begin{subfigure}[t]{\imratio\linewidth}
		\includegraphics[width=\linewidth]{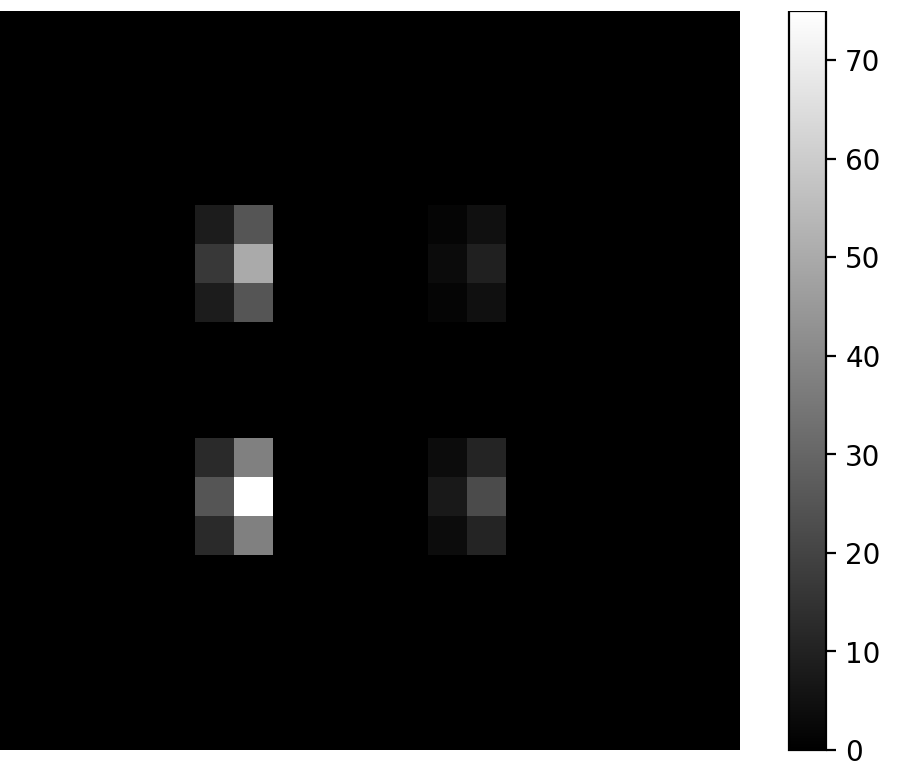}
		\caption{\centering\scriptsize CAD reference.}			
		\label{subfig:conc:gt}
	\end{subfigure}
	\hfil
	\begin{subfigure}[t]{\imratio\linewidth}
		\includegraphics[width=\linewidth]{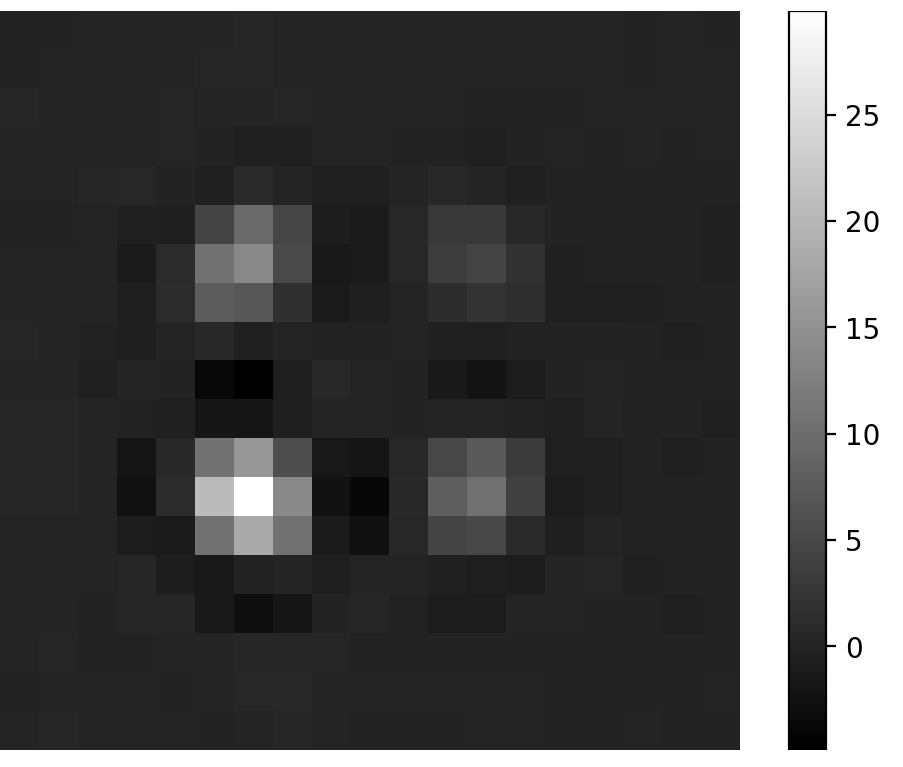}
		\caption{\centering\scriptsize Tikhonov.}			
		\label{subfig:conc:tik}
	\end{subfigure}
	\hfil
	\begin{subfigure}[t]{\imratio\linewidth}
		\includegraphics[width=\linewidth]{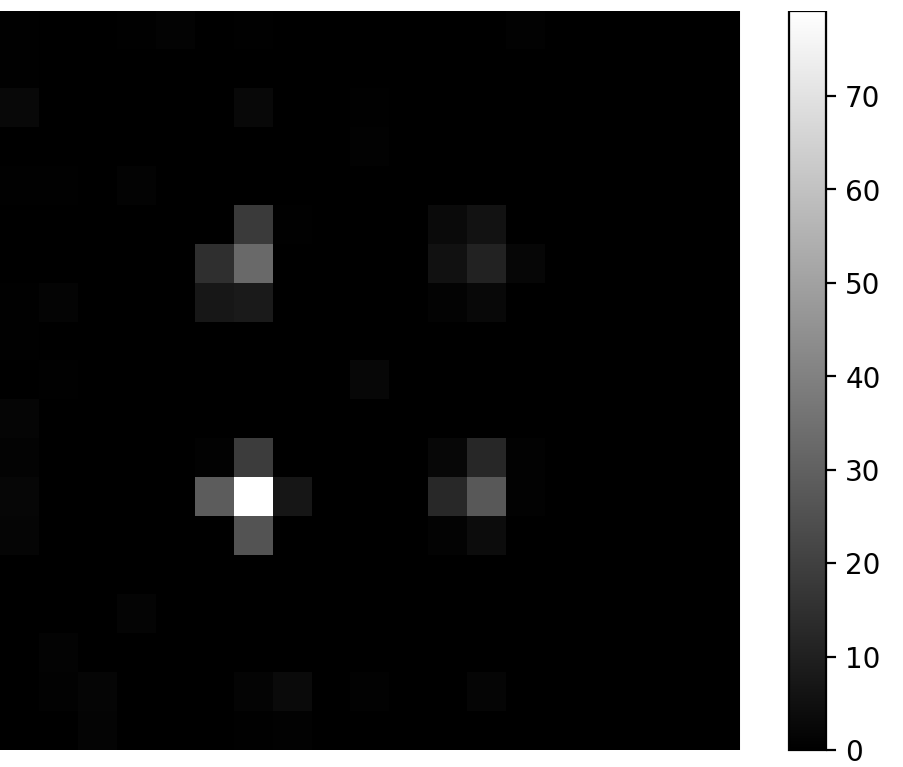}
		\caption{\centering\scriptsize ART.}			
		\label{subfig:conc:ART}
	\end{subfigure}
	\hfil
	\begin{subfigure}[t]{\imratio\linewidth}
		\includegraphics[width=\linewidth]{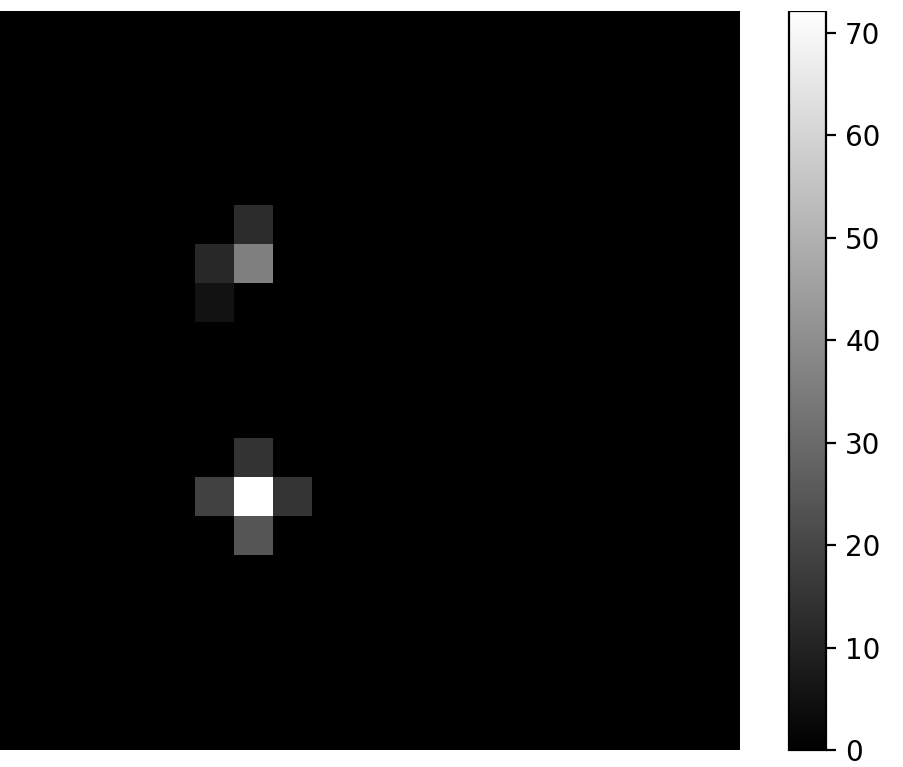}
		\caption{\centering\scriptsize DIP.}			
		\label{subfig:conc:dip}
	\end{subfigure}
	\par\medskip
	\begin{subfigure}[t]{\imratio\linewidth}
		\includegraphics[width=\linewidth]{results/empty.png}
	\end{subfigure}
	\hfil
	\begin{subfigure}[t]{\imratio\linewidth}
		\includegraphics[width=\linewidth]{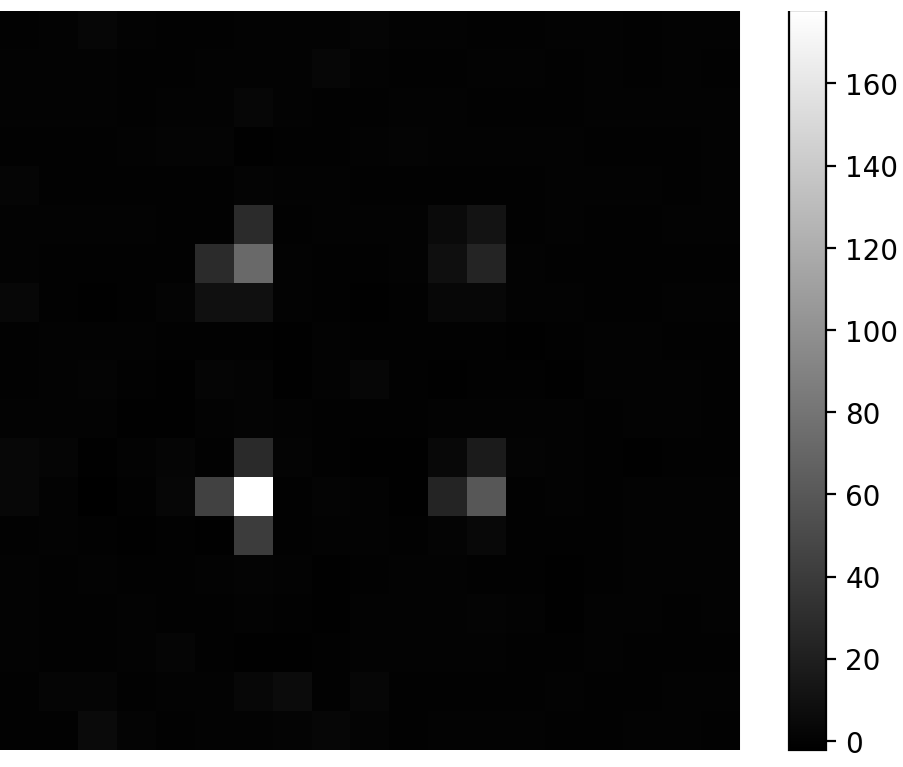}
		\caption{\centering\scriptsize PP-MPI.}			
		\label{subfig:conc:ppmpi}
	\end{subfigure}
	\hfil
	\begin{subfigure}[t]{\imratio\linewidth}
		\includegraphics[width=\linewidth]{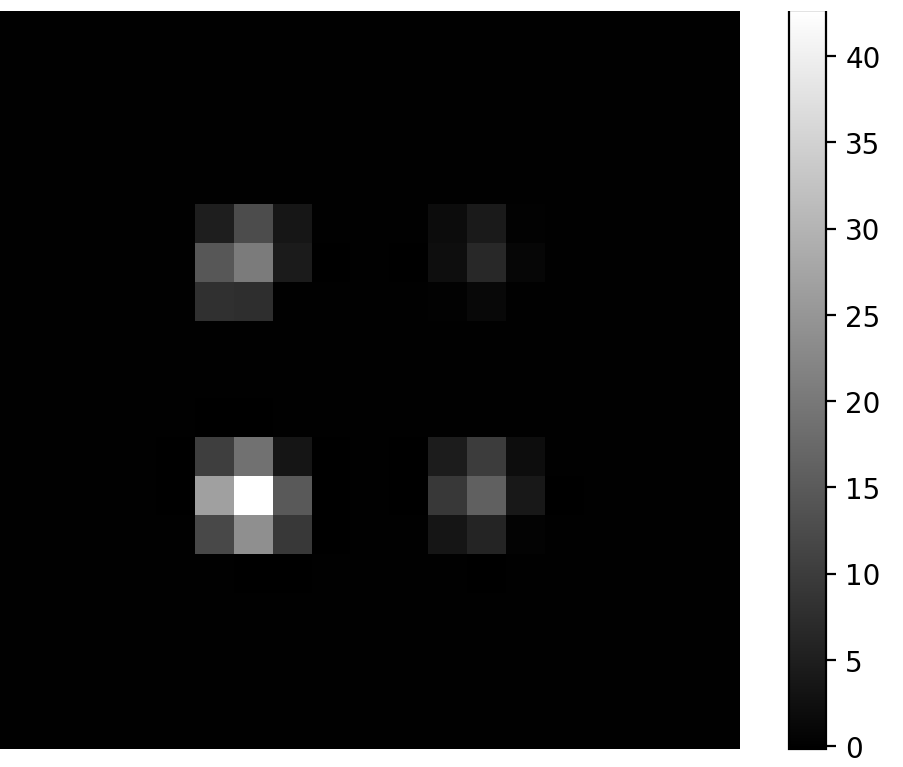}
		\caption{\centering\scriptsize ZeroShot-PnP.}			
		\label{subfig:conc:hqs}
	\end{subfigure}
	\hfil
	\begin{subfigure}[t]{\imratio\linewidth}
		\includegraphics[width=\linewidth]{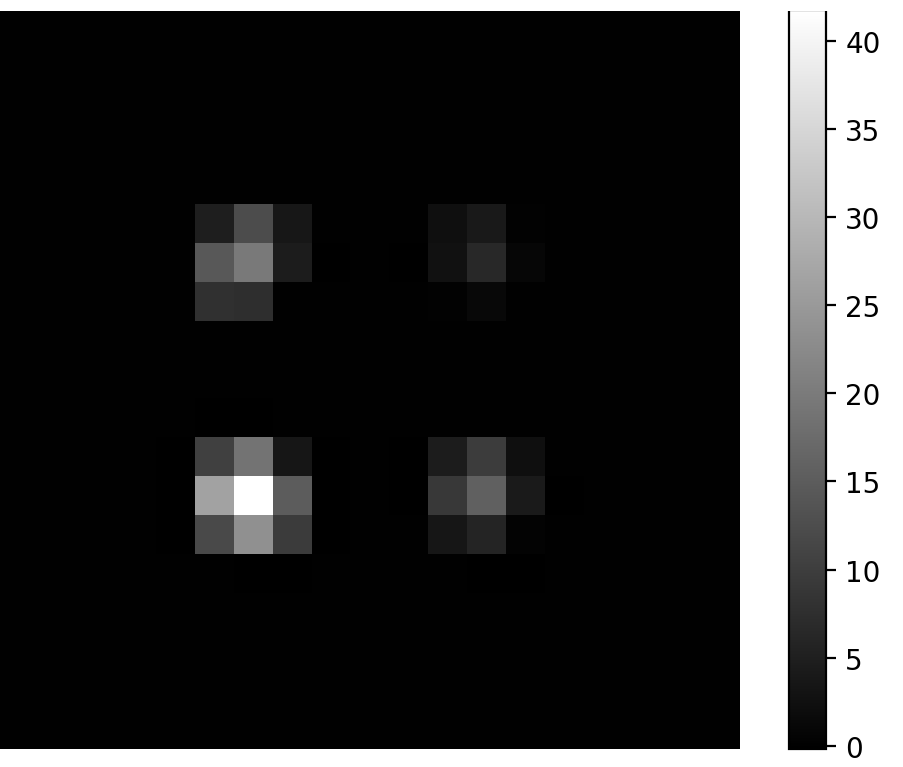}
		\caption{\centering\scriptsize ZeroShot-$\ell^1$-PnP.}			
		\label{subfig:conc:l1pnp}
	\end{subfigure}

	\caption{Reconstructions for the concentration phantom. We display the 13-th xy-slice. We observe that, even though the PSNR value of the DIP reconstruction is high (cf. Table~\ref{tab:exp1:measures}), it is possible that parts of the phantom are missing. }
	\label{fig:exp1:concentration}
\end{figure}

\paragraph{Results.} We apply the proposed methods and the comparison methods to the Open MPI Dataset. The reconstructions where performed with the parameters validated on the hybrid dataset (Table~\ref{tab:exp1:validation}). Selected slices of the reconstructions of the three phantoms are displayed in Figures~\ref{fig:exp1:shape},~\ref{fig:exp1:resolution} and~\ref{fig:exp1:concentration}.
\begin{table*}[t]
	\footnotesize{
		\begin{center}
			\begin{tabular}{ |c|c|c|c|c|c|c|c| }
				\hline
				Phantom & Metric & \multicolumn{4}{c|}{Comparison Methods} & \multicolumn{2}{c|}{Proposed ZeroShot-PnP}\\
				\hline
				& & Tikhonov & ART & DIP & PP-MPI~\cite{askin2022pnp} & with $\ell^1$-Prior & without $\ell^1$-Prior\\
				\hline
				Shape & $\mathrm{PSNR}_\mathrm{max}$ & 22.64 & 19.73 & 28.96 & 17.65 & 31.84 & 31.87 \\
				Phatom& $\mathrm{SSIM}_\mathrm{max}$ & 0.489 & 0.679 & 0.871 & 0.654 & 0.954 & 0.954 \\
				\hline
				Resolution & $\mathrm{PSNR}_\mathrm{max}$ & 30.64 & 31.80 & 32.89 & 30.69 & 32.1 & 32.09 \\
				Phantom & $\mathrm{SSIM}_\mathrm{max}$ & 0.593 & 0.695 & 0.737 & 0.658 & 0.731 & 0.730 \\
				\hline
				Concentr. & $\mathrm{PSNR}_\mathrm{max}$ & 36.14 & 39.47 & 37.78 & 35.32 & 37.54 & 37.45 \\
				Phatom& $\mathrm{SSIM}_\mathrm{max}$ & 0.518 & 0.471 & 0.441 & 0.522 & 0.578 & 0.577 \\
				\hline
			\end{tabular}
		\end{center}
	}
	\caption{Results obtained with the ZeroShot-$\ell^1$-PnP, the Tikhonov, the ART, the DIP and the PP-MPI methods on the 3D Open MPI dataset. We display the $\mathrm{PSNR}_\mathrm{max}$ and $\mathrm{SSIM}_\mathrm{max}$ scores for the reconstructions obtained using the validated parameters in Table~\ref{tab:exp1:validation}.}
	\label{tab:exp1:measures}
\end{table*}
An overview of the results obtained in terms of $\mathrm{PSNR}_\mathrm{max}$ and $\mathrm{SSIM}_\mathrm{max}$ are displayed in Tab.~\ref{tab:exp1:measures}. 
We can observe that the ZeroShot-$\ell^1$-PnP and the ZeroShot-PnP methods achieve similar results, and in some instances the additional $\ell^1$-prior can increase the PSNR score (cf. shape phantom). In general, the additional $\ell^1$ prior does not appear to play a big role in the reconstruction, suggesting that for many applications, the ZeroShot-PnP algorithm could be sufficient. The reconstruction quality according to  SSIM of ZeroShot-$\ell^1$-PnP and the ZeroShot-PnP are higher than the baseline Tikhonov, the ART method and the previous PP-MPI approach. The same holds true for the PSNR metric except for the concentration phantom, for which the ART method obtained the highest PSNR score (cf. Table~\ref{tab:exp1:measures}). Concerning comparison with the DIP method, we observe from Table~\ref{tab:exp1:measures} that the proposed methods achieve higher PSNR values on the shape phantom, whereas it achieves lower PSNR values on the resolution and concentration phantoms; the SSIM score achieved by the proposed methods are higher also on the concentration phantoms.

However, the DIP takes considerably more time to perform a reconstruction (cf. runtime in Table~\ref{tab:exp1:runtimes}) and involves a random initial state $z$ on which the quality of the reconstruction depends. Moreover, comparing the DIP reconstruction in Figure~\ref{subfig:conc:dip} with the CAD reference~\ref{subfig:conc:gt} and the ZeroShot-$\ell^1$-PnP reconstruction~\ref{subfig:conc:l1pnp}, it is possible that even if the PSNR value is higher than the value obtained with the proposed methods, some features of the phantom may not be reconstructed by the DIP.
\begin{figure}[t]
	\centering
	\begin{subfigure}[t]{\linewidth}
		\includegraphics[width=\linewidth]{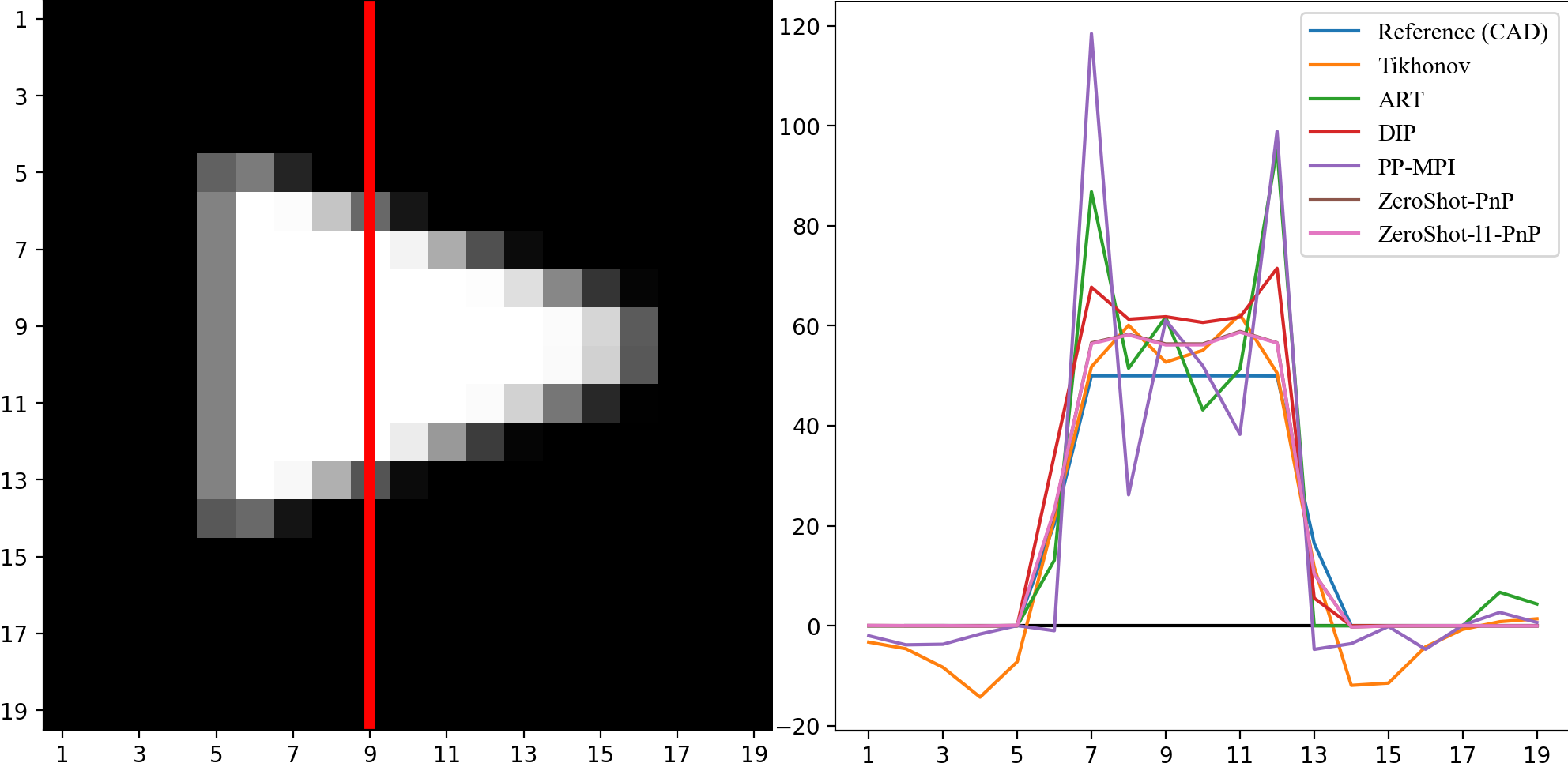}
	\end{subfigure}
	\caption{Example of a profile cut for the shape phantom. We consider the 10-th xz-slice and the cut along the 9-th x pixel. We point out that the profiles obtained with the ZeroShot-PnP and the ZeroShot-$\ell^1$-PnP overlap almost perfectly.}
	\label{fig:exp1:line}
\end{figure}
Because the $19\times 19\times 19$ resolution of the reconstructions, the visual quality assessment of the differences between reconstructions can be difficult. For a better understanding of the differences we have plotted in Figure~\ref{fig:exp1:line} a 1D line plot that compares the graph of the reconstructions and the reference CAD model. In Figure~\ref{fig:exp1:line} we can see that the Tikhonov method smooths out the reference and takes negative values, whereas the ART, the DIP and the PP-MPI tend to overshoot when compared to the ZeroShot-PnP reconstruction. The ZeroShot-$\ell^1$-PnP and the ZeroShot-PnP lines are almost completely overlapping. We observe that the DIP can present artifacts in the reconstruction (cf. Figure~\ref{subfig:conc:dip} and~\ref{subfig:conc:gt}) and that the ART method can achieve good results in some phantoms (e.g., the PSNR measure for the concentration phantom in Table~\ref{tab:exp1:measures}).

All 2D slices for each of the reconstructed phantom using the proposed and the comparison methods can be found in the Supplementary Material.
\begin{table}[ht!]
	\footnotesize{
		\begin{center}
			\begin{tabular}{ |c|c|c|c|  }
				\hline
				& Shape ph.& Res. ph.& Conc. ph.\\
				\hline
				Tikhonov & $6.25\cdot 10^{-4}$ s & $6.59\cdot 10^{-4}$ s & $6.54\cdot 10^{-4}$ s  \\
				ART & 573.90 s & 597.10  s & 574.38 s \\
				DIP & 129.93 s & 165.87 s &  165.46 s \\
				PP-MPI & 251.25 s & 251.71 s & 251.65 s  \\
				ZeroShot-PnP & 1.10 s & 1.16 s & 1.02 s \\
				ZeroShot-$\ell^1$-PnP & 0.72 s & 0.74 s &  0.74 s \\
				\hline
			\end{tabular}
		\end{center}
	}
	\caption{Wall clock runtime table of the examined methods. The times are computed by averaging 100 reconstructions on the Open MPI dataset. We observe the proposed ZeroShot-PnP and ZeroShot-$\ell^1$-PnP take the least amount of time for the reconstructions excluding the Tikhonov method.}
	\label{tab:exp1:runtimes}
\end{table}
\paragraph{Runtimes.} Finally, we have measured the wall clock reconstruction times for all methods. Because of the use of the SVD in this experiment, Tikhonov type reconstructions are very speedy on the GPU, in particular, one direct solution takes about $6\cdot 10^{-4}$ seconds.

From Table~\ref{tab:exp1:runtimes} we observe that - using the validated parameters in Table~\ref{tab:exp1:validation} on the Open MPI Dataset - apart from the Tikhonov method, the ZeroShot-PnP and ZeroShot-$\ell^1$-PnP methods are the fastest in this comparison and achieve the results in Table~\ref{tab:exp1:measures} in less than 2 seconds.

\subsection{Experiment 2: Study on the Differences with the PP-MPI Approach.}\label{sec:exp1bis}

In this experiment we study more closely the differences between the previous PP-MPI approach and the ZeroShot-$\ell^1$-PnP approaches proposed for MPI in this paper. The main difference between the PP-MPI approach and the ZeroShot-$\ell^1$-PnP lies in the denoiser employed; the employment of the $\ell^1$-HQS-splitting in Equations~\eqref{eq:PnP:split:data2}-\eqref{eq:PnP:split:st} is directly related to the choice of the ZeroShot-Denoiser~\cite{Zhang2022pnp}. Indeed, the ZeroShot-Denoiser can take as input the noise level map of the reconstruction. To show the benefit of using the ZeroShot-$\ell^1$-PnP algorithm in combination with the ZeroShot-Denoiser (as proposed), we consider in this experiment the following additional method: we substitute the PP-MPI-Denoiser~\cite{askin2022pnp} in the $\ell^1$-HQS-splitting in place of the ZeroShot-Denoiser.

\paragraph{Preprocessing.} We consider the same preprocessing as in Experiment 1.

\paragraph{Parameter Validation.} We choose the parameters according to Section~\ref{sec:param:selection}.
The validated starting parameters $\mu_0$ as well as the resulting number of iterations are displayed  in Table~\ref{tab:exp1bis:validation}. We observe from Table~\ref{tab:exp1bis:validation} that the proposed ZeroShot-PnP and ZeroShot-$\ell^1$-PnP achieve the highest PSNR and SSIM scores on the validation dataset. Moreover, we observe that, on the validation dataset, the HQS-splittings (with and without $\ell^1$-prior) combined with the PP-MPI-Denoiser have lower PSNR scores when compared to the original PP-MPI method and the proposed methods; nevertheless, the $\ell^1$-HQS-splitting with the PP-MPI-Denoiser has a lower number of validated iterations $n_{\mathrm{it}}$ and higher SSIM score when compared with the PP-MPI method.

\begin{table}[t]
	\footnotesize{Hybrid Dataset:}
	\tiny{
		\begin{center}
			\begin{tabular}{ |c|c|c|c|c|c| }
				\hline
				& \multicolumn{3}{c|}{PP-MPI-Denoiser~\cite{askin2022pnp}} & \multicolumn{2}{c|}{Proposed Denoiser~\cite{Zhang2022pnp}}\\
				\hline
				& PP-MPI & HQS-Splitting Eq.~\eqref{eq:HQS:pnp} & $\ell^1$-HQS-Splitting Eq.~\eqref{eq:PnP:split:data2}-\eqref{eq:PnP:split:st} & ZeroShot-PnP & ZeroShot-$\ell^1$-PnP \\ 
				\hline
				Param.& $\mathrm{SNR}_{\mathrm{db}} = $22 & $\mu_0 = 5\cdot 10^{11}$ & $\mu_0 = 8\cdot 10^{11}$ & $\mu_0 = 2\cdot 10^{5}$ & $\mu_0 = 3\cdot 10^{5}$\\
				Iteration & $n_\mathrm{it}=$1607  & $n_\mathrm{it}8$ & $n_\mathrm{it}=7$ & $n_\mathrm{it}=7$ & $n_\mathrm{it}=5$\\
				\hline
				PSNR &  $28.13 \pm 3.75$ & $27.06 \pm 4.80$ & $26.92 \pm 4.78$ & $30.45 \pm 5.04$ & $29.77 \pm 5.82$\\
				\hline
				SSIM &   $0.408 \pm 0.217$ & $0.483 \pm 0.208$ & $0.493 \pm 0.213$ & $0.788\pm 0.141$ & $0.766\pm 0.189$\\
				\hline
			\end{tabular}
		\end{center}
	}
	\footnotesize{Open MPI Dataset:}
	\tiny{
		\begin{center}
			\begin{tabular}{ |c|c|c|c|c|c|c| }
				\hline
				Phantom & Metric & \multicolumn{3}{c|}{PP-MPI-Denoiser~\cite{askin2022pnp}} & \multicolumn{2}{c|}{Proposed Denoiser~\cite{Zhang2022pnp}}\\
				\hline
				&  & PP-MPI~\cite{askin2022pnp} & HQS-Spl. Eq.~\eqref{eq:HQS:pnp}& $\ell^1$-HQS-Spl. Eq.~\eqref{eq:PnP:split:data2}-\eqref{eq:PnP:split:st}& ZeroShot-PnP & ZeroShot-$\ell^1$-PnP\\
				\hline
				Shape & $\mathrm{PSNR}_\mathrm{max}$ & 17.65 & 26.91 & 27.45 & 31.87 & 31.84 \\
				Phatom & $\mathrm{SSIM}_\mathrm{max}$ & 0.654 & 0.712& 0.768 & 0.954 & 0.954\\
				\hline
				Resolution & $\mathrm{PSNR}_\mathrm{max}$ & 30.69 & 27.46 & 26.44 & 32.09 & 32.1\\
				Phantom & $\mathrm{SSIM}_\mathrm{max}$ & 0.658 & 0.0003 & 0.0574 & 0.730 & 0.731\\
				\hline
				Concentr. & $\mathrm{PSNR}_\mathrm{max}$  & 35.32 & 33.43 & 32.86  & 37.45 & 37.54\\
				Phatom & $\mathrm{SSIM}_\mathrm{max}$  & 0.522 & 0.0041 & 0.0543 & 0.577 & 0.578\\
				\hline
			\end{tabular}
		\end{center}
	}
	\caption{Validation on the hybrid dataset in Section~\ref{sec:simul:dataset} and qualitative comparison on the Open MPI Dataset of the PP-MPI method, the proposed algorithms and the joint $\ell^1$-HQS splitting scheme with the PP-MPI-Denoiser. We observe that the proposed ZeroShot-PnP and ZeroShot-$\ell^1$-PnP algorithms yields the higher PSNR and SSIM score both on the hybrid dataset and on the Open MPI Dataset.
	}
	\label{tab:exp1bis:validation}
	
\end{table}
\paragraph{Results.} We have applied the algorithms with the validated parameters on the Open MPI Dataset. From the results in Table~\ref{tab:exp1bis:validation}, we observe that the proposed ZeroShot-$\ell^1$-PnP and ZeroShot-PnP algorithms achieve the highest PSNR and SSIM scores. On the validation dataset (cf. Table~\ref{tab:exp1bis:validation}) we have observed that employing the PP-MPI-Denoiser in combination with the $\ell^1$-HQS-splitting scheme leads to a higher SSIM score. This results is only partially achieved on the real Open MPI Dataset: from Table~\ref{tab:exp1bis:validation} it is visible that the $\ell^1$-HQS-splitting with the PP-MPI-Denoiser improves the SSIM of the shape phantom when compared with the PP-MPI method, but at the same time, it fails in reconstructing the resolution and concentration phantoms (the SSIM scores are low). In particular, this suggests that the proposed $\ell^1$-HQS-splitting scheme works well in combination with the ZeroShot-Denoiser, and the automatic parameter update proposed.

\subsection{Experiment 3: Reconstructions with Decreasing Levels of Preprocessing for the ZeroShot-$\ell^1$-PnP.}\label{sec:exp2}

In our third experiment we consider the shape phantom and study the proposed ZeroShot-PnP and ZeroShot-$\ell^1$-PnP algorithms with respect to different levels of preprocessing.

\paragraph{Preprocessing.} We consider three different preprocessing levels for the shape phantom data. For all three preprocessing levels we consider the frequencies between 80 and 625 kHz. At first we consider whitening and an rSVD with reduced rank $K=2000$; secondly, we do not perform whitening but still consider the rSVD with $K=2000$; finally, we do not perform whitening and consider a full rank rSVD with $K=19^3 = 6859$.

\paragraph{Parameters.} Consistently with Experiment 1, we validate the parameters using our hybrid dataset. The validated parameters are displayed in Table~\ref{tab:exp3:validation}.

\begin{table}[ht!]
	\footnotesize{
		\begin{center}
			\begin{tabular}{ |c|c|c|c|c|  }
				\hline
				Method & Quantity &  \multicolumn{3}{c|}{Pre-Processing}\\ 
				\hline
				\multicolumn{2}{c|}{}  & Whitening & No Whitening & No Whitening \\
				\multicolumn{2}{c|}{}  & rSVD K=2000 & rSVD K=2000 & rSVD K=6859\\
				\hline
				ZeroShot &  PSNR & $30.45 \pm 5.04$ & $29.73 \pm 4.97$ &  $33.90 \pm 6.24$ \\
				PnP & SSIM & $0.788 \pm 0.141$ & $0.805 \pm 0.091$ & $0.831 \pm 0.124$ \\
				\hline
				\multicolumn{1}{c|}{}& $n_\mathrm{it}$& 7 & 8  &  9\\ 
				\multicolumn{1}{c|}{}& $\mu_0$ & $2\cdot 10^5$ & $7\cdot 10^{8}$ & $10^9$ \\
				\hline
				ZeroShot & PSNR & $29.77 \pm 5.82$ & $29.06 \pm 5.83$ & $32.40 \pm 5.01$ \\
				$\ell^1$-PnP & SSIM & $0.766 \pm 0.189$ & $0.802 \pm 0.164$ & $0.810 \pm 0.161$ \\
				\hline
				\multicolumn{1}{c|}{}& $n_\mathrm{it}$ & 5 & 4  &  75\\ 
				\multicolumn{1}{c|}{}& $\mu_0$ & $3\cdot 10^{5}$ & $7\cdot 10^{8}$  & $5\cdot 10^{10}$ \\
				\cline{2-5\emph{}}
			\end{tabular}
		\end{center}
	}
	\caption{The ZeroShot-$\ell^1$-PnP and ZeroShot-PnP algorithms validated with different levels of preprocessing.}
	\label{tab:exp3:validation}
\end{table}

\begin{table}[ht!]
	\footnotesize{
		\begin{center}
			\begin{tabular}{ |c|c|c|c|c|c|  }
				\hline
				Phantom & Method & Quantity &  \multicolumn{3}{c|}{Pre-Processing}\\ 
				\hline
				\multicolumn{3}{c|}{}  & Whitening & No Whitening & No Whitening \\
				\multicolumn{3}{c|}{}  & rSVD K=2000 & rSVD K=2000 & rSVD K=6859\\
				\hline
				Shape & ZeroShot &  $\mathrm{PSNR}_{\mathrm{max}}$ & 31.84 & 31.10 & 31.30 \\
				Phantom & PnP& $\mathrm{SSIM}_{\mathrm{max}}$ & 0.954 & 0.906 & 0.950 \\
				\cline{1-6\emph{}}
				\multicolumn{1}{c|}{} & ZeroShot &  $\mathrm{PSNR}_{\mathrm{max}}$ & 31.87 & 31.14 & 29.8\\
				\multicolumn{1}{c|}{} & $\ell^1$-PnP & $\mathrm{SSIM}_{\mathrm{max}}$ & 0.954 & 0.947  & 0.956 \\
				\hline
				Resolution & ZeroShot &  $\mathrm{PSNR}_{\mathrm{max}}$ &32.10 & 32.08 & 31.91\\
				Phantom& PnP& $\mathrm{SSIM}_{\mathrm{max}}$& 0.731 & 0.275  & 0.7132 \\
				\cline{1-6\emph{}}
				\multicolumn{1}{c|}{} & ZeroShot&  $\mathrm{PSNR}_{\mathrm{max}}$& 32.09 & 32.09 & 31.04 \\
				\multicolumn{1}{c|}{} & $\ell^1$-PnP & $\mathrm{SSIM}_{\mathrm{max}}$ & 0.730 & 0.752 & 0.632 \\
				\hline
				Concentration & ZeroShot &  $\mathrm{PSNR}_{\mathrm{max}}$& 37.54 & 37.22 & 39.32 \\
				Phantom & PnP& $\mathrm{SSIM}_{\mathrm{max}}$& 0.578 & 0.308 & 0.546 \\
				\cline{1-6\emph{}}
				\multicolumn{1}{c|}{} &ZeroShot &  $\mathrm{PSNR}_{\mathrm{max}}$ & 37.45 & 37.16 & 38.78 \\
				\multicolumn{1}{c|}{} & $\ell^1$-PnP & $\mathrm{SSIM}_{\mathrm{max}}$ & 0.577 & 0.568  & 0.489\\
				\cline{2-6\emph{}}
			\end{tabular}
		\end{center}
	}
	\caption{The ZeroShot-$\ell^1$-PnP and ZeroShot-PnP reconstruction on the 3D Open MPI Dataset with the validated parameters in Table \ref{tab:exp3:validation} and metric computed with different levels of preprocessing.}
	\label{tab:exp3:recap}
\end{table}

\paragraph{Results.} In Table~\ref{tab:exp3:recap} we display the PSNR and SSIM values of the reconstructions. 
With decreased levels of preprocessing the results are competitive with the baseline methods (cf. Table~\ref{tab:exp1:measures}) 
and comparable with the reconstruction obtained with the ZeroShot-$\ell^1$-PnP and ZeroShot-PnP in Experiment 1.
\def\imratio{0.32}
\begin{figure}[t]
	\centering
	\begin{subfigure}[t]{\imratio\linewidth}
		\includegraphics[width=\linewidth]{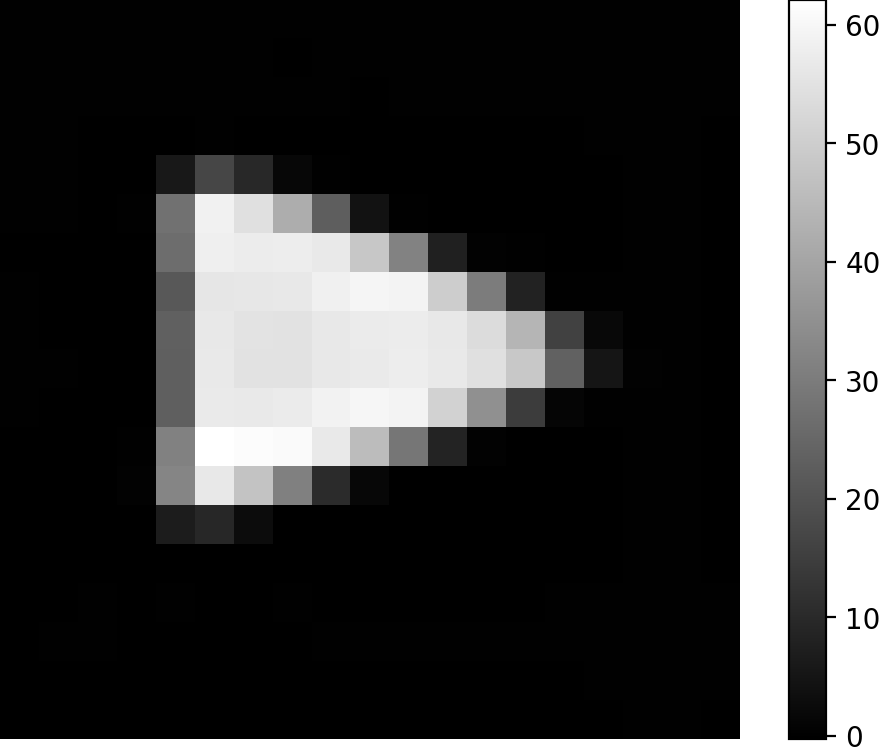}
		\caption{\centering\scriptsize ZeroShot-PnP whitened, rSVD K=2000.}
		\label{subfig:exp3zs:allpre}
	\end{subfigure}
	\hfil
	\begin{subfigure}[t]{\imratio\linewidth}
		\includegraphics[width=\linewidth]{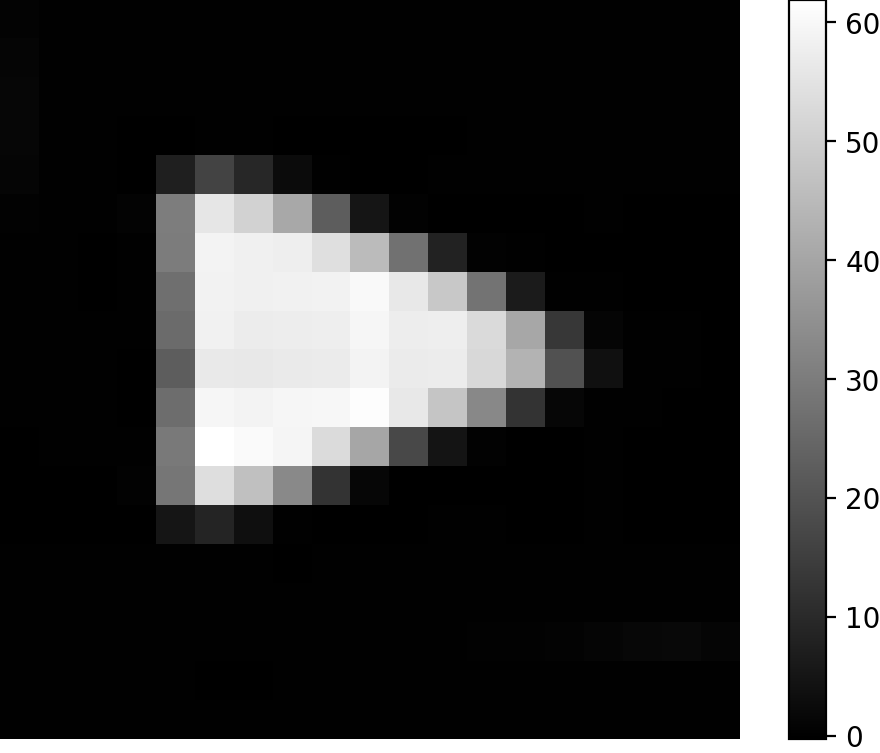}
		\caption{\centering\scriptsize ZeroShot-PnP not whitened, rSVD K=2000.}
		\label{subfig:exp3zs:nowhite}
	\end{subfigure}
	\hfil
	\begin{subfigure}[t]{\imratio\linewidth}
		\includegraphics[width=\linewidth]{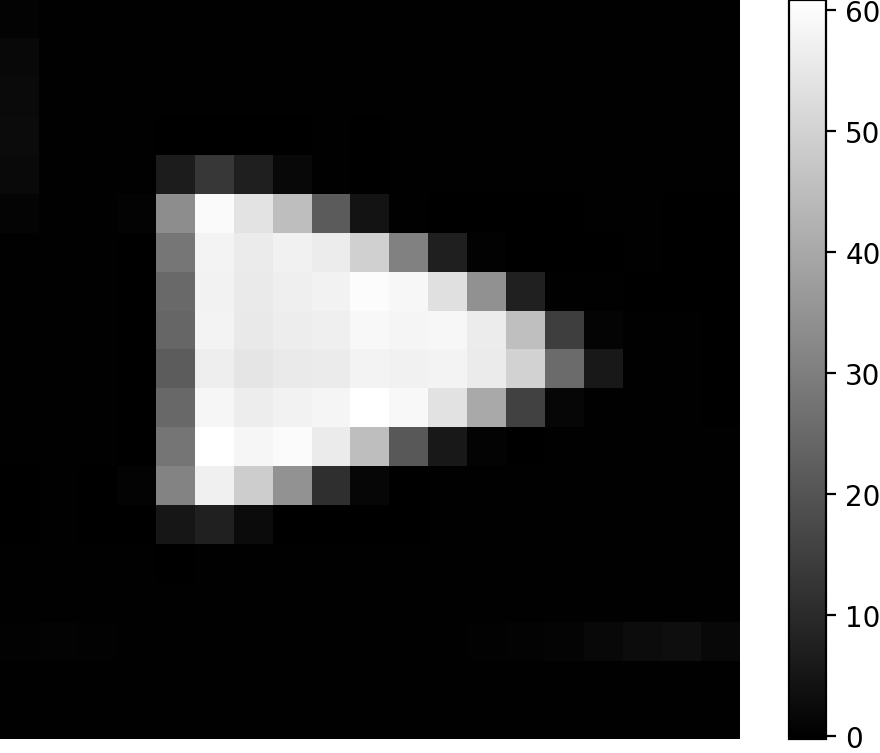}
		\caption{\centering\scriptsize ZeroShot-PnP not whitened, rSVD K=6859.}			
		\label{subfig:exp3zs:nowhite:fullrank}
	\end{subfigure}
	\par\medskip
	\begin{subfigure}[t]{\imratio\linewidth}
		\includegraphics[width=\linewidth]{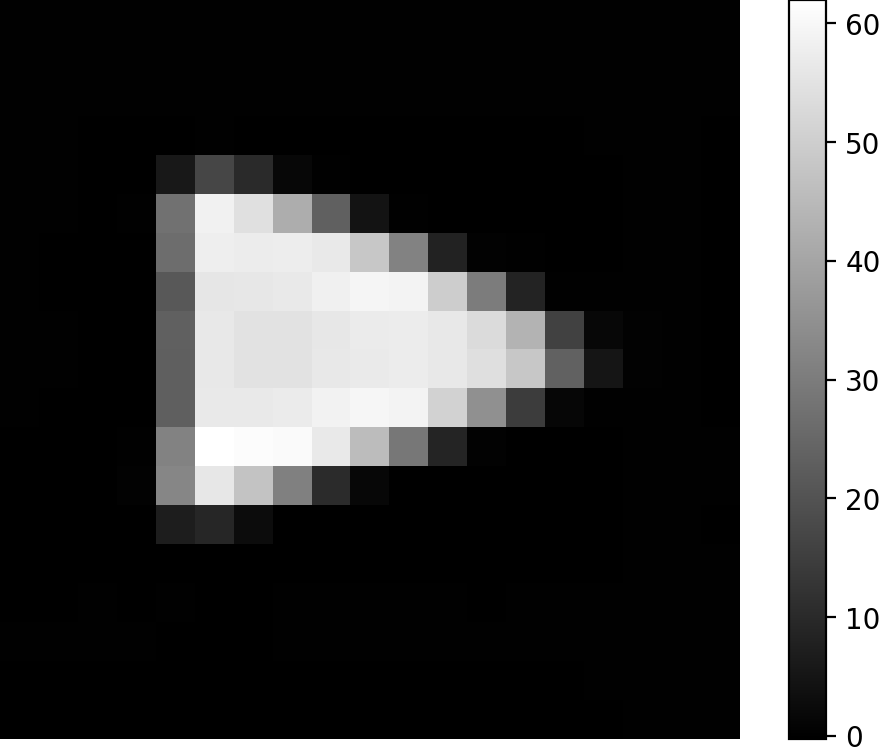}
		\caption{\centering\scriptsize ZeroShot-$\ell^1$-PnP whitened, rSVD K=2000.}
		\label{subfig:exp3l1:allpre}
	\end{subfigure}
	\hfil
	\begin{subfigure}[t]{\imratio\linewidth}
		\includegraphics[width=\linewidth]{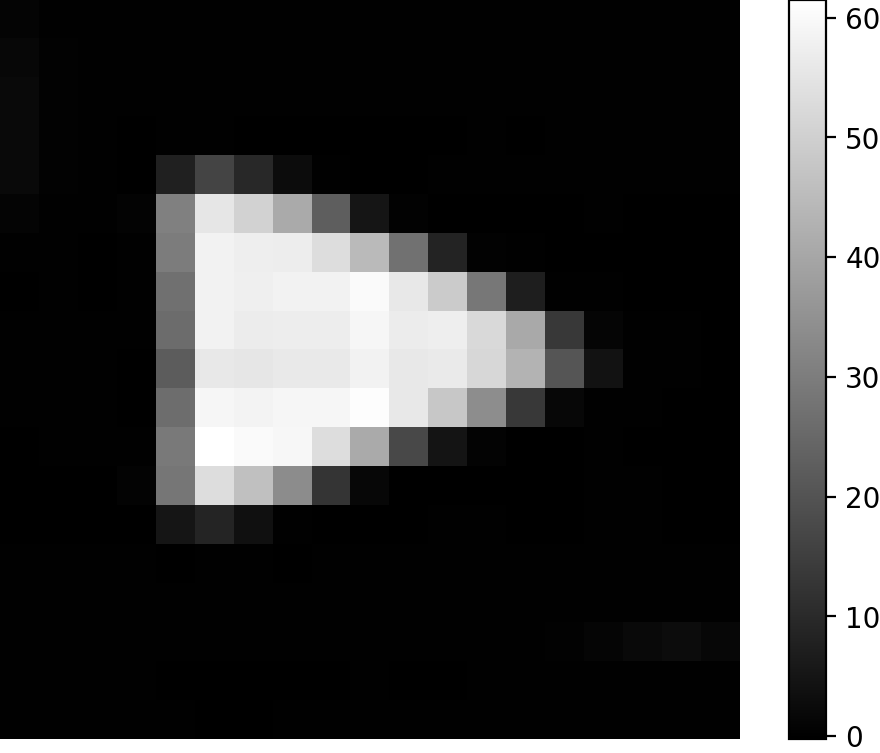}
		\caption{\centering\scriptsize ZeroShot-$\ell^1$-PnP not whitened, rSVD K=2000.}
		\label{subfig:exp3l1:nowhite}
	\end{subfigure}
	\hfil
	\begin{subfigure}[t]{\imratio\linewidth}
		\includegraphics[width=\linewidth]{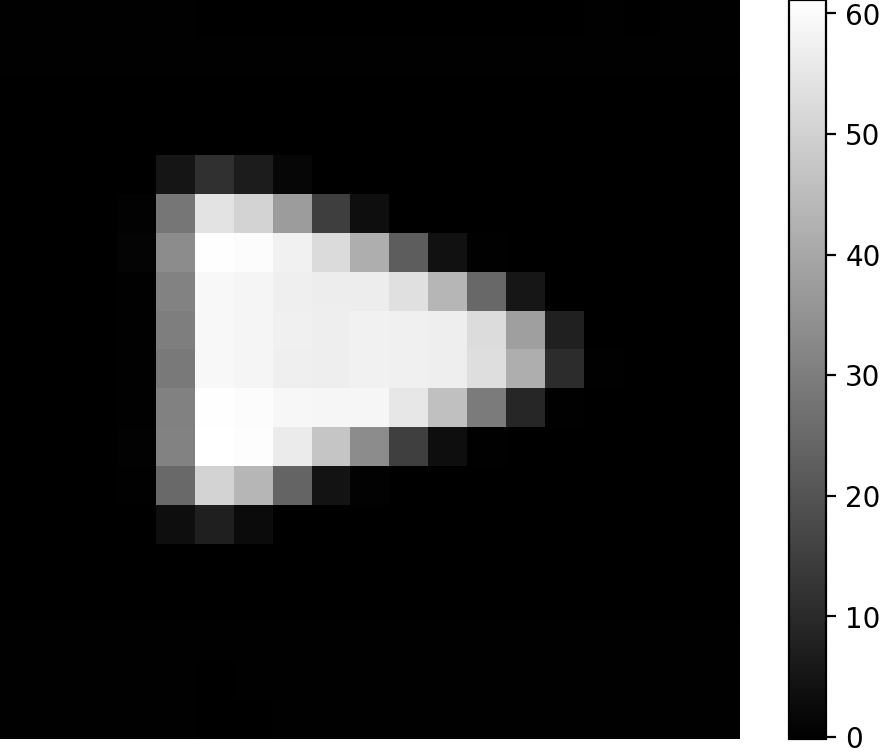}
		\caption{\centering\scriptsize ZeroShot-$\ell^1$-PnP not whitened, rSVD K=6859.}			
		\label{subfig:exp3l1:nowhite:fullrank}
	\end{subfigure}
	\caption{Reconstructions of the shape phantom using the ZeroShot-PnP and ZeroShot-$\ell^1$-PnP methods with increasingly lower levels of preprocessing. To properly compare the results we display the 10-th xz-slice as in Figure~\ref{fig:exp1:shape}.}
	\label{fig:exp3}
\end{figure}
In Figure~\ref{fig:exp3} a selected slice of the reconstructed shape phantom with different levels of preprocessing are displayed.
We observe that the proposed ZeroShot-$\ell^1$-PnP and ZeroShot-PnP methods are rather stable with respect to different preprocessing choices. Further, the results suggest the possibility of employing the proposed methods with less preprocessed data. In the next experiment we go a step further, we consider only the basic preprocessing by backround correcting and suppressing excitation crosstalk and do not rely on the SVD.

\subsection{Experiment 4: Reconstruction With neither SNR Thresholding, Whitening nor SVD.}\label{sec:exp4}

In our fourth and last experiment we study the proposed ZeroShot-$\ell^1$-PnP and ZeroShot-PnP algorithms on the OpenMPI data set with the basic preprocessing consisting of background removal and suppression of excitation crosstalk. In contrast to Experiment 2, we do not rely on any kind of SVD, neither for preprocessing, parameter selection, nor in the algorithmic part (Tikhonov type problem \eqref{eq:PnP:split:data2}). Employing methods not based on an SVD is important for future application scenarios 
since the computation of an SVD will become incrementally cumbersome with increasing problem size. The dimension of the system matrix in this scenario is $151230\times 6859$. In particular, without the SVD we cannot solve the Euler-Lagrange equation~\eqref{eq:PnP:split:data2} directly, so we use the Conjugated Gradient algorithm with a tolerance of $10^{-12}$ and a number of maximum iterations of $10000.$ 

\paragraph{Preprocessing.} We do not perform any SNR thresholding, we do not whiten the data nor perform a rSVD. (We remind that the frequencies below 80 kHz are to be excluded because a hardware filtering is in place to suppress excitation crosstalk and reduces the SNR of frequencies below 80 kHz~\cite{Rahmer_etal2012}, cf. Section~\ref{sec:Preproc}.)

\paragraph{Parameters.} We choose the parameters according to Section~\ref{sec:param:selection} and validate the number of iterations $n_\mathrm{it}$ and the parameter $\mu_0$. For the ZeroShot-PnP method the validated parameters are $n_\mathrm{it}=9$ and $\mu_0 = 10^9$ whereas for the ZeroShot-$\ell^1$-PnP $n_\mathrm{it}=48$ and $\mu_0 = 5\cdot 10^{10}$. The overview of the validated parameters can be found together with the reconstruction metrics in Table \ref{tab:exp4}.

\def\imratio{0.32}
\begin{figure}[t]
	\centering
	\begin{subfigure}[t]{\imratio\linewidth}
		\includegraphics[width=\linewidth]{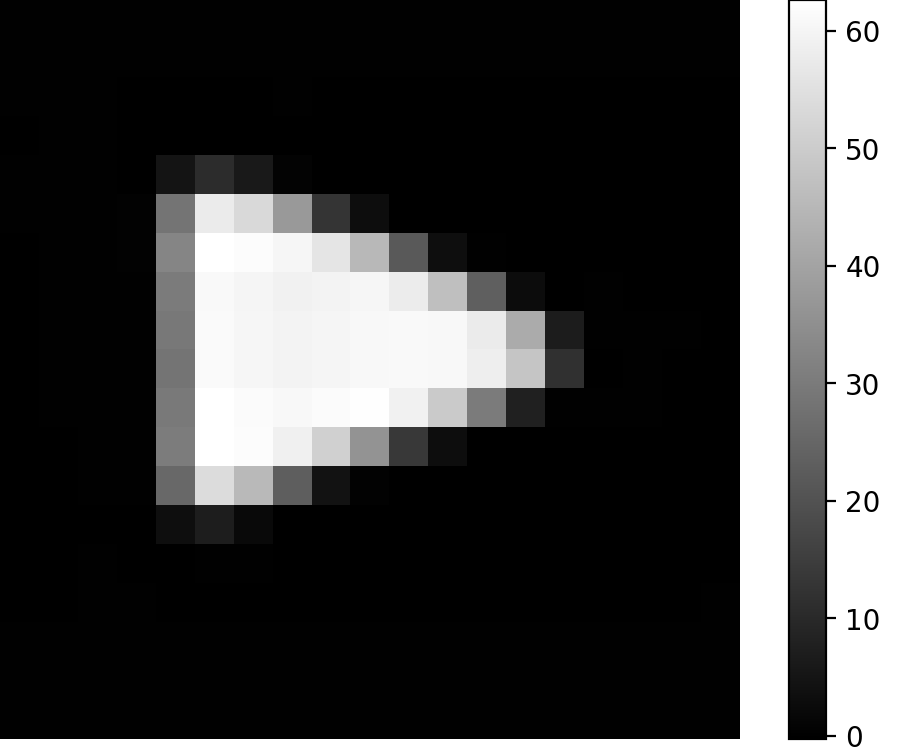}
		\caption{\centering\scriptsize ZeroShot-$\ell^1$-PnP Shape 10-th xz-slice.}
		\label{subfig:exp4:shape:l1}
	\end{subfigure}
	\hfil
	\begin{subfigure}[t]{\imratio\linewidth}
		\includegraphics[width=\linewidth]{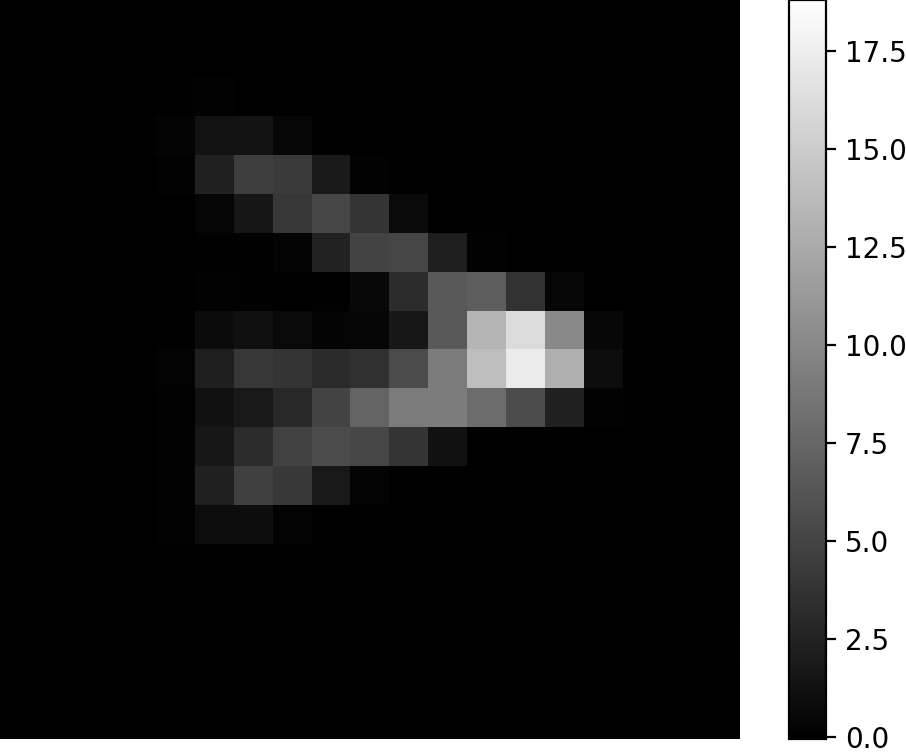}
		\caption{\centering\scriptsize ZeroShot-$\ell^1$-PnP Resolution 10-th xy-slice.}
		\label{subfig:exp4:res:l1}
	\end{subfigure}
	\hfil
	\begin{subfigure}[t]{\imratio\linewidth}
		\includegraphics[width=\linewidth]{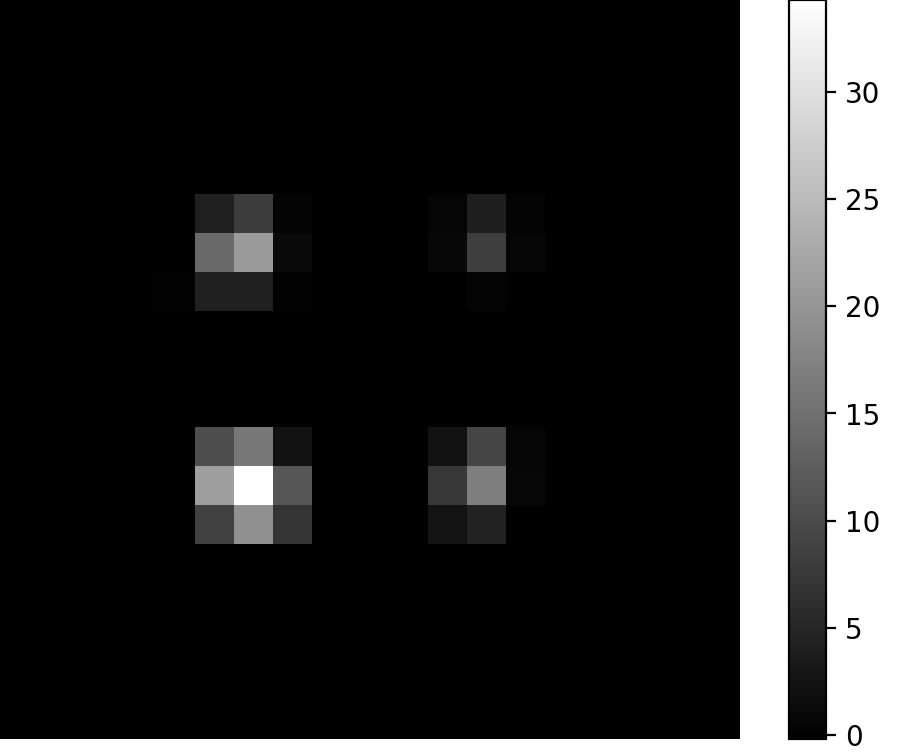}
		\caption{\centering\scriptsize ZeroShot-$\ell^1$-PnP Concentration 13-th xy-slice.}			
		\label{subfig:exp4:conc:l1}
	\end{subfigure}
	\par\medskip
	\begin{subfigure}[t]{\imratio\linewidth}
		\includegraphics[width=\linewidth]{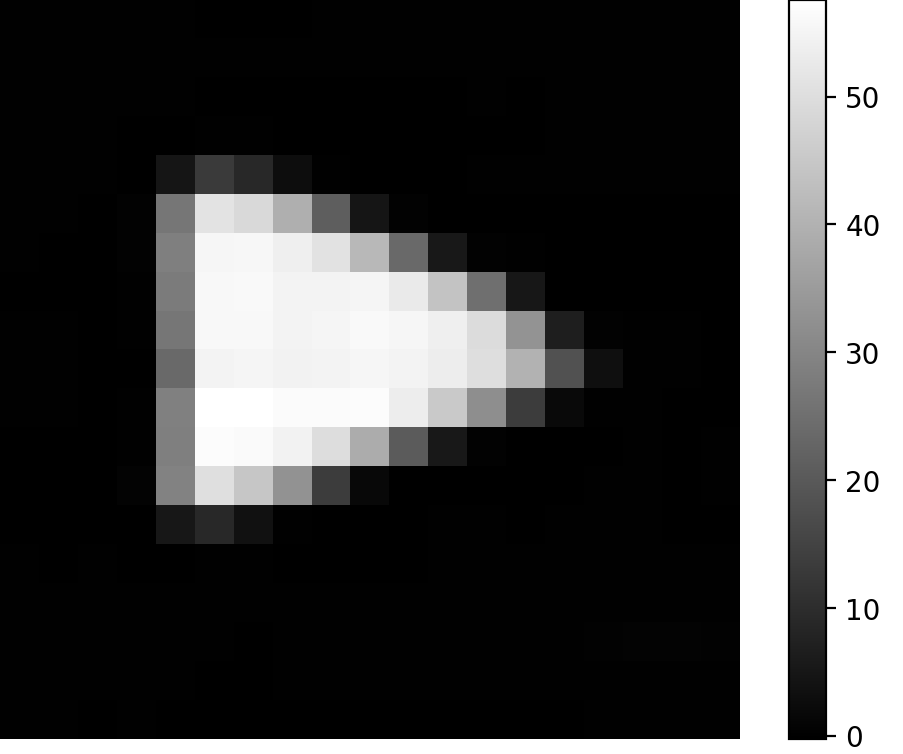}
		\caption{\centering\scriptsize ZeroShot-PnP Shape 10-th xz-slice.}
		\label{subfig:exp4:shape:pnp}
	\end{subfigure}
	\hfil
	\begin{subfigure}[t]{\imratio\linewidth}
		\includegraphics[width=\linewidth]{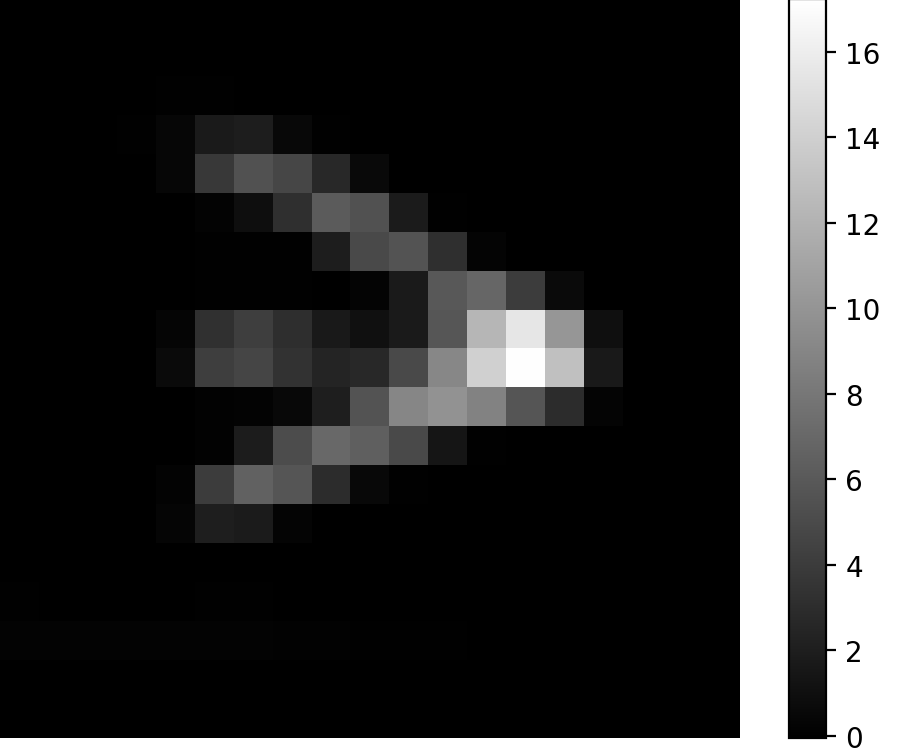}
		\caption{\centering\scriptsize ZeroShot-PnP Resolution 10-th xy-slice.}
		\label{subfig:exp4:res:pnp}
	\end{subfigure}
	\hfil
	\begin{subfigure}[t]{\imratio\linewidth}
		\includegraphics[width=\linewidth]{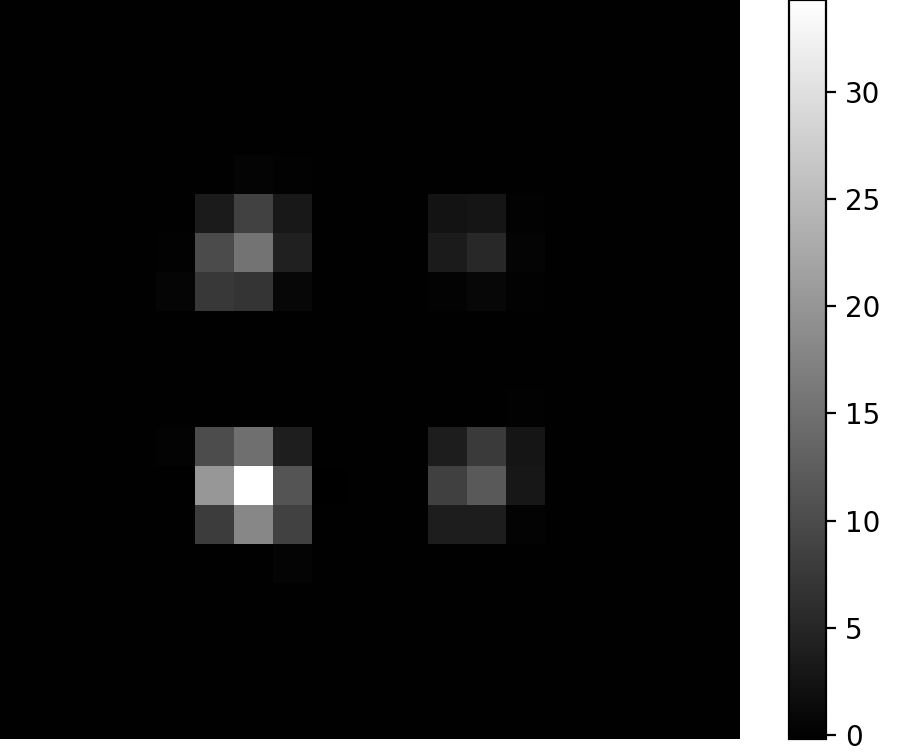}
		\caption{\centering\scriptsize ZeroShot-PnP Concentration 13-th xy-slice.}			
		\label{subfig:exp4:conc:pnp}
	\end{subfigure}
	\caption{Selected slices of the reconstructions obtained using the proposed ZeroShot-$\ell^1$-PnP method (upper row) and with the ZeroShot-PnP method (lower row) with the validated parameters and the basic preprocessing consisting of background correction and cutoff of the frequencies belo 80 kHz to suppress excitation crosstalk).}
	\label{fig:exp4:slices}
\end{figure}
\begin{table}[ht!]
	\footnotesize{
		\begin{center}
			\begin{tabular}{ |c|c|c| }
				\hline
				& ZeroShot-PnP &  ZeroShot-$\ell^1$-PnP\\
				\hline
				Param.& $\mu_0 = 10^9$ & $\mu_0 = 5\cdot 10^{10}$\\
				Iteration& $n_\mathrm{it}=9$ & $n_\mathrm{it}=48$\\
				\hline
				PSNR &  $33.87 \pm 6.22$ & $32.20 \pm 5.08$\\
				\hline
				SSIM &  $0.831 \pm 0.124$ & $0.824 \pm 0.135$\\
				\hline
			\end{tabular}
		\end{center}
		\begin{center}
			\begin{tabular}{ |c|c|c|c|c|  }
				\hline
				Method& Metric & Shape Ph. & Resolution Ph.& Concentration Ph.\\ 
				\hline
				ZeroShot-PnP& $\mathrm{PSNR}_\mathrm{max}$ & 31.16 & 32.06 & 36.97 \\
				& $\mathrm{SSIM}_\mathrm{max}$ & 0.953 & 0.750 & 0.551 \\
				\hline
				ZeroShot-$\ell^1$-PnP& $\mathrm{PSNR}_\mathrm{max}$ & 29.26 & 31.27 & 37.36 \\
				& $\mathrm{SSIM}_\mathrm{max}$ & 0.942 & 0.648 & 0.576 \\
				\hline
			\end{tabular}
		\end{center}
	}
	\caption{Validated parameters and evaluation of the reconstructions  on the Open MPI dataset with only the cut-off of frequencies below 80 kHz as a preprocessing step.}
	\label{tab:exp4}
\end{table}

\paragraph{Results.}
Selected slices for each phantom in the Open MPI Dataset are shown in Figure~\ref{fig:exp4:slices} and the quality evaluation with $\mathrm{PSNR}_\mathrm{max}$ and $\mathrm{SSIM}_\mathrm{max}$ is reported in Table~\ref{tab:exp4}. We observe that the result obtained in this experiments with the minimal preprocessing (background correction and cutoff below 80 kHz) are comparable with the results obtained with the preprocessing and the knowledge of the SVD in Experiment 1 (cf. Table~\ref{tab:exp1:measures}). The reconstruction time takes about 90 seconds, 30 seconds of which are taken to move the full matrix $A$ to the GPU and to form the matrix $A^{T} A$. Finally, the reconstruction with the full matrix with the ZeroShot-PnP (with and without the $\ell^1$-prior) approach here presented lasts about 1 minute including the computation of the quantity $A^{T} f$.

\section{Discussion and Conclusion}
\label{sec:discussionAndconclusion}

\subsection{Discussion}
\label{sec:discussion}

We start out to discuss relations to the paper \cite{askin2022pnp} which is to our knowledge the work closest to ours in the literature. The authors of \cite{askin2022pnp} were the first to use a PnP method for MPI reconstruction.
They train their denoiser on ``MPI-friendly'' MRI data using data of images of vessels,
more precisely, they 
derived their training data from time-of-flight cerebral magnetic resonance angiograms (MRA) to closely mimic the vascular structures targeted in their MPI scans.
In contrast to~\cite{askin2022pnp}, we use a different approach for denoising.
We employ a larger size neural network which has been trained on a larger size data set consisting of natural images.
The employed denoiser is not learned or transfer learned on MPI or similar MRI data. This is particularly appealing due to the very limited amount of (publicly) available MPI data.
The authors of \cite{askin2022pnp} provide a qualitative and quantitative evaluation on simulated data, and show the benefit of their PnP approach by qualitative evaluation on the resolution phantom of the Open MPI dataset. In contrast we provide a quantitative and qualitative evaluation of the proposed ZeroShot-$\ell^1$-PnP and ZeroShot-PnP approach on the full 3D Open MPI dataset.
Moreover, we use a different iterative scheme, namely the HQS, as a basis contrasting the ADMM type approach in \cite{askin2022pnp}. The simplicity of the HQS splitting allows to easily formulate variants of it, including for example the $\ell^1$ norm, which can be substituted with no additional effort by another convex prior and its proximal map. An interesting observation from the results is that the additional $\ell^1$-prior does not play a prominent role in the reconstruction results, suggesting that the employment of the zero-shot denoiser and the parameter selection scheme are the main contributors to the reconstruction quality. Using HQS is appealing also due to relationship between the parameters and the Gaussian noise levels. Additionally, differently form the previous PP-MPI method, we employ a continuation strategy that allows, together with the interpretability of the parameters as Tikhonov regularization parameter and noise level, to produce an automatic parameter update at each iterations that feeds the neural network with the additional estimated noise level of the reconstruction at each iteration. Finally, we validate the parameters on a hybrid validation dataset and apply them to real data  -- in our case the Open MPI Dataset. In particular, for the validation dataset we consider generated phantoms whose maximum concentration are rescaled between 50 mmol and 150 mmol, which is the correct magnitude of the phantoms in the Open MPI dataset. We have observed that this validated dataset produced parameters that produced good quality results also for the resolution phantom, whose maximum concentration lies below 50 mmol. This suggests that a rough estimate of the concentration inside a target phantom seems to be enough to produce a reconstruction parameter choice using the same validation procedure on hybrid dataset but different rescalings.

In the previous PP-MPI approach the denoiser has been trained on MPI-friendly images of size $32\times 32$ pixels and with training data ranging between 0 and $1.5$. This means that, if the complexity of the MPI images to be reconstructed and the level of their details increases, it is reasonable to expect that another \emph{ad hoc} denoiser shall be trained on a new dataset in order to be able to denoise more complicated images. The usage of a zero-shot denoiser pre-trained on a wide range of natural images might enable the employment of the suggested ZeroShot-PnP algorithm with little additional changes. 

One interesting aspect of the proposed method is the automatic parameter selection at each iteration based on the estimation of the noise level and its input into the denoiser. The noise level estimation in this paper has been performed with the simple estimation formula in Eq.\eqref{eq:noiselevel} which corresponds to the noise level of the iterate if the phantom is constant. This is of course not true for most phantoms and the formula in Equation~\eqref{eq:noiselevel} offers only an upper bound for the noise; in this direction, the employment of more sophisticated noise level estimation formulae appear to be effective.

Finally, we discuss relations with the DIP approach.
The DIP approach stands out since, to our knowledge, it is the only machine learning approach for which a qualitative and quantitative evalution has been carried out on the Open MPI dataset \cite{dittmer2020deep}. 
The DIP approach is an unsupervised approach acting on a single image only.  
This is on the one hand appealing since other training data cannot influence the reconstruction 
and on the other hand no training data is needed 
which is particularly appealing in the context of MPI.  
Although not an unsupervised approach, the proposed method shares the property that we need no training data (due to using the denoiser in a zero shot fashion). However, the training data of the denoiser may influence the reconstruction. 
Moreover, the DIP approach suffers from high computational times whereas the ZeroShot-$\ell^1$-PnP can converge in a few seconds for preprocessed data and in about one minute for not preprocessed data. 
Finally, the benefit of using the zero-shot denoiser in the context of MPI is not only suggested by the quality of both the qualitative and quantitative results but also by its flexibility.

\subsection{Conclusion}
\label{sec:conclusion}
In this work, we have developed a PnP approach (with $\ell^1$-prior) for MPI reconstruction using a generic zero-shot denoiser. In particular, we employed a larger size DNN, namely the DRUnet architecture of~\cite{Zhang2022pnp} which has been trained on a larger size data set consisting of natural images. In order to apply the image denoiser on volumetric data, we employ slicing along the axes and averaging~\cite{askin2022pnp}. We have created a hybrid dataset that we use to estimate the reconstruction parameters; then, we applied the estimated parameters on the 3D Open MPI Dataset. We have quantitatively and qualitatively evaluated our approach 
on the full 3D Open MPI data set \cite{knopp2020openmpidata}, 
which presently represents the only publicly available 3D Field-Free Point MPI data set. 
To that end, we have employed suitable preprocessing,  
and we have compared our scheme quantitatively and qualitatively with the DIP approach of \cite{dittmer2020deep} as well as the Tikhonov regularization, the ART method and the previous Plug-and-Play algorithm (PP-MPI) with trained denoiser~\cite{askin2022pnp}. We have shown that the proposed ZeroShot(-$\ell^1$)-PnP algorithmic scheme works well in combination with the ZeroShot-Denoiser~\cite{Zhang2022pnp} because of the ability of the ZeroShot-Denoiser to take into account noise level estimations of the reconstruction at each iteration of the algorithm; in relation to this, we have proposed an automatic parameter update strategy based on noise estimation for the ZeroShot-$\ell^1$-PnP algorithm. Moreover, we have shown that the proposed ZeroShot-($\ell^1$-)PnP approach is able to perform fast reconstructions with different levels of preprocessing. In particular, we show that comparable results can be obtained even only removing frequencies below 80 kHz and in particular without relying on the SVD of the system matrix.

\section*{Acknowledgment}
This research was supported by the Hessian Ministry of Higher Education, Research, Science and the Arts within the Framework of the ``Programm zum Aufbau eines akademischen Mittelbaus an hessischen Hochschulen" and by the German Science Fonds DFG under grant INST 168/4-1.
We would like to thank Tobias Kluth for valuable discussions and support concerning the preprocessing of MPI data and the DIP method.

\bibliographystyle{siam}
\bibliography{literature}

\newpage

\section*{Supplementary Material}

\beginsupplement

\def\imratio{0.15}
\def\phant{shape}
\def\Phant{Shape}

\foreach \ax in {x,y,z} {
	\begin{figure*}[!b]
		\centering
		\foreach \i in {1,...,5} {%
			\begin{subfigure}[t]{\imratio\linewidth}
				\includegraphics[width=\linewidth]{supplementary_material/exp1/\phant/pnp_l1/CAD_gt_\phant_0_\ax slice_\i.png}
				\caption{\footnotesize CAD, slice {\i} along {\ax}.}
			\end{subfigure}
			\begin{subfigure}[t]{\imratio\linewidth}
				\includegraphics[width=\linewidth]{supplementary_material/exp1/\phant/tik/tik_\phant_0_\ax slice_\i.png}
				\caption*{\footnotesize Tikhonov, slice {\i} along {\ax}.}
			\end{subfigure}
			\begin{subfigure}[t]{\imratio\linewidth}
				\includegraphics[width=\linewidth]{supplementary_material/exp1/\phant/dip/dip_\phant_0_\ax slice_\i.png}
				\caption*{\footnotesize DIP, slice {\i} along {\ax}.}
			\end{subfigure}
			\begin{subfigure}[t]{\imratio\linewidth}
				\includegraphics[width=\linewidth]{supplementary_material/exp1/\phant/ppmpi/ppmpi_\phant_0_\ax slice_\i.png}
				\caption*{\footnotesize PP-MPI, slice {\i} along {\ax}.}
			\end{subfigure}
			\begin{subfigure}[t]{\imratio\linewidth}
				\includegraphics[width=\linewidth]{supplementary_material/exp1/\phant/pnp_l1_zero/pnp_l1_\phant_0_\ax slice_\i.png}
				\caption*{\footnotesize ZeroShot-PnP, slice {\i} along {\ax}.}
			\end{subfigure}
			\begin{subfigure}[t]{\imratio\linewidth}
				\includegraphics[width=\linewidth]{supplementary_material/exp1/\phant/pnp_l1/pnp_l1_\phant_0_\ax slice_\i.png}
				\caption*{\footnotesize ZeroShot-$\ell^1$-PnP, slice {\i} along {\ax}.}
			\end{subfigure}
		}
		\caption{{\Phant} phantom in Experiment 1, slices along the {\ax}-axis from 1 to 5.}
	\end{figure*}
	\clearpage
	\begin{figure*}
		\centering
		\foreach \i in {6,...,10} {%
			\begin{subfigure}[t]{\imratio\linewidth}
				\includegraphics[width=\linewidth]{supplementary_material/exp1/\phant/pnp_l1/CAD_gt_\phant_0_\ax slice_\i.png}
				\caption{\footnotesize CAD, slice {\i} along {\ax}.}
			\end{subfigure}
			\begin{subfigure}[t]{\imratio\linewidth}
				\includegraphics[width=\linewidth]{supplementary_material/exp1/\phant/tik/tik_\phant_0_\ax slice_\i.png}
				\caption*{\footnotesize Tikhonov, slice {\i} along {\ax}.}
			\end{subfigure}
			\begin{subfigure}[t]{\imratio\linewidth}
				\includegraphics[width=\linewidth]{supplementary_material/exp1/\phant/dip/dip_\phant_0_\ax slice_\i.png}
				\caption*{\footnotesize DIP, slice {\i} along {\ax}.}
			\end{subfigure}
			\begin{subfigure}[t]{\imratio\linewidth}
				\includegraphics[width=\linewidth]{supplementary_material/exp1/\phant/ppmpi/ppmpi_\phant_0_\ax slice_\i.png}
				\caption*{\footnotesize PP-MPI, slice {\i} along {\ax}.}
			\end{subfigure}
			\begin{subfigure}[t]{\imratio\linewidth}
				\includegraphics[width=\linewidth]{supplementary_material/exp1/\phant/pnp_l1_zero/pnp_l1_\phant_0_\ax slice_\i.png}
				\caption*{\footnotesize ZeroShot-PnP, slice {\i} along {\ax}.}
			\end{subfigure}
			\begin{subfigure}[t]{\imratio\linewidth}
				\includegraphics[width=\linewidth]{supplementary_material/exp1/\phant/pnp_l1/pnp_l1_\phant_0_\ax slice_\i.png}
				\caption*{\footnotesize ZeroShot-$\ell^1$-PnP, slice {\i} along {\ax}.}
			\end{subfigure}
		}
		\caption{{\Phant} phantom in Experiment 1, slices along the {\ax}-axis from 6 to 10.}
	\end{figure*}
	\clearpage
	\begin{figure*}
		\centering
		\foreach \i in {11,...,15} {%
			\begin{subfigure}[t]{\imratio\linewidth}
				\includegraphics[width=\linewidth]{supplementary_material/exp1/\phant/pnp_l1/CAD_gt_\phant_0_\ax slice_\i.png}
				\caption{\footnotesize CAD, slice {\i} along {\ax}.}
			\end{subfigure}
			\begin{subfigure}[t]{\imratio\linewidth}
				\includegraphics[width=\linewidth]{supplementary_material/exp1/\phant/tik/tik_\phant_0_\ax slice_\i.png}
				\caption*{\footnotesize Tikhonov, slice {\i} along {\ax}.}
			\end{subfigure}
			\begin{subfigure}[t]{\imratio\linewidth}
				\includegraphics[width=\linewidth]{supplementary_material/exp1/\phant/dip/dip_\phant_0_\ax slice_\i.png}
				\caption*{\footnotesize DIP, slice {\i} along {\ax}.}
			\end{subfigure}
			\begin{subfigure}[t]{\imratio\linewidth}
				\includegraphics[width=\linewidth]{supplementary_material/exp1/\phant/ppmpi/ppmpi_\phant_0_\ax slice_\i.png}
				\caption*{\footnotesize PP-MPI, slice {\i} along {\ax}.}
			\end{subfigure}
			\begin{subfigure}[t]{\imratio\linewidth}
				\includegraphics[width=\linewidth]{supplementary_material/exp1/\phant/pnp_l1_zero/pnp_l1_\phant_0_\ax slice_\i.png}
				\caption*{\footnotesize ZeroShot-PnP, slice {\i} along {\ax}.}
			\end{subfigure}
			\begin{subfigure}[t]{\imratio\linewidth}
				\includegraphics[width=\linewidth]{supplementary_material/exp1/\phant/pnp_l1/pnp_l1_\phant_0_\ax slice_\i.png}
				\caption*{\footnotesize ZeroShot-$\ell^1$-PnP, slice {\i} along {\ax}.}
			\end{subfigure}
		}
		\caption{{\Phant} phantom in Experiment 1, slices along the {\ax}-axis from 11 to 15.}
	\end{figure*}
	\clearpage
	\begin{figure*}
		\centering
		\foreach \i in {16,...,19} {%
			\begin{subfigure}[t]{\imratio\linewidth}
				\includegraphics[width=\linewidth]{supplementary_material/exp1/\phant/pnp_l1/CAD_gt_\phant_0_\ax slice_\i.png}
				\caption{\footnotesize CAD, slice {\i} along {\ax}.}
			\end{subfigure}
			\begin{subfigure}[t]{\imratio\linewidth}
				\includegraphics[width=\linewidth]{supplementary_material/exp1/\phant/tik/tik_\phant_0_\ax slice_\i.png}
				\caption*{\footnotesize Tikhonov, slice {\i} along {\ax}.}
			\end{subfigure}
			\begin{subfigure}[t]{\imratio\linewidth}
				\includegraphics[width=\linewidth]{supplementary_material/exp1/\phant/dip/dip_\phant_0_\ax slice_\i.png}
				\caption*{\footnotesize DIP, slice {\i} along {\ax}.}
			\end{subfigure}
			\begin{subfigure}[t]{\imratio\linewidth}
				\includegraphics[width=\linewidth]{supplementary_material/exp1/\phant/ppmpi/ppmpi_\phant_0_\ax slice_\i.png}
				\caption*{\footnotesize PP-MPI, slice {\i} along {\ax}.}
			\end{subfigure}
			\begin{subfigure}[t]{\imratio\linewidth}
				\includegraphics[width=\linewidth]{supplementary_material/exp1/\phant/pnp_l1_zero/pnp_l1_\phant_0_\ax slice_\i.png}
				\caption*{\footnotesize ZeroShot-PnP, slice {\i} along {\ax}.}
			\end{subfigure}
			\begin{subfigure}[t]{\imratio\linewidth}
				\includegraphics[width=\linewidth]{supplementary_material/exp1/\phant/pnp_l1/pnp_l1_\phant_0_\ax slice_\i.png}
				\caption*{\footnotesize ZeroShot-$\ell^1$-PnP, slice {\i} along {\ax}.}
			\end{subfigure}
		}
		\caption{{\Phant} phantom in Experiment 1, slices along the {\ax}-axis from 16 to 19.}
	\end{figure*}
}

\def\phant{resolution}
\def\Phant{Resolution}

\foreach \ax in {x,y,z} {
	\begin{figure*}
		\centering
		\foreach \i in {1,...,5} {%
			\begin{subfigure}[t]{\imratio\linewidth}
				\includegraphics[width=\linewidth]{supplementary_material/exp1/\phant/pnp_l1/CAD_gt_\phant_0_\ax slice_\i.png}
				\caption{\footnotesize CAD, slice {\i} along {\ax}.}
			\end{subfigure}
			\begin{subfigure}[t]{\imratio\linewidth}
				\includegraphics[width=\linewidth]{supplementary_material/exp1/\phant/tik/tik_\phant_0_\ax slice_\i.png}
				\caption*{\footnotesize Tikhonov, slice {\i} along {\ax}.}
			\end{subfigure}
			\begin{subfigure}[t]{\imratio\linewidth}
				\includegraphics[width=\linewidth]{supplementary_material/exp1/\phant/dip/dip_\phant_0_\ax slice_\i.png}
				\caption*{\footnotesize DIP, slice {\i} along {\ax}.}
			\end{subfigure}
			\begin{subfigure}[t]{\imratio\linewidth}
				\includegraphics[width=\linewidth]{supplementary_material/exp1/\phant/ppmpi/ppmpi_\phant_0_\ax slice_\i.png}
				\caption*{\footnotesize PP-MPI, slice {\i} along {\ax}.}
			\end{subfigure}
			\begin{subfigure}[t]{\imratio\linewidth}
				\includegraphics[width=\linewidth]{supplementary_material/exp1/\phant/pnp_l1_zero/pnp_l1_\phant_0_\ax slice_\i.png}
				\caption*{\footnotesize ZeroShot-PnP, slice {\i} along {\ax}.}
			\end{subfigure}
			\begin{subfigure}[t]{\imratio\linewidth}
				\includegraphics[width=\linewidth]{supplementary_material/exp1/\phant/pnp_l1/pnp_l1_\phant_0_\ax slice_\i.png}
				\caption*{\footnotesize ZeroShot-$\ell^1$-PnP, slice {\i} along {\ax}.}
			\end{subfigure}
		}
		\caption{{\Phant} phantom in Experiment 1, slices along the {\ax}-axis from 1 to 5.}
	\end{figure*}
	\clearpage
	\begin{figure*}
		\centering
		\foreach \i in {6,...,10} {%
			\begin{subfigure}[t]{\imratio\linewidth}
				\includegraphics[width=\linewidth]{supplementary_material/exp1/\phant/pnp_l1/CAD_gt_\phant_0_\ax slice_\i.png}
				\caption{\footnotesize CAD, slice {\i} along {\ax}.}
			\end{subfigure}
			\begin{subfigure}[t]{\imratio\linewidth}
				\includegraphics[width=\linewidth]{supplementary_material/exp1/\phant/tik/tik_\phant_0_\ax slice_\i.png}
				\caption*{\footnotesize Tikhonov, slice {\i} along {\ax}.}
			\end{subfigure}
			\begin{subfigure}[t]{\imratio\linewidth}
				\includegraphics[width=\linewidth]{supplementary_material/exp1/\phant/dip/dip_\phant_0_\ax slice_\i.png}
				\caption*{\footnotesize DIP, slice {\i} along {\ax}.}
			\end{subfigure}
			\begin{subfigure}[t]{\imratio\linewidth}
				\includegraphics[width=\linewidth]{supplementary_material/exp1/\phant/ppmpi/ppmpi_\phant_0_\ax slice_\i.png}
				\caption*{\footnotesize PP-MPI, slice {\i} along {\ax}.}
			\end{subfigure}
			\begin{subfigure}[t]{\imratio\linewidth}
				\includegraphics[width=\linewidth]{supplementary_material/exp1/\phant/pnp_l1_zero/pnp_l1_\phant_0_\ax slice_\i.png}
				\caption*{\footnotesize ZeroShot-PnP, slice {\i} along {\ax}.}
			\end{subfigure}
			\begin{subfigure}[t]{\imratio\linewidth}
				\includegraphics[width=\linewidth]{supplementary_material/exp1/\phant/pnp_l1/pnp_l1_\phant_0_\ax slice_\i.png}
				\caption*{\footnotesize ZeroShot-$\ell^1$-PnP, slice {\i} along {\ax}.}
			\end{subfigure}
		}
		\caption{{\Phant} phantom in Experiment 1, slices along the {\ax}-axis from 6 to 10.}
	\end{figure*}
	\clearpage
	\begin{figure*}
		\centering
		\foreach \i in {11,...,15} {%
			\begin{subfigure}[t]{\imratio\linewidth}
				\includegraphics[width=\linewidth]{supplementary_material/exp1/\phant/pnp_l1/CAD_gt_\phant_0_\ax slice_\i.png}
				\caption{\footnotesize CAD, slice {\i} along {\ax}.}
			\end{subfigure}
			\begin{subfigure}[t]{\imratio\linewidth}
				\includegraphics[width=\linewidth]{supplementary_material/exp1/\phant/tik/tik_\phant_0_\ax slice_\i.png}
				\caption*{\footnotesize Tikhonov, slice {\i} along {\ax}.}
			\end{subfigure}
			\begin{subfigure}[t]{\imratio\linewidth}
				\includegraphics[width=\linewidth]{supplementary_material/exp1/\phant/dip/dip_\phant_0_\ax slice_\i.png}
				\caption*{\footnotesize DIP, slice {\i} along {\ax}.}
			\end{subfigure}
			\begin{subfigure}[t]{\imratio\linewidth}
				\includegraphics[width=\linewidth]{supplementary_material/exp1/\phant/ppmpi/ppmpi_\phant_0_\ax slice_\i.png}
				\caption*{\footnotesize PP-MPI, slice {\i} along {\ax}.}
			\end{subfigure}
			\begin{subfigure}[t]{\imratio\linewidth}
				\includegraphics[width=\linewidth]{supplementary_material/exp1/\phant/pnp_l1_zero/pnp_l1_\phant_0_\ax slice_\i.png}
				\caption*{\footnotesize ZeroShot-PnP, slice {\i} along {\ax}.}
			\end{subfigure}
			\begin{subfigure}[t]{\imratio\linewidth}
				\includegraphics[width=\linewidth]{supplementary_material/exp1/\phant/pnp_l1/pnp_l1_\phant_0_\ax slice_\i.png}
				\caption*{\footnotesize ZeroShot-$\ell^1$-PnP, slice {\i} along {\ax}.}
			\end{subfigure}
		}
		\caption{{\Phant} phantom in Experiment 1, slices along the {\ax}-axis from 11 to 15.}
	\end{figure*}
	\clearpage
	\begin{figure*}
		\centering
		\foreach \i in {16,...,19} {%
			\begin{subfigure}[t]{\imratio\linewidth}
				\includegraphics[width=\linewidth]{supplementary_material/exp1/\phant/pnp_l1/CAD_gt_\phant_0_\ax slice_\i.png}
				\caption{\footnotesize CAD, slice {\i} along {\ax}.}
			\end{subfigure}
			\begin{subfigure}[t]{\imratio\linewidth}
				\includegraphics[width=\linewidth]{supplementary_material/exp1/\phant/tik/tik_\phant_0_\ax slice_\i.png}
				\caption*{\footnotesize Tikhonov, slice {\i} along {\ax}.}
			\end{subfigure}
			\begin{subfigure}[t]{\imratio\linewidth}
				\includegraphics[width=\linewidth]{supplementary_material/exp1/\phant/dip/dip_\phant_0_\ax slice_\i.png}
				\caption*{\footnotesize DIP, slice {\i} along {\ax}.}
			\end{subfigure}
			\begin{subfigure}[t]{\imratio\linewidth}
				\includegraphics[width=\linewidth]{supplementary_material/exp1/\phant/ppmpi/ppmpi_\phant_0_\ax slice_\i.png}
				\caption*{\footnotesize PP-MPI, slice {\i} along {\ax}.}
			\end{subfigure}
			\begin{subfigure}[t]{\imratio\linewidth}
				\includegraphics[width=\linewidth]{supplementary_material/exp1/\phant/pnp_l1_zero/pnp_l1_\phant_0_\ax slice_\i.png}
				\caption*{\footnotesize ZeroShot-PnP, slice {\i} along {\ax}.}
			\end{subfigure}
			\begin{subfigure}[t]{\imratio\linewidth}
				\includegraphics[width=\linewidth]{supplementary_material/exp1/\phant/pnp_l1/pnp_l1_\phant_0_\ax slice_\i.png}
				\caption*{\footnotesize ZeroShot-$\ell^1$-PnP, slice {\i} along {\ax}.}
			\end{subfigure}
		}
		\caption{{\Phant} phantom in Experiment 1, slices along the {\ax}-axis from 16 to 19.}
	\end{figure*}
}

\def\phant{concentration}
\def\Phant{Concentration}

\foreach \ax in {x,y,z} {
	\begin{figure*}
		\centering
		\foreach \i in {1,...,5} {%
			\begin{subfigure}[t]{\imratio\linewidth}
				\includegraphics[width=\linewidth]{supplementary_material/exp1/\phant/pnp_l1/CAD_gt_\phant_0_\ax slice_\i.png}
				\caption{\footnotesize CAD, slice {\i} along {\ax}.}
			\end{subfigure}
			\begin{subfigure}[t]{\imratio\linewidth}
				\includegraphics[width=\linewidth]{supplementary_material/exp1/\phant/tik/tik_\phant_0_\ax slice_\i.png}
				\caption*{\footnotesize Tikhonov, slice {\i} along {\ax}.}
			\end{subfigure}
			\begin{subfigure}[t]{\imratio\linewidth}
				\includegraphics[width=\linewidth]{supplementary_material/exp1/\phant/dip/dip_\phant_0_\ax slice_\i.png}
				\caption*{\footnotesize DIP, slice {\i} along {\ax}.}
			\end{subfigure}
			\begin{subfigure}[t]{\imratio\linewidth}
				\includegraphics[width=\linewidth]{supplementary_material/exp1/\phant/ppmpi/ppmpi_\phant_0_\ax slice_\i.png}
				\caption*{\footnotesize PP-MPI, slice {\i} along {\ax}.}
			\end{subfigure}
			\begin{subfigure}[t]{\imratio\linewidth}
				\includegraphics[width=\linewidth]{supplementary_material/exp1/\phant/pnp_l1_zero/pnp_l1_\phant_0_\ax slice_\i.png}
				\caption*{\footnotesize ZeroShot-PnP, slice {\i} along {\ax}.}
			\end{subfigure}
			\begin{subfigure}[t]{\imratio\linewidth}
				\includegraphics[width=\linewidth]{supplementary_material/exp1/\phant/pnp_l1/pnp_l1_\phant_0_\ax slice_\i.png}
				\caption*{\footnotesize ZeroShot-$\ell^1$-PnP, slice {\i} along {\ax}.}
			\end{subfigure}
		}
		\caption{{\Phant} phantom in Experiment 1, slices along the {\ax}-axis from 1 to 5.}
	\end{figure*}
	\clearpage
	\begin{figure*}
		\centering
		\foreach \i in {6,...,10} {%
			\begin{subfigure}[t]{\imratio\linewidth}
				\includegraphics[width=\linewidth]{supplementary_material/exp1/\phant/pnp_l1/CAD_gt_\phant_0_\ax slice_\i.png}
				\caption{\footnotesize CAD, slice {\i} along {\ax}.}
			\end{subfigure}
			\begin{subfigure}[t]{\imratio\linewidth}
				\includegraphics[width=\linewidth]{supplementary_material/exp1/\phant/tik/tik_\phant_0_\ax slice_\i.png}
				\caption*{\footnotesize Tikhonov, slice {\i} along {\ax}.}
			\end{subfigure}
			\begin{subfigure}[t]{\imratio\linewidth}
				\includegraphics[width=\linewidth]{supplementary_material/exp1/\phant/dip/dip_\phant_0_\ax slice_\i.png}
				\caption*{\footnotesize DIP, slice {\i} along {\ax}.}
			\end{subfigure}
			\begin{subfigure}[t]{\imratio\linewidth}
				\includegraphics[width=\linewidth]{supplementary_material/exp1/\phant/ppmpi/ppmpi_\phant_0_\ax slice_\i.png}
				\caption*{\footnotesize PP-MPI, slice {\i} along {\ax}.}
			\end{subfigure}
			\begin{subfigure}[t]{\imratio\linewidth}
				\includegraphics[width=\linewidth]{supplementary_material/exp1/\phant/pnp_l1_zero/pnp_l1_\phant_0_\ax slice_\i.png}
				\caption*{\footnotesize ZeroShot-PnP, slice {\i} along {\ax}.}
			\end{subfigure}
			\begin{subfigure}[t]{\imratio\linewidth}
				\includegraphics[width=\linewidth]{supplementary_material/exp1/\phant/pnp_l1/pnp_l1_\phant_0_\ax slice_\i.png}
				\caption*{\footnotesize ZeroShot-$\ell^1$-PnP, slice {\i} along {\ax}.}
			\end{subfigure}
		}
		\caption{{\Phant} phantom in Experiment 1, slices along the {\ax}-axis from 6 to 10.}
	\end{figure*}
	\clearpage
	\begin{figure*}
		\centering
		\foreach \i in {11,...,15} {%
			\begin{subfigure}[t]{\imratio\linewidth}
				\includegraphics[width=\linewidth]{supplementary_material/exp1/\phant/pnp_l1/CAD_gt_\phant_0_\ax slice_\i.png}
				\caption{\footnotesize CAD, slice {\i} along {\ax}.}
			\end{subfigure}
			\begin{subfigure}[t]{\imratio\linewidth}
				\includegraphics[width=\linewidth]{supplementary_material/exp1/\phant/tik/tik_\phant_0_\ax slice_\i.png}
				\caption*{\footnotesize Tikhonov, slice {\i} along {\ax}.}
			\end{subfigure}
			\begin{subfigure}[t]{\imratio\linewidth}
				\includegraphics[width=\linewidth]{supplementary_material/exp1/\phant/dip/dip_\phant_0_\ax slice_\i.png}
				\caption*{\footnotesize DIP, slice {\i} along {\ax}.}
			\end{subfigure}
			\begin{subfigure}[t]{\imratio\linewidth}
				\includegraphics[width=\linewidth]{supplementary_material/exp1/\phant/ppmpi/ppmpi_\phant_0_\ax slice_\i.png}
				\caption*{\footnotesize PP-MPI, slice {\i} along {\ax}.}
			\end{subfigure}
			\begin{subfigure}[t]{\imratio\linewidth}
				\includegraphics[width=\linewidth]{supplementary_material/exp1/\phant/pnp_l1_zero/pnp_l1_\phant_0_\ax slice_\i.png}
				\caption*{\footnotesize ZeroShot-PnP, slice {\i} along {\ax}.}
			\end{subfigure}
			\begin{subfigure}[t]{\imratio\linewidth}
				\includegraphics[width=\linewidth]{supplementary_material/exp1/\phant/pnp_l1/pnp_l1_\phant_0_\ax slice_\i.png}
				\caption*{\footnotesize ZeroShot-$\ell^1$-PnP, slice {\i} along {\ax}.}
			\end{subfigure}
		}
		\caption{{\Phant} phantom in Experiment 1, slices along the {\ax}-axis from 11 to 15.}
	\end{figure*}
	\clearpage
	\begin{figure*}
		\centering
		\foreach \i in {16,...,19} {%
			\begin{subfigure}[t]{\imratio\linewidth}
				\includegraphics[width=\linewidth]{supplementary_material/exp1/\phant/pnp_l1/CAD_gt_\phant_0_\ax slice_\i.png}
				\caption{\footnotesize CAD, slice {\i} along {\ax}.}
			\end{subfigure}
			\begin{subfigure}[t]{\imratio\linewidth}
				\includegraphics[width=\linewidth]{supplementary_material/exp1/\phant/tik/tik_\phant_0_\ax slice_\i.png}
				\caption*{\footnotesize Tikhonov, slice {\i} along {\ax}.}
			\end{subfigure}
			\begin{subfigure}[t]{\imratio\linewidth}
				\includegraphics[width=\linewidth]{supplementary_material/exp1/\phant/dip/dip_\phant_0_\ax slice_\i.png}
				\caption*{\footnotesize DIP, slice {\i} along {\ax}.}
			\end{subfigure}
			\begin{subfigure}[t]{\imratio\linewidth}
				\includegraphics[width=\linewidth]{supplementary_material/exp1/\phant/ppmpi/ppmpi_\phant_0_\ax slice_\i.png}
				\caption*{\footnotesize PP-MPI, slice {\i} along {\ax}.}
			\end{subfigure}
			\begin{subfigure}[t]{\imratio\linewidth}
				\includegraphics[width=\linewidth]{supplementary_material/exp1/\phant/pnp_l1_zero/pnp_l1_\phant_0_\ax slice_\i.png}
				\caption*{\footnotesize ZeroShot-PnP, slice {\i} along {\ax}.}
			\end{subfigure}
			\begin{subfigure}[t]{\imratio\linewidth}
				\includegraphics[width=\linewidth]{supplementary_material/exp1/\phant/pnp_l1/pnp_l1_\phant_0_\ax slice_\i.png}
				\caption*{\footnotesize ZeroShot-$\ell^1$-PnP, slice {\i} along {\ax}.}
			\end{subfigure}
		}
		\caption{{\Phant} phantom in Experiment 1, slices along the {\ax}-axis from 16 to 19.}
	\end{figure*}
}

\clearpage

\def\imratio{0.15}
\def\phant{shape}
\def\Phant{Shape}

\foreach \ax in {x,y,z} {
	\begin{figure*}
		\centering
		\foreach \i in {1,...,10} {%
			\begin{subfigure}[t]{\imratio\linewidth}
				\includegraphics[width=\linewidth]{supplementary_material/exp3/\phant_l1/CAD_gt_\phant_0_\ax slice_\i.png}
				\caption{\footnotesize CAD, slice {\i} along {\ax}.}
			\end{subfigure}
			\begin{subfigure}[t]{\imratio\linewidth}
				\includegraphics[width=\linewidth]{supplementary_material/exp3/\phant_l1/pnp_l1_\phant_0_\ax slice_\i.png}
				\caption*{\footnotesize ZeroShot-$\ell^1$-PnP, slice {\i} along {\ax}.}
			\end{subfigure}
			\begin{subfigure}[t]{\imratio\linewidth}
				\includegraphics[width=\linewidth]{supplementary_material/exp3/\phant_zero/pnp_l1_\phant_0_\ax slice_\i.png}
				\caption*{\footnotesize ZeroShot-PnP, slice {\i} along {\ax}.}
			\end{subfigure}
		}
		\caption{{\Phant} phantom in Experiment 4, minimum level of preprocessing, no SVD, slices along the {\ax}-axis from 1 to 10.}
	\end{figure*}
	
	\begin{figure*}
		\foreach \i in {11,...,19} {%
			\begin{subfigure}[t]{\imratio\linewidth}
				\includegraphics[width=\linewidth]{supplementary_material/exp3/\phant_l1/CAD_gt_\phant_0_\ax slice_\i.png}
				\caption{\footnotesize CAD, slice {\i} along {\ax}.}
			\end{subfigure}
			\begin{subfigure}[t]{\imratio\linewidth}
				\includegraphics[width=\linewidth]{supplementary_material/exp3/\phant_l1/pnp_l1_\phant_0_\ax slice_\i.png}
				\caption*{\footnotesize ZeroShot-$\ell^1$-PnP, slice {\i} along {\ax}.}
			\end{subfigure}
			\begin{subfigure}[t]{\imratio\linewidth}
				\includegraphics[width=\linewidth]{supplementary_material/exp3/\phant_zero/pnp_l1_\phant_0_\ax slice_\i.png}
				\caption*{\footnotesize ZeroShot-PnP, slice {\i} along {\ax}.}
			\end{subfigure}
		}
		\caption{{\Phant} phantom in Experiment 4, minimum level of preprocessing, no SVD, slices along the {\ax}-axis from 11 to 19.}
	\end{figure*}
}

\def\phant{resolution}
\def\Phant{Resolution}

\foreach \ax in {x,y,z} {
	\begin{figure*}
		\centering
		\foreach \i in {1,...,10} {%
			\begin{subfigure}[t]{\imratio\linewidth}
				\includegraphics[width=\linewidth]{supplementary_material/exp3/\phant_l1/CAD_gt_\phant_0_\ax slice_\i.png}
				\caption{\footnotesize CAD, slice {\i} along {\ax}.}
			\end{subfigure}
			\begin{subfigure}[t]{\imratio\linewidth}
				\includegraphics[width=\linewidth]{supplementary_material/exp3/\phant_l1/pnp_l1_\phant_0_\ax slice_\i.png}
				\caption*{\footnotesize ZeroShot-$\ell^1$-PnP, slice {\i} along {\ax}.}
			\end{subfigure}
			\begin{subfigure}[t]{\imratio\linewidth}
				\includegraphics[width=\linewidth]{supplementary_material/exp3/\phant_zero/pnp_l1_\phant_0_\ax slice_\i.png}
				\caption*{\footnotesize ZeroShot-PnP, slice {\i} along {\ax}.}
			\end{subfigure}
		}
		\caption{{\Phant} phantom in Experiment 4, minimum level of preprocessing, no SVD, slices along the {\ax}-axis from 1 to 10.}
	\end{figure*}
	
	\begin{figure*}
		\foreach \i in {11,...,19} {%
			\begin{subfigure}[t]{\imratio\linewidth}
				\includegraphics[width=\linewidth]{supplementary_material/exp3/\phant_l1/CAD_gt_\phant_0_\ax slice_\i.png}
				\caption{\footnotesize CAD, slice {\i} along {\ax}.}
			\end{subfigure}
			\begin{subfigure}[t]{\imratio\linewidth}
				\includegraphics[width=\linewidth]{supplementary_material/exp3/\phant_l1/pnp_l1_\phant_0_\ax slice_\i.png}
				\caption*{\footnotesize ZeroShot-$\ell^1$-PnP, slice {\i} along {\ax}.}
			\end{subfigure}
			\begin{subfigure}[t]{\imratio\linewidth}
				\includegraphics[width=\linewidth]{supplementary_material/exp3/\phant_zero/pnp_l1_\phant_0_\ax slice_\i.png}
				\caption*{\footnotesize ZeroShot-PnP, slice {\i} along {\ax}.}
			\end{subfigure}
		}
		\caption{{\Phant} phantom in Experiment 4, minimum level of preprocessing, no SVD, slices along the {\ax}-axis from 11 to 19.}
	\end{figure*}
}

\def\phant{concentration}
\def\Phant{Concentration}

\foreach \ax in {x,y,z} {
	\begin{figure*}
		\centering
		\foreach \i in {1,...,10} {%
			\begin{subfigure}[t]{\imratio\linewidth}
				\includegraphics[width=\linewidth]{supplementary_material/exp3/\phant_l1/CAD_gt_\phant_0_\ax slice_\i.png}
				\caption{\footnotesize CAD, slice {\i} along {\ax}.}
			\end{subfigure}
			\begin{subfigure}[t]{\imratio\linewidth}
				\includegraphics[width=\linewidth]{supplementary_material/exp3/\phant_l1/pnp_l1_\phant_0_\ax slice_\i.png}
				\caption*{\footnotesize ZeroShot-$\ell^1$-PnP, slice {\i} along {\ax}.}
			\end{subfigure}
			\begin{subfigure}[t]{\imratio\linewidth}
				\includegraphics[width=\linewidth]{supplementary_material/exp3/\phant_zero/pnp_l1_\phant_0_\ax slice_\i.png}
				\caption*{\footnotesize ZeroShot-PnP, slice {\i} along {\ax}.}
			\end{subfigure}
		}
		\caption{{\Phant} phantom in Experiment 4, minimum level of preprocessing, no SVD, slices along the {\ax}-axis from 1 to 10.}
	\end{figure*}
	
	\begin{figure*}
		\foreach \i in {11,...,19} {%
			\begin{subfigure}[t]{\imratio\linewidth}
				\includegraphics[width=\linewidth]{supplementary_material/exp3/\phant_l1/CAD_gt_\phant_0_\ax slice_\i.png}
				\caption{\footnotesize CAD, slice {\i} along {\ax}.}
			\end{subfigure}
			\begin{subfigure}[t]{\imratio\linewidth}
				\includegraphics[width=\linewidth]{supplementary_material/exp3/\phant_l1/pnp_l1_\phant_0_\ax slice_\i.png}
				\caption*{\footnotesize ZeroShot-$\ell^1$-PnP, slice {\i} along {\ax}.}
			\end{subfigure}
			\begin{subfigure}[t]{\imratio\linewidth}
				\includegraphics[width=\linewidth]{supplementary_material/exp3/\phant_zero/pnp_l1_\phant_0_\ax slice_\i.png}
				\caption*{\footnotesize ZeroShot-PnP, slice {\i} along {\ax}.}
			\end{subfigure}
		}
		\caption{{\Phant} phantom in Experiment 4, minimum level of preprocessing, no SVD, slices along the {\ax}-axis from 11 to 19.}
	\end{figure*}
}

\end{document}